\documentclass[12pt,a4paper]{report}
\usepackage{geometry}
\usepackage{setspace}

\usepackage[utf8]{inputenc}
\usepackage{amssymb}
\usepackage{amsmath}
\usepackage{amsfonts}
\usepackage{amsthm}
\usepackage{graphicx}
\graphicspath{ {figures/} }
\usepackage{array}
\usepackage{dsfont}
\usepackage{epstopdf}
\usepackage{tkz-euclide}
\usepackage{bbold}
\usepackage{placeins}
\usepackage{color}
\usepackage{float}

\usepackage{tikz}
\usetikzlibrary{arrows.meta}
\usetikzlibrary{bending}
\usepackage{dynkin-diagrams}
\usepackage[vcentermath]{youngtab}
\usepackage{ytableau}

\newcommand{\beq}{\begin{equation}}
	\newcommand{\eeq}{\end{equation}}

\begin{document}

\thispagestyle{empty}
\def\thefootnote{\fnsymbol{footnote}}
\vspace*{0.5cm}
\begin{center}
{\large Alikhanyan National Science Laboratory \\
(Yerevan Physics Institute)}\\
\vspace*{5cm}
{\bf\Large Mane Avetisyan}\\
\vspace{1cm}
{\bf\Large Vogel's Universality and its Applications }\\
\vspace{1cm}
{\Large Ph.D. Thesis 
}\\
\vspace{1cm}
{\large  Supervisor: Ruben Mkrtchyan}\\
\vspace{5cm}
{\large Yerevan-2022}
\end{center}
\newpage

\tableofcontents

\newpage

\thispagestyle{empty}
\listoffigures
\listoftables

\newpage

\section*{Acknowledgements}

The philosopher Simone Weil, the sister of famous André Weil, said, {\it “Attention is the rarest and purest form of generosity.”}

I am grateful to everyone who has shown any bit of attention to the work presented in this thesis.

\newpage
\section*{Abstract}

The present thesis represents developments in two main directions related to the simple Lie algebras. 
The first one is devoted to the representation theory of the simple Lie algebras. Specifically we present
recent results, which include new universal formulae in Vogel's universal description, as well as the discovery of additional properties of those formulae.
In the second part of the thesis we demonstrate applications of Vogel's description to the study of a physical theory.
 Namely, we explicitly formulate the { \it refined} Chern-Simons theories on $S^3$ for each of the simple gauge groups, including the exceptional ones.
\vspace{0.5 cm}

{\bf Relevance of the scientific research.} Vogel's universal approach to simple Lie algebras is a powerful and attractive tool both for 
mathematicians and theoretical physicists. First of all, it allows unifying innately discrete objects such as different simple Lie algebras into analytical functions defined in Vogel’s plane.
This is indeed a remarkable phenomenon in science. On the other hand, the possibility of treating different algebras on an equal footing 
provides a new possibility for physicists to work with the gauge theories built upon all simple gauge groups. These arguments motivate the
relevance of developing Vogel's approach and investigating its applications to physical gauge theories. 
\vspace{0.5 cm}

{\bf Purpose of the work.} One of the aims of this work is the deeper understanding of Vogel's universal description of simple Lie algebras. Another
one is opening a new door to the possibility of setting up a duality between the refined Chern-Simons theories on $S^3$ built upon the exceptional gauge algebras 
and some (refined) topological strings living on specific Calabi-Yau manifolds.
\vspace{0.5 cm}

{\bf The novelty of the work.} The research presented develops Vogel's universal approach to simple Lie algebras by expanding the
list of universal representations which has remained unchanged since 2005. It also presents an explicit expression for the partition function
of the refined Chern-Simons on $S^3$ for all simple gauge groups.
\vspace{0.5 cm}

{\bf Results submitted for defense:} 

1. Derivation of universal dimension and quantum dimension formulae for Cartan products of arbitrary powers of the adjoint $g$ and
$X_2$ representations ($X_2^k g^n$, $k,n \in Z_+$) of the simple Lie algebras. Study of these formulae under permutations of universal parameters and
demonstration that in their stable limits the outputs are quantum dimensions of some representations of the corresponding algebras. 

2. Definition of the {\it linear resolvability} feature of the universal formulae. 
Proof that the all known quantum dimension formulae are linearly resolvable.

3. Derivation of universal eigenvalues of the second Casimir operator on the Cartan products of arbitrary powers of the adjoint $g$ and $X_2$
 representations.
 
4. Geometrical interpretation of the universal formulae.
Establishment of correspondence between non-uniqueness factors of universal formulae and geometrical configurations of
points and lines. Derivation of a four-by-four non-uniqueness factor using this correspondence. 

5. Refinement of the Kac-Peterson identity for the determinant of the symmetrized Cartan matrix. Derivation of an explicit formula for the partition functions of the 
refined Chern-Simons theory on $S^3$ with an arbitrary simple gauge group.

6. Universal-like representation of all these partition functions of the refined Chern-Simons theory on $S^3$ with an arbitrary simple gauge group. This representation
aims at a further check of possible Chern-Simons/topological strings dualities for all gauge groups.
 
 \vspace{0.5 cm}

{\bf The current work is based on the following articles:}

1. M.Y. Avetisyan and R.L. Mkrtchyan, 
$X_2$ Series of Universal Quantum Dimensions, arXiv:1812.07914, 
J. Phys. A: Math. Theor. Volume 53, Number 4, 045202, (2020) 

doi:10.1088/1751-8121/ab5f4d

2. M.Y. Avetisyan and R.L. Mkrtchyan, 
On $(ad)^n (X2)^k$ series of universal quantum dimensions,
arXiv:1909.02076, 
J. Math. Phys. 61, 101701 (2020)

doi:10.1063/5.0007028
 
3. M. Y. Avetisyan, 
On universal eigenvalues of the Casimir operator,
arXiv:1908.08794,
Phys. Part. Nucl., Lett. 17(5), pp 779-783 (2020)

doi:10.1134/S1547477120050039

4. M.Y.Avetisyan and R.L.Mkrtchyan, 
Universality and Quantum Dimensions,
Phys. Part. Nucl., Lett. 17(5), pp784-788 (2020),

doi:10.1134/S1547477120050040

5. M.Y. Avetisyan, 
Universal dimensions of simple Lie algebras and configurations of points and lines,
Proceedings of Science, Vol 394, (2021)

doi:10.22323/1.394.0005

6. M.Y. Avetisyan and R.L. Mkrtchyan, 
On partition functions of refined Chern-Simons theories on $S^3$,
arxiv:2107.08679, JHEP 10 (2021) 033, 

https://doi.org/10.1007/JHEP10(2021)033

7. M.Y. Avetisyan and R.L. Mkrtchyan,
On linear resolvability of universal quantum dimensions,
Journal of Knot Theory and its Ramifications, Vol. 31, No. 2 (2022) 2250014,

https://doi.org/10.1142/S0218216522500146

8\footnote{The results presented in this paper are not submitted for defence for timing reasons.}. M.Y. Avetisyan and R.L. Mkrtchyan, 
Uniqueness of universal dimensions and configurations of points and lines,
arxiv:2101.10860v3, Geometriae Dedicata, (2022) 216:41, 

https://doi.org/10.1007/s10711-022-00699-2

\vspace{0.5 cm}

This thesis is organized as follows:

Chapter 1 is introductory notions describing Vogel's universality and its state-of-the-art.

Chapter 2 is devoted to the presentation of the new universal formulae, derived in the scope of the representation theory of the simple Lie algebras.

Chapter 3 focuses on the revelation of a non-trivial property of our universal formulae, which we call "linear resolvability", and provides the proof that all known universal
quantum dimensions are { \it linearly resolvable}.

Chapter 4 presents the establishment of a connection between simple Lie algebras and geometrical configurations of points and lines, by proposing a problem
of the uniqueness of the universal formulae describing the representations of the algebras.

Chapter 5 addresses the applications of Vogel's universality to physical problems and presents an explicit expression
for the partition function of the refined Chern-Simons theory on $S^3$.

Chapter 6 is the summary of the work and discusses the possible directions of research springing out of it.

\newpage

\chapter{Introduction}

\subsection{On Vogel’s universal approach to the simple Lie algebras}
The universal description of the simple Lie algebras was first introduced by P. Vogel in his Universal Lie Algebra \cite{V0,V}. 
He was aiming at a derivation of the most general weight system for Vassiliev's finite knot invariants. For some unpredicted difficulties 
this project in fact was not a success. However, a uniform parameterization of the simple Lie algebras appeared as a byproduct of it, (see Table \ref{tab:V1} and Table \ref{tab:V2}).
 
 \begin{table}[ht]

\caption{Vogel's parameters for simple Lie algebras} \label{tab:V1}
 \centering
\begin{tabular}{|c|c|c|c|c|c|}
\hline
Root system & Lie algebra  & $\alpha$ & $\beta$ & $\gamma$  & $t=h^\vee$\\   
\hline    
$A_n$ &  $\mathfrak {sl}_{n+1}$     & $-2$ & 2 & $(n+1) $ & $n+1$\\
$B_n$ &   $\mathfrak {so}_{2n+1}$    & $-2$ & 4& $2n-3 $ & $2n-1$\\
$C_n$ & $ \mathfrak {sp}_{2n}$    & $-2$ & 1 & $n+2 $ & $n+1$\\
$D_n$ &   $\mathfrak {so}_{2n}$    & $-2$ & 4 & $2n-4$ & $2n-2$\\
$G_2$ &  $\mathfrak {g}_{2}  $    & $-2$ & $10/3 $& $8/3$ & $4$ \\
$F_4$ & $\mathfrak {f}_{4}  $    & $-2$ & $ 5$& $ 6$ & $9$\\
$E_6$ &  $\mathfrak {e}_{6}  $    & $-2$ & $ 6$& $ 8$ & $12$\\
$E_7$ & $\mathfrak {e}_{7}  $    & $-2$ & $ 8$& $ 12$ & $18$ \\
$E_8$ & $\mathfrak {e}_{8}  $    & $-2$ & $ 12$& $20$ & $30$\\
\hline  
\end{tabular}
\end{table}

 \begin{table}[ht] \label{tab:V2}
	\caption{Vogel's parameters and distinguished lines}
	\centering
	\begin{tabular}{|r|r|r|r|r|r|} 
		\hline Algebra/Parameters & $\alpha$ &$\beta$  &$\gamma$  & $t$ & Line \\ 
		\hline  $\mathfrak{sl}_N$  & $-2$ & 2 & $N$ & $N$ & $\alpha+\beta=0, sl$ \\ 
		\hline $\mathfrak{so}_N$ & $-2$  & 4 & $N-4$ & $N-2$ & $ 2\alpha+\beta=0, so$ \\ 
		\hline  $ \mathfrak{sp}_N$ & $-2$  & 1 & $N/2+2$ & $N/2+1$ & $ \alpha +2\beta=0,sp$ \\ 
		\hline $exc(n)$ & $-2$ &  $n+4$ &  $2n+4$& $3n+6$ & $\gamma=2(\alpha+\beta), exc$\\ 
		\hline 
	\end{tabular}
	
	{For the exceptional 
		line $n=-2/3,0,1,2,4,8$ for $\mathfrak {g}_{2}, \mathfrak {so}_{8}, \mathfrak{f}_{4}, \mathfrak{e}_{6}, \mathfrak {e}_{7},\mathfrak {e}_{8} $, 
		respectively.} \label{tab:V2}
\end{table}

To give an idea of the origin of these tables we write the following universal (i.e. valid for any simple Lie algebra) decomposition of the symmetric square of the adjoint representation \cite{V0}:

\begin{eqnarray}\label{sad}
S^2 \mathfrak{g}=1 \oplus Y_2(\alpha) \oplus Y_2(\beta) \oplus Y_2(\gamma)
\end{eqnarray}

Let $2t$ denote the eigenvalue of the second Casimir operator on the adjoint representation $\mathfrak{g}$ and the eigenvalues of the same operator on representations in 
(\ref{sad}) be $4t-2\alpha, 4t-2\beta, 4t-2\gamma$, correspondingly. In this way we define $\alpha, \beta, \gamma$ (Vogel's) parameters.
 It can be proved \cite{V0} that with these definitions  $\alpha+\beta+\gamma=t$.
 
\begin{figure}[htp]
    \centering
    \includegraphics[width=15cm]{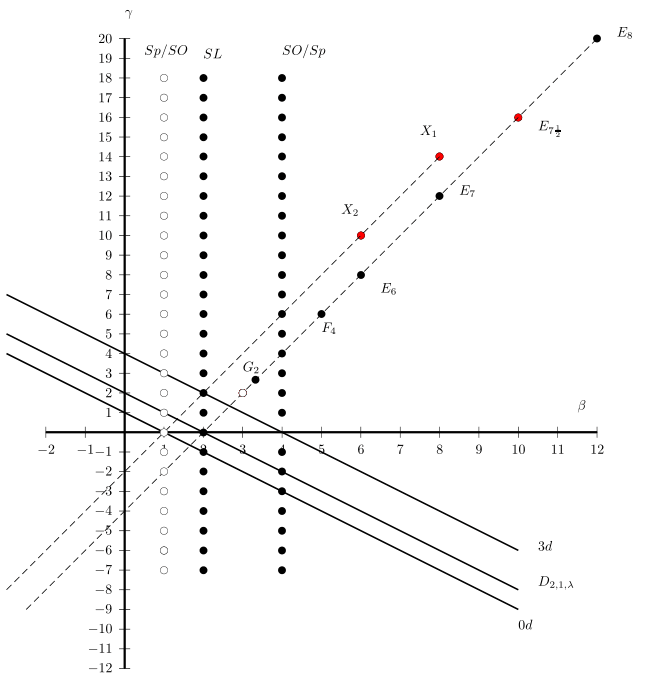}
    \caption{Vogel's plane}
    \label{fig:Vplane}
\end{figure}

According to the definitions, the entire theory is invariant with respect to a rescaling of the parameters (which corresponds to the rescaling of the invariant scalar product in algebra), 
and with respect to the permutation of the universal (or, Vogel's) parameters  $\alpha, \beta, \gamma$.
In essence, these parameters belong to a projective plane, which is factorized w.r.t. its homogeneous coordinates and is called Vogel's plane, 
see Figure \ref{fig:Vplane}. 

As is seen, it demonstrates the points from Vogel's table. Also, it includes some additional points and lines studied by Landsberg, Manivel, Westbury, and Mkrtchyan,
namely, the line corresponding to $D_{(2,1,\lambda)}$ superalgebras, the $3d$ line, which passes through the $sl(2)$ point, etc.
This parameterization of the simple Lie algebras happens to be very convenient and useful. In particular, the
existence of the so-called universal formulae for several objects appearing both in the representation theory of 
the simple Lie algebras and physical theories built upon the symmetries corresponding to the simple groups is made possible due to this
parametrization.

As typical examples of universal formulae, those for the dimensions of representations from (\ref{sad}) are presented below:

\begin{eqnarray} \label{ad}
\text{dim}\, \mathfrak{g} &=&-\frac{(2t-\alpha)(2t-\beta)(2t-\gamma)}{\alpha\beta\gamma} \\
 \label{Y}
 \text{dim} \, Y_2(\alpha)&=& \frac{\left(  2t  - 3\alpha \right) \,\left( \beta - 2t \right) \,\left( \gamma - 2t \right) \,t\,\left( \beta + t \right) \,
      \left( \gamma + t \right) }{\alpha^2\,\left( \alpha - \beta \right) \,\beta\,\left( \alpha - \gamma \right) \,\gamma}
\end{eqnarray}
and the other two (\ref{Y}) representations are obtained by permutations of the parameters \cite{V0,LM1}. 

\subsection{A bird's-eye view on the state of play}

There are a number of universal formulae for different objects in the theory and applications of simple Lie algebras. E.g. Vogel \cite{V0} found a complete decomposition of the third power of the 
adjoint representation in terms of  Universal Lie Algebra, defined by him, and universal dimension formulae for all representations involved. 
Landsberg and Manivel \cite{LM1} presented a method that allows the derivation of certain universal dimension formulae for simple Lie algebras and derived those for the 
Cartan powers of the adjoint, $Y_2(.)$, and their Cartan products. 
A universal formula for the quantum dimension of the adjoint representation has been found by Westbury \cite{W3}. 
Sergeev, Veselov, and Mkrtchyan have derived \cite{MSV} a universal formula for generating function for the eigenvalues of higher Casimir operators on the adjoint representation. 

In subsequent works, applications to physics were developed, particularly the universality of the partition function of Chern-Simons theory on a $S^3$ sphere \cite{MV,M13,KhM16-1}. 
Its connection with q-dimension of the $k\Lambda_0$ representation of affine Kac-Moody algebras \cite{M17} was shown, and
 the universal knot polynomials for 2- and 3-strand torus knots \cite{W15, MMM, MM, M16QD} were calculated.

Another application of universal formulae is the derivation of non-perturbative corrections to Gopakumar-Vafa partition function \cite{KM, KS} by gauge/string duality from the universal partition
 function of Chern-Simons theory. This shows the relevance of the "analytical continuation" of the universal formulae from the points of Vogel's table (\ref{tab:V1}) to the entire Vogel's plane.   

A completely different direction of development, the Diophantine classification of simple Lie algebras \cite{M16} and its connection with the McKay correspondence \cite{KhM16-2}, is also worth mentioning.

\subsection{Results presented in this work}
Our first achievement is embodied in the extension of the list of universal quantum dimension formulae for the representations of the simple Lie algebras.

The initial list of universal formulae derived by Vogel was first expanded by Landsberg and Manivel \cite{LM1}. They in fact proved that the arbitrary Cartan powers
of the representations appearing in the symmetric square of the adjoint (\ref{sad}), and the Cartan products of the powers of any two of them are universal. 

It seemed natural to ask if the same is true for the representations appearing in the antisymmetric square of the adjoint:
\begin{eqnarray}\label{aad}
\wedge^2 \mathfrak{g}=\mathfrak{g} \oplus X_2
\end{eqnarray}

And the answer happened to be "yes"! In \cite{AM} we first derived a universal quantum dimension formula for the Cartan products of $X_2$ representation.
Soon we managed to generalize this result by deriving universal quantum dimensions for the Cartan products of arbitrary powers of $X_2$ and the adjoint $\mathfrak{g}$, \cite{X2kng}.

Another achievement is the derivation of the universal eigenvalues of the second Casimir operator on the same representations, \cite{Casimir}.

Chapter 2 is devoted to the detailed presentation of these three results.

The next attainment relates to the discovery of a remarkable property, which we call {\it linear resolvability}, of the quantum dimension formulae derived in \cite{AM, X2kng}.

The seeds of this discovery have been sowed in \cite{LM1} where the authors examined universal dimensions at the points, corresponding to
the permuted universal parameters. 
They noticed that for some of these points their formula has singularities and called them {\it indeterminacy locis}.

After derivation of the universal quantum dimensions for the Cartan products of arbitrary powers of $X_2$ and the adjoint $\mathfrak{g}$ \cite{AM, X2kng}, 
we carried out a similar examination of these formulae and encountered analogous singularities for them too.
Actually, we succeeded in understanding these singularities better. 
Namely, we showed that for all possible singular points those new formulae admit radial limit in all but a finite number of directions. 
We called such formulae {\it linearly resolvable} (LR) and claimed that the new universal quantum dimension formulae derived in Chapter 2 are LR.

Chapter 3 is focused on the discussion of this property.

Chapter 4 describes how we connected the theory of simple Lie algebras with the theory of {\it geometrical configurations of points and lines}  \cite{GB}.

This achievement is rooted in the question of whether there is more than one universal dimension formula yielding the same outputs at the
distinguished points and sharing the same structure with a particular known universal representation. Or, equivalently, are the known universal dimension formulae unique?

To answer this question we search for the so-called {\it non-uniqueness factors} – non-trivial functions, that yield $1$ at the points, corresponding to the simple Lie algebras. 
We notice that an equivalent geometrical formulation of this question shows that the existence of such non-uniqueness factors is directly dependent on the existence of
particular types of configurations of points and lines. 

The final achievement of the present work is the generalization of the partition function of the {\it refined} Chern-Simons
theory on $S^3$ to all simple gauge algebras.

Vogel's universal description of the simple Lie algebras soaks into physical theories, based on gauge groups corresponding to these algebras.
Particularly, quantities, appearing in these theories, such as the central charge \cite{MV} in the Chern-Simons theory on $S^3$, the partition function of the same theory,
the volume of a group, etc., were shown to be expressed in terms of Vogel's universal parameters \cite{MV, M13, KhM16-1}. In fact, this means that there appears a 
possibility for treating the physical theories with the classical and the exceptional gauge groups on an equal footing. 
This possibility has shown itself as a valuable tool for establishing and/or investigating dualities between theories, in particular, the Chern-Simons/topological strings duality \cite{MV, KS, KM, RM20}.

In this work we make the first step towards understanding the {\it refined} Chern-Simons/topolo\-gi\-cal strings dualities for each of the simple gauge groups.
Particularly, we succeeded in generalizing the Kac-Peterson formula for the volume of the fundamental domain of the coroot lattice of a Lie algebra, which leads us
to the presentation of a partition function of the refined Chern-Simons theories for all simple gauge groups at once.
This presentation makes it possible to derive each of the refined partition functions in a form, suitable for comparing it with the Gopakumar-Vafa partition functions for topological strings.

Chapter 5 is devoted to a detailed description of these procedures.

Finally, the summary of this work and the vision of the future directions of development are presented in Chapter 6.

\newpage
\chapter{Extending the “universality island”. Derivation of the universal quantum dimensions for $(X_2)^k(g)^n$ 
and the universal second Casimir on them}
After the first universal dimension formulae, derived by Vogel in 1999 \cite{V0, V}, the uncertainty, caused by 
the revelation of a zero divisor in the $\Lambda$ algebra was still unanswered. The question of whether the initial list of those formulae can
be further extended was uncertain until the publication \cite{LM1} in 2005, where Landsberg and Manivel presented a universal 
expression for the dimensions of arbitrary Cartan powers of the adjoint $\mathfrak{g}$ and the $Y_2$ representations, which appear in the universal decomposition
of the symmetric square of the adjoint representation:
\begin{equation}
S^2 \mathfrak{g}=1 \oplus Y_2(\alpha) \oplus Y_2(\beta) \oplus Y_2(\gamma)
\end{equation}
Their technique of derivation of that new universal formula differed from the one Vogel had used: it was essentially based on the examination of the
root systems and some important properties of the Weyl dimension formula, following from the structure of the root systems.

The results, in this chapter are obtained using a similar technique to Landsberg's and Manivel's. 

At first, we derive a universal dimension as well as a quantum dimension for the Cartan powers of $X_2$ representation, 
which appears in the decomposition of the antisymmetric square of the adjoint:
\begin{eqnarray}
\wedge^2 \mathfrak{g}=\mathfrak{g} \oplus X_2
\end{eqnarray}

Then we manage to generalize that formula by derivation of a universal (quantum) dimension for the Cartan products of
arbitrary powers of $g$ and $X_2$ representations: $(X_2)^k(g)^n$.

Finally, we show that the eigenvalues of the second Casimir operator on these representations are also universal and present the corresponding
universal expression.
\newpage
\section{$X_2^k$}
In this section we present the derivation of universal formulae for quantum dimensions for an arbitrary Cartan power of the  $X_2$ representation, appearing in the following decomposition 

\begin{eqnarray}
\wedge^2 \mathfrak{g}=\mathfrak{g}+X_2
\end{eqnarray}

The $k$-th Cartan power of a representation with the highest weight $\lambda$ is that with the highest weight  $k\lambda$. Note that for  $sl(n)$ algebras $X_2$ is not an irreducible representation until one considers the Lie algebra's semidirect product with the automorphism group of its Dynkin diagram (instead of the algebra itself), as suggested and implemented in \cite{Del,DM,Cohen} for the exceptional algebras. Particularly, in  $sl(n)$  case one has $Z_2$ as an automorphism group and $X_2$ is the sum of representations with highest weights $2\omega_1+\omega_{n-2}$ and $\omega_2+2\omega_{n-1}$. Its Cartan power we consider to be the sum of Cartan powers of these two representations. More generally, any irrep of simple Lie algebras below is considered to be extended by the automorphism group of their Dynkin diagram. We shall see, that the universal formulae yield answers for irreps of such extended Lie algebras, i.e. if there appears an irrep which is not invariant under automorphism, then it appears in the sum with his automorphism-transformed version(s), so that the invariance is recovered. 

For $k=1$ the universal quantum dimension of $X_2$ has been given in \cite{D13}:

\begin{eqnarray} \nonumber
D_Q^{X_2}=\frac{\sinh\left(\frac{x}{4}(2t-\alpha)\right)\sinh\left(\frac{x}{4}
	(2t-\beta)\right)\sinh\left(\frac{x}{4}(2t-\gamma)\right)}{\sinh\left(\frac{\alpha x}{4}\right)\sinh\left(\frac{\beta x}{4}\right)\sinh\left(\frac{\gamma x}{4}\right)}
\times \\ \nonumber
\frac{\sinh\left(\frac{x}{4}(t+\alpha)\right)\sinh\left(\frac{x}{4}(t+\beta)\right)\sinh\left(\frac{x}{4}(t+\gamma)\right)}{\sinh\left(\frac{\alpha x}{2}\right)\sinh\left(\frac{\beta x}{2}\right)\sinh\left(\frac{\gamma x}{2}\right)} \times \\ \frac{\sinh\left(\frac{x}{2}(t-\alpha)\right)\sinh\left(\frac{x}{2}(t-\beta)\right)\sinh\left(\frac{x}{2}(t-\gamma)\right)}{\sinh\left(\frac{x}{4}(t-\alpha)\right)\sinh\left(\frac{x}{4}(t-b)\right)\sinh\left(\frac{x}{4}(t-\gamma)\right)} \label{k=1}
\end{eqnarray}

Below we generalize this formula for $k>1$ cases and discuss its properties.  

Note that $X_2$ had remained the only representation from the square of the adjoint, which had not had a universal formula for (quantum) dimensions of its Cartan powers. 
For powers of other representations, i.e. $Y_2(.)$, both usual and quantum dimensions are given in \cite{LM1,MMM,M17}.

\subsection{Technique}

There is no regular way of obtaining universal formulae (and their very existence is not guaranteed). Vogel's approach gave unique answers for dimensions, but it was
 based on the calculation with ring $\Lambda$, which appears to have \cite{V} divisors of zero, so that approach is not self-consistent if one does not handle that issue carefully. 
 In fact, in  \cite{LM1} (and in the present work) the restricted definition of universal formulae is adopted, namely- they have to give correct answers for true simple Lie algebras at
  the corresponding points of Vogel's table \ref{tab:V1}. 

That allows one to use the Weyl formula for characters, restricted to the Weyl line, i.e. for quantum dimensions (see e.g. \cite{DiF}, 13.170): 

\begin{eqnarray}\label{W}
D_Q^\lambda= 
\chi_{\lambda}(x\rho)= \prod_{\mu >0} \frac{\sinh(\frac{x}{2}(\mu,\lambda+\rho))}{\sinh(\frac{x}{2}(\mu,\rho))}
\end{eqnarray}

where $\lambda$ is the highest root of the given irreducible representation, $\rho$ is the Weyl's vector, the sum of the fundamental weights. The usual dimensions are obtained in
 the $x \rightarrow 0$ limit of the quantum ones. Both sides of this formula are invariant w.r.t. the simultaneous rescaling (in "opposite directions") of the scalar product in algebra and the parameter $x$.The automorphism of the Dynkin diagram leads to the equality of quantum dimensions for representations with the highest weights connected by an automorphism.

Evidently, only the roots with a non-zero scalar product with  $\lambda$ contribute. So, one has to express the scalar product of such roots with  $\lambda$ and $\rho$ in terms of the universal parameters, and that has to be done in a uniform way for all simple Lie algebras. Then one may hope to get a universal expression for $D_\lambda$. 

To describe the technique, consider, e.g. the case of $\lambda=\theta$, the highest weight of the adjoint representation. As it is shown in \cite{LM1}, the values of scalar product of roots with $\theta$ are either $2$ (for root $\theta$ itself) or $1$. These last roots  can be organized into three "segments" (see definition below) with unit spacing of $(\rho,\alpha)$ (we normalize the scalar product as in \cite{LM1} and table \ref{tab:V1} by $\alpha=-2$), which we present below for $E_7$ as an example:

\begin{table}[h]
\caption{Height $ht=(\rho,\mu)$ and $n_{ht}$ for all roots $\mu$ with $(\theta,\mu)=1$ for $E_7$}
	\centering
		\begin{tabular}{|l|r|r|r|r|r|r|r|r|r|r|r|r|r|r|r|r|}
		\hline
		$ht$&1&2&3&4&5&6&7&8&9&10&11&12&13&14&15&16\\
		\hline
				$n_{ht}$ &1&1&1&2&2&3&3&3&3&3&3&2&2&1&1&1\\
				\hline
		\end{tabular}
		\label{tab:E7-black}
		\end{table}

where there are the values of scalar products with $\rho$, i.e. the heights $ht$ of roots in the first line, and in the second line -  $n_{ht}$ - the number of roots on that height (remember we consider the roots 
$\mu$ with $(\mu,\theta)$, only). So, we see, that roots with $(\theta,\mu)=1$ can be organized into three sets of roots, which we shall call "segments of roots", or simply segments. A segment of roots is the
 finite number of roots with equidistant values of heights including exactly one root for any given height from that equidistant sequence of heights. The first, the longest segment, has length $t-2=16$, 
 with heights from $1$ to $16$, the second is in the center of the first, is of length $\gamma-2=10$ (we order universal parameters as $\gamma \geq \beta >-2)$, and the third segment, again in 
 the center of the first (and the second) segments, has length $\beta-2$. The same pattern of segments is observed for most of the simple Lie algebras. 

With this data, it is easy to obtain universal formulae for dimensions \cite{LM1} and quantum dimensions \cite{M17} for $k$-th Cartan power of the adjoint representation. Namely, numerators and denominators of consecutive roots of the given segment of roots cancel (\ref{D}), so for each segment there remains a number of the first denominators and the same number of the last numerators, which finally lead to the universal formulae.

These results have been proven in \cite{LM1} partially by "general" considerations, restricted, however, to the algebras of the rank at least three, and partially by case-by-case considerations for each algebra separately.

The description above reflects the advantage of the approach - the possibility of using the Weyl formula, as a basis of calculations, and shortcomings, which come from the use of very restricted sets of truly existing simple Lie algebras, see more on that below. Particularly, one can add an arbitrary polynomial to the results, which accepts zero values on the lines of the simple Lie algebras (tables \ref{tab:V1}, \ref{tab:V2}). Such "minimal" symmetric polynomial can easily be written:

\begin{eqnarray}
(\alpha+\beta)(\beta+\gamma)(\gamma+\alpha)(2\alpha+\beta)(2\beta+\alpha)(2\alpha+\gamma)(2\gamma+\alpha)\cdot \\\nonumber
(2\beta+\gamma)(2\gamma+\beta)(2\gamma+2\alpha-\beta)(2\gamma+2\beta-\alpha)(2\alpha+2\beta-\gamma)
\end{eqnarray}

However, one can require that, first, the formula should be presented as a ratio of products of linear functions over universal parameters (and not the sum of such expressions), and, second, that Deligne hypothesis \cite{D13} should be satisfied. Deligne assumes that the standard relations of characters (recall that quantum dimensions are characters on the Weyl line) namely, the product of characters of two representations is equal to the sum of characters of their decomposition, should be satisfied on the entire Vogel's plane (and not on the points of Vogel's table, only). Deligne's hypothesis is checked in some cases \cite{M16QD}, particularly for the symmetric cube of the adjoint representation. At this time it is not known whether it is possible to satisfy one or both of these requirements, as well as the very existence of universal formulae, is not guaranteed. So, we do not worry about this problem further in this paper and present the new universal formulae in the natural way we found them. 

So, below we use this approach to obtain the universal formulae for quantum dimensions of $k$-th Cartan powers of  $X_2$ representation.

Next, we present data for $E_n$ algebras and try to rewrite them in the universal form. It appears that it is not sufficient for derivation of the general formula, due to the ambiguities of rewriting the answers in the universal form. We use two additional ideas: first is that the results should not be singular for $sl(n)$ algebra, and, second, that the answer should be invariant w.r.t. the permutation of two parameters. In that way, we obtain the final formula (\ref{D}) below. All this, however, does not combine into formal derivation and all together should be considered as an educated guess. The formal proof is carried out in the Appendix, for all algebras. We nevertheless outline these steps to show how we came to the final, sufficiently complicated formula. The development of a general method for derivation of universal formulae still remains an open problem.

\subsection{$E_n$ data}

It appears that $E_n$ are the only algebras, which can hint at a universal form of non-trivial contributions to the Weyl formula (\ref{D}) for $X_2$ representation. So below we present relevant roots and their contributions. 

\subsection*{$E_8$}

Dimension of $E_8$=248, number of positive roots $|\Delta_+|=120$, Vogel's parameters $(\alpha,\beta,\gamma)=(-2,12,20)$.
For $E_8$ the highest weight of $X_2$ is $\lambda=\omega_7$, in Dynkin's numeration of roots (for reader's convenience we give it below): 

\begin{tikzpicture}
\dynkin[label,ordering=Dynkin]{E}{8}
\end{tikzpicture}

The number of positive roots $\mu$ with $(\lambda,\mu)=0$ is $1+|\Delta_+|_{E_6}=1+36=37$.
 
The number of positive roots $\mu$ with $(\lambda,\mu)=1$ is 54 and is given in table \ref{tab:E8-1} with numbers $n$.

\FloatBarrier
\begin{table}[ht]
\caption{Number $n_{ht}$ vs height $ht=(\rho,\mu)$ for roots $\mu$ with $(\lambda,\mu)=1$ for $E_8$}
	\centering
		\begin{tabular}{|l|r|r|r|r|r|r|r|r|r|r|r|r|r|r|r|r|r|r|}
		\hline
		$ht$&1&2&3&4&5&6&7&8&9&10&11&12&13&14&15&16&17&18\\
		\hline
		$n_{ht}$ &1&2&2&2&3&4&4&4&5&5&4&4&4&3&2&2&2&1 \\
		\hline
		\end{tabular}
		\label{tab:E8-1}
		\end{table}
\FloatBarrier

So, here we have 5 segments of roots.

The number of positive roots $\mu$ with $(\lambda,\mu)=2$ is 27 and is given in table \ref{tab:E8-2} with numbers $n_{ht}$.

\FloatBarrier
\begin{table}[h]	
\caption{Number $n_{ht}$ vs height $ht=(\rho,\mu)$ for roots $\mu$ with $(\lambda,\mu)=2$ for $E_8$}
	\centering
		\begin{tabular}{|l|r|r|r|r|r|r|r|r|r|r|r|r|r|r|r|r|r|}
		\hline
		$ht$&11&12&13&14&15&16&17&18&19&20&21&22&23&24&25&26&27 \\
		\hline
		$n_{ht}$ &1&1&1&1&2&2&2&2&3&2&2&2&2&1&1&1&1 \\
		\hline
		\end{tabular}	
		\label{tab:E8-2}
		\end{table}
\FloatBarrier

So, here we have 3 segments of roots. 

The number of positive roots $\mu$ with $(\lambda,\mu)=3$ is 2 and is given in table \ref{tab:E8-3}

\FloatBarrier
\begin{table}[h]
\caption{Number $n_{ht}$ vs height $ht=(\rho,\mu)$ for roots $\mu$ with $(\lambda,\mu)=3$ for $E_8$}
	\centering
		\begin{tabular}{|l|r|r|}
		\hline
		$ht$&28&29 \\
		\hline
		$n_{ht}$ &1&1 \\
		\hline
		\end{tabular}
		\label{tab:E8-3}
		\end{table}
\FloatBarrier

So, here we have 1 segment, consisting of two roots. 

Check the total number of roots: 37+54+27+2=120, as it should be.

\subsection*{$E_7$}

Dimension $E_7$=133, number of positive roots $|\Delta_+|=63$, Vogel's parameters $(\alpha,\beta,\gamma,t)=(-2,8,12,18)$.

For $E_7$ $\lambda=\omega_3$, in Dynkin's numeration of roots: 

\begin{tikzpicture}
\dynkin[label,ordering=Dynkin]{E}{7}
\end{tikzpicture}

The number of positive roots $\mu$ with $(\lambda,\mu)=0$ is $1+|\Delta_+|_{A_5}=1+15=16$.
 
The number of positive roots $\mu$ with $(\lambda,\mu)=1$ is 30 and is given in table \ref{tab:E7-1} with multiplicities.

\FloatBarrier
\begin{table}[h]
\caption{Number $n_{ht}$ vs height $ht=(\rho,\mu)$ for roots $\mu$ with $(\lambda,\mu)=1$ for $E_7$}
	\centering
		\begin{tabular}{|l|r|r|r|r|r|r|r|r|r|r|}
		\hline
		$ht$&1&2&3&4&5&6&7&8&9&10\\
		\hline
		$n_{ht}$ &1&2&3&4&5&5&4&3&2&1 \\
		\hline
		\end{tabular}
		\label{tab:E7-1}
		\end{table}
\FloatBarrier

So, here we have 5 segments of roots, i.e. sequences with a unit distance between consecutive roots.

The number of positive roots $\mu$ with $(\lambda,\mu)=2$ is 15 and is given in table \ref{tab:E7-2} with multiplicities.

\FloatBarrier
\begin{table}[h]
\caption{Number $n_{ht}$ vs height $ht=(\rho,\mu)$ for roots $\mu$ with $(\lambda,\mu)=2$ for $E_7$}
	\centering
		\begin{tabular}{|l|r|r|r|r|r|r|r|r|r|}
		\hline
		$ht$&7&8&9&10&11&12&13&14&15 \\
		\hline
		$n_{ht}$ &1&1&2&2&3&2&2&1&1 \\
		\hline
		\end{tabular}
		\label{tab:E7-2}
		\end{table}
\FloatBarrier

So, here we have 3 segments of roots. 

The number of positive roots $\mu$ with $(\lambda,\mu)=3$ is 2 and is given in table \ref{tab:E7-3} with multiplicities.

\FloatBarrier
\begin{table}[h]
\caption{Number $n_{ht}$ vs height $ht=(\rho,\mu)$ for roots $\mu$ with $(\lambda,\mu)=3$ for $E_7$}
	\centering
		\begin{tabular}{|l|r|r|}
		\hline
		$ht$&16&17 \\
		\hline
		$n_{ht}$ &1&1 \\
		\hline
		\end{tabular}
		\label{tab:E7-3}
		\end{table}
\FloatBarrier

So, here we have 1 segment of roots. 

Check the total number of roots: 16+30+15+2=63.

\subsection*{$E_6$}

dim$E_6$=78, $|\Delta_+|=36$, $(\alpha,\beta,\gamma,t)=(-2,6,8,12)$.

For $E_6$ $\lambda=\omega_4$, in Dynkin's numeration of roots.

\begin{tikzpicture}
\dynkin[label,ordering=Dynkin]{E}{6}
\end{tikzpicture}

The number of positive roots $\mu$ with $(\lambda,\mu)=0$ is $7$.
 
The number of positive roots $\mu$ with $(\lambda,\mu)=1$ is 18 and is given in table \ref{tab:E6-1} with numbers $n_{ht}$.

\FloatBarrier
\begin{table}[h]
\caption{Number $n_{ht}$ vs height $ht=(\rho,\mu)$ for roots $\mu$ with $(\lambda,\mu)=1$ for $E_6$}
	\centering
		\begin{tabular}{|l|r|r|r|r|r|r|}
		\hline
		$ht$&1&2&3&4&5&6\\
		\hline
		$n_{ht}$ &1&3&5&5&3&1 \\
		\hline
		\end{tabular}
		\label{tab:E6-1}
		\end{table}
\FloatBarrier
So, here we have 5 segments of roots, i.e. sequences with a unit distance between consecutive roots.

The number of positive roots $\mu$ with $(\lambda,\mu)=2$ is 9 and is given in table \ref{tab:E6-2} with multiplicities.
\FloatBarrier
\begin{table}[h]
\caption{Number $n_{ht}$ vs height $ht=(\rho,\mu)$ for roots $\mu$ with $(\lambda,\mu)=2$ for $E_6$}
	\centering
		\begin{tabular}{|l|r|r|r|r|r|}
		\hline
		$ht$&5&6&7&8&9 \\
		\hline
		$n_{ht}$ &1&2&3&2&1 \\
		\hline
		\end{tabular}
		\label{tab:E6-2}
		\end{table}
\FloatBarrier
So, here we have 3 segments of roots. 

The number of positive roots $\mu$ with $(\lambda,\mu)=3$ is 2 and is given in table \ref{tab:E6-3} with multiplicities.
\FloatBarrier
\begin{table}[h]
\caption{Number $n_{ht}$ vs height $ht=(\rho,\mu)$ for roots $\mu$ with $(\lambda,\mu)=3$ for $E_6$}
	\centering
		\begin{tabular}{|l|r|r|}
		\hline
		$ht$&10&11 \\
		\hline
		$n_{ht}$ &1&1 \\
		\hline
		\end{tabular}
		\label{tab:E6-3}
		\end{table}
\FloatBarrier
So, here we have 1 segment of roots. 

Check the total number of roots: 7+18+9+2=36.

\subsection{Quantum dimensions}

Now we calculate the contributions of roots with $(\lambda,\mu) \neq 0$ in the Weyl formula for quantum dimension. 

The contribution of roots with $(\lambda,\mu)=3$ comes from two roots of heights $t-1,t-2$ (recall the normalization $\alpha=-2$):  

\begin{eqnarray} 
L_3=\frac{\text{sinh}\left(\frac{x}{2}(t+1)\right)\text{sinh}\left(\frac{x}{2}(t+2)\right)}{\text{sinh}\left(\frac{x}{2}(t-2)\right)\text{sinh}\left(\frac{x}{2}(t-1)\right)}
\end{eqnarray} 

Due to the rescaling invariance, mentioned after (\ref{W}), we can recover the parameter $\alpha$ in this formula in explicit form by substitution  
\begin{eqnarray}
\beta \rightarrow -2\beta/\alpha, \gamma \rightarrow -2\gamma/\alpha, t \rightarrow -2t/\alpha,  x\rightarrow -x\alpha/2
\end{eqnarray}
Then $L_3$ accepts the form
\begin{eqnarray} 
L_{3}=\frac{\sinh
\left(\frac{x}{4}
(3 \alpha (k-1)-2 (\beta+\gamma))\right) \sinh
\left(\frac{x}{4} (\alpha (3
k-4)-2
(\beta+\gamma))\right)}{\sinh\left(\frac{x}{2} (2
\alpha+\beta+\gamma)\right)
\sinh\left(\frac{x}{4}  
(3 \alpha+2 (\beta+\gamma))\right)}
\end{eqnarray}

Below we skip the intermediate formulae in normalization $\alpha=-2$ and present the final ones with explicit $\alpha$ recovered.

Next consider roots with $(\lambda,\mu)=2$. There are three segments, the first (longest) one starts at height $\beta-1$ and ends at height $t-3$, its contribution in the Weyl formula is 

\begin{eqnarray} 
L_{21}=\prod _{i=1}^{2 k} \frac{\sinh \left(\frac{1}{4}
x (\alpha (i-5)-2
(\beta+\gamma))\right)}{\sinh\left(\frac{1}{4} x
(\alpha
(i-2)-2
\beta)\right)}
\end{eqnarray}

The second segment starts at height $t/2$ and ends at height $(t+\gamma-4)/2$, the contribution is 
\begin{eqnarray} 
L_{22}=\prod _{i=1}^{2 k} \frac{\sinh \left(\frac{1}{4} x
(-\alpha
(i-3)+\beta+2 \gamma)\right)}{\sinh\left(\frac{1}{4} x
(-\alpha
(i-2)+\beta+\gamma)\right)}
\end{eqnarray} 

The third segment includes one root at height $(\gamma+2\beta-6)/2$ and it's contribution is
\begin{eqnarray} 
L_{23}=\frac{\sinh
\left(\frac{1}{4} x (\alpha (3-2 k)+2
\beta+\gamma)\right)}{\sinh\left(\frac{1}{4} x (3 \alpha+2
\beta+\gamma)\right)} 
\end{eqnarray}

Next are the roots with $(\lambda,\mu)=1$. There are five segments, the first (longest) one starts at height $1$ and ends at height $(\gamma+2\beta-8)/2$, its contribution in the Weyl formula is 

\begin{equation}
L_{11}=\prod _{i=1}^k \frac{\sinh
\left(\frac{1}{4} x (-\alpha (i-4)+2
\beta+\gamma)\right)}{\sinh\left(\frac{\alpha i x}{4}\right)}
\end{equation}

The second segment starts at height $2$ and ends at height $\gamma-2$, contributing 

\begin{eqnarray} 
L_{12}=\prod _{i=1}^k \frac{\sinh
\left(\frac{1}{4} x (\alpha (i-3)-2 \gamma )\right)}{\sinh
\left(\frac{1}{4} \alpha (i+1) x\right)}
\end{eqnarray} 

The third segment starts at height $(\beta-2)/2$ and ends at $(\gamma+\beta-4)/2$, contributing

\begin{eqnarray} \label{L13}
L_{13}=\prod _{i=1}^k \frac{\sinh \left(\frac{1}{4} x
(-\alpha
(i-2)+\beta+\gamma)\right)}{\sinh\left(\frac{1}{4} x (\beta-
\alpha
(i-2))\right)}
\end{eqnarray} 

The fourth segment is similar to the third one but shorter by one element on each end:

\begin{eqnarray} 
L_{14}=\prod _{i=1}^k\frac{\sinh \left(\frac{1}{4} x
(-\alpha (i-3)+\beta+\gamma)\right)}
{\sinh \left(\frac{1}{4} x (\alpha
(-i)+\alpha+\beta)\right)} 
\end{eqnarray} 

The fifth segment consists of two roots, starting at height $(\gamma-2)/2$, and contribution will be

\begin{eqnarray}\label{L15exp}
\frac{\text{sinh}\left(\frac{x}{2}(\beta-3+k)\right)}{\text{sinh}\left(\frac{x}{2}\left(\frac{\gamma-2}{2}\right)\right)}\frac{\text{sinh}\left(\frac{x}{2}(\beta-2+k)\right)}{\text{sinh}\left(\frac{x}{2}\left(\frac{\gamma-2}{2}+1\right)\right)}
\end{eqnarray}

This contribution is appropriate at $k=1$, in a sense that all contributions together - the product of all $L$-s - form the corrects answer (\ref{k=1}). However, for $k>1$ and for $sl(n)$ algebras (i.e. on the line $\alpha+\beta=0$) one loses the zero of (\ref{L15exp}) on that line which at $k=1$ cancels out with the zero in denominator of (\ref{L13}), also on the same line. So,
in analogy with other contributions above, we simply change this contribution to other one, namely $L_{15}$, written below. It cancels mentioned singularity for an arbitrary $k$, coincides with (\ref{L15exp}) on the exceptional line $\gamma=2(\alpha+\beta)$, but differs in other points:

\begin{eqnarray} \label{L15}
L_{15}=\prod _{i=1}^k\frac{\sinh \left(\frac{1}{4} x (\alpha
(i-3)-2
\beta)\right) \sinh
\left(\frac{1}{4} x (\alpha (i-2)-2 \beta)\right)
}{\sinh\left(\frac{1}{4}
x (\gamma-\alpha
(i-2))\right) \sinh\left(\frac{1}{4} x (\alpha
(-i)+\alpha+\gamma)\right)} 
\end{eqnarray}

However, this is not the end of the story. We expect that our final formula should be invariant under switch of the $\beta$ and
$\gamma$ parameters, in analogy with the universal formula (\ref{sad}) for  $Y_2(\alpha)$ . So we add a new multiplier, which  in some "minimal" way symmetrizes the product of all $L_?$ multipliers above w.r.t. the switch $\beta \leftrightarrow \gamma$:
\begin{eqnarray} \label{Lc}
L_{corr}=\prod _{i=1}^k\frac{\sinh \left(\frac{1}{4} x (\alpha
(-(i+k-4))+2
\beta+\gamma)\right)}{\sinh\left(\frac{1}{4} x (\alpha
(i+k-2)-2
\gamma)\right)}
\end{eqnarray}

Finally, our main result is 

{\bf Proposition 2.1}

{\it The function 
\begin{eqnarray} \label{D}
X_2(x,k,\alpha,\beta,\gamma)\equiv X_2(k,\alpha)= L_3L_{21}L_{22}L_{23}L_{11}L_{12}L_{13}L_{14}L_{15}L_{corr}
\end{eqnarray}
is equal, besides exceptions, to the quantum dimensions of $k$-th Cartan power of above defined $X_2$ representation for any given simple Lie algebra on corresponding point of Vogel's table \ref{tab:V1}. Exceptions are: $sp(2n)$, for which the formula gives the quantum dimensions of $X_2$ at $k=1$, and zero otherwise, and the $B_2$ algebra. Exact details are given in the tables \ref{tab:x2acl} and \ref{tab:x2aex}.}

The case by case proof of {\bf Proposition 2.1} will be given in the next section after the generalized formula, when $k>1$, will be presented.

 \FloatBarrier
 \begin{table}[h]
   \caption{$X_2(k,\alpha)$ for classical algebras}
    \centering
     \begin{tabular}{|c|p{2cm}|p{2cm}|c|}
     \hline
          $k$&1&2&$\geq3$ \\
          \hline
        $A_1$&0&0&0\\  
        \hline
        $A_2$&$3\omega_1\oplus 3\omega_2$&$
        6\omega_1\oplus 6\omega_2$& $3k\omega_1\oplus 3k\omega_2$
        \\  
        \hline
         $A_n,n\geq3$&$(2\omega_1+\omega_{n-1})\oplus 
       (\omega_2+2\omega_{n})$&$2(2\omega_1+\omega_{n-1})\oplus
       2(\omega_2+2\omega_{n})$&$k(2\omega_1+\omega_{n-1})\oplus
       k(\omega_2+2\omega_{n})$\\
       \hline
        $B_2$&$\omega_1+2\omega_2$&0&0\\  
        \hline
        $B_3$&$\omega_1+2\omega_3$&$2\omega_1+4\omega_3$
        &
        $k(\omega_1+2\omega_3)$\\  
        \hline
        $B_n,n\geq4$&$\omega_1+\omega_3$&$2(\omega_1+
       \omega_3)$&$k(\omega_1+
       \omega_3)$\\
       \hline
        $C_n,n\geq3$&$2\omega_1+\omega_2$&0&0\\
       \hline
        $D_4$&$\omega_1+\omega_3+\omega_4$&$
        2(\omega_1+\omega_3+\omega_4)$&
        $k(\omega_1+\omega_3+\omega_4)$\\
        \hline
        $D_n,n\geq5$&$\omega_1+\omega_3$&$2(\omega_1+
       \omega_3)$&$k(\omega_1+ \omega_3)$\\
       \hline
       \end{tabular}
       \label{tab:x2acl}
         \end{table}     
         \FloatBarrier

 \FloatBarrier
 \begin{table}[h]
 \caption{$X_2(k,\alpha)$ for $Exc$ line}
  \centering
     \begin{tabular}{|c|c|c|}
     \hline
          $k$&1&$\geq2$\\
          \hline
       $G_2$&$3\omega_1$&$3k\omega_1$\\
       \hline
       $F_4$&$\omega_2$&$k\omega_2$\\
       \hline
       $E_6$&$\omega_3$&$k\omega_3$\\
       \hline
       $E_7$&$\omega_2$&$k\omega_2$\\
       \hline
    $E_8$&$\omega_6$&$k\omega_6$\\
       \hline
       $D_4$&$\omega_1+\omega_3+\omega_4$&
       $k(\omega_1+\omega_3+\omega_4)$\\
       \hline
       \end{tabular}
       \label{tab:x2aex}
    \end{table}   
\FloatBarrier

{\bf Remark 1, on $sl(n)$ case.} In the case of $sl(n)$ line  denominator of $L_{13}$ and numerator of $L_{15}$ both contain a zero multiplier, which however cancel out, i.e. one can continuously extend $X_2(k,\alpha)$ function on that line. In more detail: for the $\alpha+\beta=0$ line the mentioned fraction is 
\begin{eqnarray}
\frac{\sinh((2\beta+2\alpha)x/4)}{\sinh((\beta+\alpha)x/4)} 
\end{eqnarray}

and evidently tends to $2$ in the limit $\alpha+\beta \rightarrow 0$ independent on the direction of approaching the given point on the line on Vogel's plane. Of course, one can simply substitute the expression 
\begin{eqnarray}
\frac{\sinh((2\beta+2\alpha)x/4)}{\sinh((\beta+\alpha)x/4)}=2\cosh((\beta+\alpha)x/4)
\end{eqnarray}

in the formula (\ref{D}) for $X_2(k,\alpha)$  from the very beginning and avoid the questions about continuity of the function. 

{\bf Remark 2, on tables.} The entries of the tables \ref{tab:x2acl} and \ref{tab:x2aex}  for a given algebra and $k$ are the  representation(s), denoted by highest weight, the quantum dimension of which is given by our main formula (\ref{D}). 

{\bf Remark 3, on the connection with the dimension formulae \cite{Cohen}}. In the $x\rightarrow 0 \,\,\,$ limit $X_2(x,k,\alpha,\beta,\gamma)$ gives the universal dimension formulae. When considered on the exceptional line by taking $\alpha=y,\beta=1-y,\gamma=2$ and for $k=2$, in the $x\rightarrow 0 \,\,\,$ limit the expression for $X_2(x,k,\alpha,\beta,\gamma)$ gives the following dimension formula 

\begin{eqnarray} \label{HH}
-\frac{10 (y-6) (y-5) (y+3) (y+4) (y+5) (2 y-5) (3 y-4) (5 y-6)}{(1-2 y)^2 (y-1)^3 y^4 (3 y-2)}
\end{eqnarray}

which coincides exactly with the universal formula on the exceptional line of \cite{Cohen} for representation $H$.

{\bf Remark 4, on $sp(2n)$ case.} We assume the following interpretation of this case. The point is that Vogel's parameters for $sp(2n)$ algebras can be obtained from those of $so(2n)$ by transformation 
$(\alpha,\beta,\gamma) \rightarrow (-1/2)(\beta,\alpha,-\gamma)$, which includes transposition of $\alpha$ and $\beta$. And indeed, we see in the table \ref{tab:x2bcl}, that our formula gives quantum dimensions of some sequence of 
representations of $sp(2n)$, although not the Cartan powers of its $X_2$ representation. Simultaneously, table \ref{tab:x2bcl} doesn't give quantum dimensions of new representations of $so$. We conclude, that for $sp$ the role 
of $X_2$ sequence of representation in our formulae is played by other series, given in table \ref{tab:x2bcl}. 

  \FloatBarrier
 \begin{table}
 \caption{$X_2(k,\beta)$ for the classical algebras for sufficiently large $n$ (depends on $k$)}
  \centering
     \begin{tabular}{|c|p{2.1cm}|p{2cm}|p{2cm}|p{2cm}|c|}
     \hline
          $k$&1&2&3&4&$\geq5$\\
          \hline
       $A_n$&$ (2\omega_1+\omega_{n-1}) \oplus (2\omega_{n}+\omega_{2})$ &$ (2\omega_2+\omega_{n-3})\oplus (2\omega_{n-1}+\omega_{4})$&$(2\omega_3+\omega_{n-5})\oplus (2\omega_{n-2}+\omega_{6})$&$(2\omega_4+\omega_{n-7})\oplus (2\omega_{n-3}+\omega_{8})$&$\cdots$\\
       \hline
       $B_n$&$\omega_1+\omega_3$&0  & $0$&$0$&$0$\\
       \hline
       $C_n$&$2\omega_1+\omega_2$&$2\omega_2+\omega_4$&$2\omega_3+\omega_6$&$2\omega_4+\omega_8$&$\cdots$\\
       \hline
       $D_n$&$\omega_1+\omega_3$&0&$0$&$0$&$0$\\
       \hline
    \end{tabular}
    \label{tab:x2bcl}
    \end{table}

\section{$(g)^n(X_2)^k$}
Consider the antisymmetric square of the adjoint representation. In \cite{V0} it is shown that its decomposition 
 can be presented in a uniform way (i.e. for all simple Lie algebras) as

\begin{eqnarray}
\wedge^2 \mathfrak{g}=\mathfrak{g} \oplus X_2
\end{eqnarray}

The representation $X_2$ is irreducible w.r.t. the semidirect product of simple Lie algebra and the group of automorphisms of the corresponding Dynkin diagram (see \cite{DM,Cohen}) 
and its highest weights are given in the table \ref{hw}  in terms of fundamental ones. Here we refer to the enumeration of nodes of the Dynkin diagram, used by Dynkin \cite{Dyn52}, 
with two corrections: enumeration of nods of $E_8$ starts from the shorter wing, as in $E_7$, and enumeration of nods for $G_2$ is opposite. This enumeration coincides with that 
 used in the Wolfram Mathematica package LieART, given in its description (\cite{FK}, page 11), corrected for the enumeration of nods for $G_2$ to the opposite to that given in \cite{FK}. 
 For a definition of fundamental weights of simple Lie algebras see e.g. \cite{DiF}. 

\FloatBarrier
\begin{table}[h] \label{hw}
	\caption{Highest weights of the adjoint ($\lambda_{ad}$) and $X_2$ ($\lambda_{X_2}$) representations in terms of fundamental weights for simple Lie algebras}
	 \centering
	\begin{tabular}{|c|c|c|}
		\hline
		&$\lambda_{ad}$&$\lambda_{X_2}$\\
		\hline
		$G_2$&$\omega_2$&$3\omega_1$\\
		\hline
		$F_4$&$\omega_1$&$\omega_2$\\
		\hline
		$E_6$&$\omega_6$&$\omega_3$\\
		\hline
		$E_7$&$\omega_1$&$\omega_2$\\
		\hline
		$E_8$&$\omega_7$&$\omega_6$\\
		\hline
		$A_1$&$2\omega$& 0 \\
		\hline
		$A_N, \,\, N>1$& $\omega_1+\omega_N$&$(2\omega_1+\omega_{N-1}) \oplus (\omega_2+2\omega_N)$ \\
		\hline
		$B_2$&$ 2\omega_2$&$\omega_1+2\omega_2$\\
		\hline
		$B_3$&$\omega_2$&$\omega_1+2\omega_3$\\
		\hline
		$B_N, \, N>3$&$\omega_2$&$\omega_1+\omega_3$\\
		\hline
		$C_N$&$2\omega_1$&$2\omega_1+\omega_2$\\
		\hline
		$D_4$&$\omega_2$&
		$\omega_1+\omega_3+\omega_4$\\
		\hline
		$D_N, \,\, N>4$&$\omega_2$&$\omega_1+\omega_3$\\
		\hline
	\end{tabular}
	\label{tab:hw}
\end{table} 
\FloatBarrier

Table \ref{tab:hw} needs a comment for the $A_N$ case. In that case, $\lambda_{X_2}$ is not the highest weight, but a pair of highest weights of the direct sum of
the corresponding representations, shown in the table. This is because 
the representation $X_2$ is the direct sum of two irreducible representations of $A_N$, their highest weights being connected by the automorphism of the Dynkin diagram. 
In that case the sum e.g. $\lambda_{X_2}+\lambda_{ad}$ should be understood as a pair of two highest weights, each element of pair is the sum of the  $\lambda_{ad}$ and one of the highest weights of $\lambda_{X_2}$ pair.

According to \cite{Cohen}, \cite{Del} universal formulae give answers for the semidirect product of simple Lie algebra on the group of automorphisms of their Dynkin diagrams. 
It will be observed below that it happens in all cases we consider. 

The main object of our consideration will be the quantum dimensions of (some)  representations of simple Lie algebras. Quantum dimension of representation is character of that representation, restricted to Weyl line, i.e. the argument of character is taken to be $x\rho$, where $x$ is an arbitrary parameter and $\rho$ is the Weyl vector, i.e. the half of the sum of all positive roots. See formula (\ref{W}) below for expression of quantum dimension of irreducible representations in terms of highest weight of representation.

Next we present our main result - the universal formula for
 the quantum dimension of irreps with the highest weights $k\lambda_{X_2}+n\lambda_{ad}, \, k,n \in Z_+$:

\begin{multline}  \label{main}
   X(x,k,n,\alpha,\beta,\gamma)=\\
   L_{31}\cdot L_{32}\cdot L_{21s1}\cdot L_{21s2}\cdot L_{21s3}\cdot L_{10s1}\cdot
L_{10s2}\cdot L_{10s3}\cdot L_{11s1}\cdot L_{11s2}\cdot L_{11s3}\cdot L_{01}\cdot L_{c2}
\end{multline}
where the multipliers $L_?$ look as follows\footnote{Below we omit the numerous $\sinh$ signs and 
 use the following notation instead:

\begin{eqnarray}
a\sinh\left[x: \right. \,\, \frac{A\cdot B...}{M\cdot N...}\equiv a\frac{\sinh(xA)\sinh(xB)...}{\sinh(xM)\sinh(xN)...}
\end{eqnarray}
where $x,a,A,B,...,M,N,...$ are numbers (dots between are not necessary, provided no ambiguity arises). For example

\begin{eqnarray}
2\sinh\left[\frac{x}{4}: \right. \,\, \frac{1\cdot 4}{2}\equiv 2\frac{\sinh(\frac{x}{4})\sinh(\frac{4x}{4})}{\sinh(\frac{2x}{4})}
\end{eqnarray}

One can derive simple rules which this notation obeys. E.g. 

\begin{eqnarray}
\left( \sinh\left[x: \right. \,\, A\cdot B \right) \left( \sinh \left[x: \right. M\cdot N\right) = \sinh\left[x: \right. \,\, A\cdot B \cdot M\cdot N
\end{eqnarray}

Of course, our notation belongs to the field of q-calculus, however, we didn't find this or similar convenient notation, perhaps missed
 that.

Evidently, one gets the universal dimension formulae, just by omitting the front $\sinh$ sign for $L_?$-s and $X_2(x,k,\alpha,\beta,\gamma)$ in formulae below. 
}  (see the definition of the symbol  $\sinh\left[\frac{x}{4}: \right.$ in Appendix B):

$$
L_{31}=\sinh\left[\frac{x}{4}: \right. \frac{
-2 (\beta + \gamma)+\alpha(-4+3k+n)}{ 
4\alpha +2\beta+2\gamma}$$

$$L_{32}=\sinh\left[\frac{x}{4}: \right. \frac{
-2 (\beta + \gamma)+\alpha(-3+3k+2n)}{ 
3 \alpha+2 (\beta+\gamma)}
$$

$$L_{21s1}=\sinh\left[\frac{x}{4}: \right. \prod _{i=1}^{2k+n}  \frac{
-2(\beta+\gamma)+\alpha
(-5+i)}{-2\beta+
\alpha
(i-2)}
$$

$$L_{21s2}=\sinh\left[\frac{x}{4}: \right.\prod _{i=1}^{2k+n} \frac{
\beta+2\gamma-\alpha(-3+i)}
{\beta+\gamma -\alpha(i-2)}
$$

$$L_{21s3}=\sinh\left[\frac{x}{4}: \right. \frac{
2 \beta + \gamma+\alpha(3-2k-n)}{ 
3 \alpha+2 \beta+\gamma}
 $$

$$L_{10s1}=\sinh\left[\frac{x}{4}: \right.\prod _{i=1}^{k} \frac{  
   2\gamma-\alpha (i-3)}
{ -\alpha i}
$$

$$L_{10s2}=\sinh\left[\frac{x}{4}: \right.\prod _{i=1}^{ k} \frac{     
\beta+ \gamma-\alpha
(i-3)}
{ \beta-\alpha(i-2)}$$

$$L_{10s3}=\sinh\left[\frac{x}{4}: \right.\prod _{i=1}^{k} \frac{-2\beta+\alpha(i-3)}
{ \gamma-\alpha(i-2)}$$

$$L_{11s1}=\sinh\left[\frac{x}{4}: \right.\prod _{i=1}^{k+n} \frac{2\beta+\gamma-\alpha(i-4)}
{ \alpha(i+2)} $$

$$L_{11s2}=\sinh\left[\frac{x}{4}: \right.\prod _{i=1}^{k+n} \frac{\beta+\gamma-\alpha(i-2)}
{ \beta-\alpha (i-1)} $$

$$L_{11s3}=\sinh\left[\frac{x}{4}: \right.\prod _{i=1}^{k+n} \frac{-2\beta+\alpha(i-2)}
{ \gamma+\alpha(1-i)}$$

$$L_{01}=\sinh\left[\frac{x}{4}: \right. \frac{(\alpha(1+n))}{ 
\alpha }$$

$$L_{c2}=\sinh\left[\frac{x}{4}: \right.\prod _{i=1}^k\frac{\gamma+2\beta-\alpha(i+k+n-4))}
{\alpha(i+k+n-2)-2\gamma}$$

We do not present any derivation of this formula. It is obtained in a way, similar to the universal formula (\ref{D}) for the quantum dimension of the Cartan powers 
of $X_2$ representation 
(which is a particular case of (\ref{main})). However, even in that simpler case, that formula is obtained by an "educated guess", and not exactly derived, as we mentioned in above. 
In the present case, that remark is even more relevant, so we don't bring any incomplete "derivation", but simply present the following {\bf Proposition 2.2}  and its proof.

{\bf Proposition 2.2}

{\it The function (\ref{main}) $X(x,k,n,\alpha,\beta,\gamma)$ at the points from the 
Vogel's table is equal to the quantum dimensions of representations of simple Lie algebras presented in the tables \ref{tab:xkna}, \ref{tab:xknae} ($k,n=0,1...$)}

{\bf Remark 1.} The main formula (\ref{main}) is symmetric w.r.t. the switch of $\beta$ and $\gamma$ parameters. 
This feature becomes evident after rewriting (\ref{main}) in the following form. 

\begin{multline}
X(x,k,n,\alpha,\beta,\gamma)=\\ 
\sinh\left[\frac{x}{4}: \right.\prod _{i=0}^{k-1} \frac{(\alpha  (i-2)-2 \beta )^2(\alpha  (i-2)-2 \gamma )^2(\beta +\gamma +\alpha  (-(i-2))^2}{(\alpha  (i+1) )^2
	(\beta -\alpha  (i-1))^2(\gamma-\alpha  (i-1))^2}\times \\
\times \prod _{i=0}^{n} \frac{(\alpha  (i+k-2)-2 \beta )(\alpha  (i+k-2)-2 \gamma )(\beta +\gamma +\alpha  (-(i+k-2))}
{(\alpha (i+k+1))(\beta -\alpha  (i+k-1))(\gamma -\alpha(i+k-1))} \times \\
\times \prod _{i=1}^{2k+n} \frac{(-\beta -2 \gamma +\alpha  (i-3))(-2 \beta -\gamma +\alpha  (i-3))(\alpha 
	(i-5)-2 (\beta +\gamma ))}
{(\alpha  (i-2)-2 \beta )(\alpha  (i-2)-2 \gamma )(\beta +\gamma -\alpha  (i-2))} \times \\
\times \frac{(\alpha +\beta)(\alpha +\gamma)(\alpha  (n+1))}{(2 \alpha +2 \beta )(2 \alpha +2 \gamma )(2 \alpha +\beta
	+\gamma )}\times \\
\times \frac{(\alpha  (3 k+n-4)-2 (\beta +\gamma))(\alpha  (3 k+2 n-3)-2
	(\beta +\gamma ))}{(3 \alpha +2 \beta +2 \gamma )(4 \alpha +2 \beta +2
	\gamma )}
\end{multline}
We do not have a clear explanation for this feature, though.

{\bf Remark 2.} Formula (\ref{main}) is valid for $k=0$ and/or $n=0$ provided one assumes $\prod _{i=1}^0=1$. 
In that cases it coincides with the results of \cite{M16QD}, for $k=0$, and of \cite{AM} for $n=0$. 

{\bf Remark 3.}
The proof of the {\bf Proposition 2.2}
 is carried out case by case in Appendix B. I.e. for each set of the parameters 
$\alpha, \beta, \gamma$ from Vogel's table 
the expression (\ref{main}) is compared with the Weyl formula for the quantum
dimension (\ref{W}) of the corresponding algebra. The latter is the Weyl formula
 for the characters, restricted to the Weyl line $x\rho$ (see e.g. \cite{DiF}, 13.170): 

\begin{eqnarray}\label{W}
D_Q^\lambda= 
\chi_{\lambda}(x\rho)= \prod_{\mu >0} \frac{\sinh(\frac{x}{2}(\mu,\lambda+\rho))}{\sinh(\frac{x}{2}(\mu,\rho))}
\end{eqnarray}

Here $\lambda$ is the highest weight of the given irreducible representation, $\rho$ is the Weyl vector, the sum of the fundamental weights. 
This formula is invariant w.r.t. the simultaneous rescaling
 of the scalar product in algebra and the parameter $x$ "in the opposite directions". Note, that the automorphism of the Dynkin diagram leads to the 
 equality of quantum dimensions for representations with the highest weights connected by the automorphism.
 
  {\bf Remark 4.} The formula (\ref{main}) is not unique in the sense that one can write another similar expression - a product of (sines of) linear functions over universal parameters, yielding
  the same values on points from Vogel's table. This follows from the existence of the following expression
  
  \begin{eqnarray}\label{unit}
  \frac{(2 \alpha +\beta +\gamma ) (7 \alpha +4 \beta +\gamma ) (8 \alpha +6 \beta +\gamma )}{(3 \alpha +2 \beta +\gamma ) (4 \alpha +2 \beta +\gamma ) (10 \alpha +7 \beta +\gamma )}
  \end{eqnarray}
  
  This function is equal to $1$ on the lines $\alpha+\beta=0, 2\alpha+\beta=0, \gamma=2(\alpha+\beta)$ and evidently is not constant, so it may be used to rewrite another similar expression for (\ref{main}), without changing its outputs at the points from Vogel's table. E.g. one may ask about some "minimal" expression of (\ref{main}). However, the features under permutation of parameters might be violated. Particularly, (\ref{unit}) is not symmetric under $\beta \leftrightarrow \gamma$.  These problems are out of scope of the present paper, we hope to clarify them in the future.

 {\bf Remark 5.} Both formulae for $X(x,k,n,\alpha,\beta,\gamma)$ are complicated, so we are willing to provide the Wolfram Mathematica notebook file with these as well as other universal formulae
  under a request.

\FloatBarrier
\begin{table}[h]
	\caption{$X(x,k,n,\alpha,\beta,\gamma)$ for the classical algebras}
	 \centering
	\begin{tabular}{|c|p{2cm}|p{2cm}|c|}
		\hline
		$k,n$&$0,n$&$1,n$ &$k,n \,\,(k>1)$ \\
		\hline
		$A_1$&$n \lambda_{ad}$&0&0\\  
		\hline
		$A_N,N\geq2$&$n\lambda_{ad}$&$\lambda_{X_2}+n\lambda_{ad} $&$k\lambda_{X_2}+n\lambda_{ad}$\\
		\hline
		$B_2$&$n\lambda_{ad}$&$\lambda_{X_2}+n\lambda_{ad}$&0\\  
		\hline
		$B_N, \, N>2$&$n\lambda_{ad}$&$\lambda_{X_2}+n\lambda_{ad} $&$k\lambda_{X_2}+n\lambda_{ad}$\\
		\hline
		$C_N, \, N>2$&$n\lambda_{ad}$&$\lambda_{X_2}+n\lambda_{ad} $&0\\
		\hline
		$D_N, \, N>3$&$n\lambda_{ad}$&$\lambda_{X_2}+n\lambda_{ad} $&$k\lambda_{X_2}+n\lambda_{ad}$\\
		\hline
	\end{tabular}
	\label{tab:xkna}
\end{table}     
\FloatBarrier

\FloatBarrier
\begin{table}[h]
	\caption{$X(x,k,n,\alpha,\beta,\gamma)$ for the exceptional algebras}
	 \centering
	\begin{tabular}{|c|c|}
		\hline
		$k,n$&$k,n$\\
		\hline
		$L$&$k\lambda_{X_2}+n\lambda_{ad}$\\
		\hline
	\end{tabular}
	$L$ is any of the exceptional simple Lie algebras.
	\label{tab:xknae}
\end{table}

\section{$X_2(x,k,\alpha,\beta,\gamma)$ and $X(x,k,n,\alpha,\beta,\gamma)$ Under Permutations of Universal Parameters}

All universal dimension formulae known so far share a notable feature. It consists in yielding reasonable outputs even when considering them at the points connected 
with the initial ones via {\it permutation} of the coordinates.
The word r{\it easonable} in this context means that these outputs also correspond to (quantum) dimensions of some other representations of a given Lie algebra. 
In some cases a minus sign appears in front of the (quantum) dimensions. We refer to such output as corresponding to a {\it virtual representation}. 
In this section we show that the newly-derived quantum dimension formulae do have this notable feature. 
Our check mainly extends to the level of dimensions of representations.
The behavior of the formulae at the permuted coordinates is presented in a sequence of tables where the corresponding highest weights are listed.
The cases when a virtual representation appears are also denoted by highest weights with minus sign put in front of them. 

The values of $X_2(k,\beta)$ for algebras on the exceptional line are presented in the table \ref{tab:x2bex}. Here we see a new phenomenon: the value of $X_2(k,\beta)$ on, 
say, point $k=2$ for $D_4$ algebra  (i.e. $\alpha=-2, \beta=4, \gamma=4$) is not defined, since the limit of $0/0$ ambiguity is dependent on the direction of approaching that point. 
However, if one approaches that point by one of the relevant lines, e.g. $exc$ or $so$, reasonable results are obtained. 

We see that $X_2(3,\beta)$ for $D_4$ gives pure number $3$, independent on $x$. This should be interpreted as quantum dimension of some representation 
of semidirect product of $D_4$ and its Dynkin diagram's automorphism group $S_3$. We assume that the corresponding representation is the trivial one for $D_4$ factor 
and the non-trivial reducible three-dimensional  permutation representation of $S_3$ factor. 
  
   \begin{table} 
   \caption{$X_2(k,\beta,\alpha,\gamma)$ for the exceptional algebras} \centering
       \begin{tabular}{|c|c|c|c|c|c|}
       \hline
            $k$&1&2&3&4&$\geq5$\\
            \hline
         $G_2$&$3\omega_1$&0&0&0&0\\
         \hline
         $F_4$&$\omega_2$&$3\omega_4$&0&0&0\\
         \hline
         $E_6$&$\omega_3$&$3\omega_1\oplus 3\omega_5$&$-(\omega_1+
         \omega_5)$&0&0\\
         \hline
         $E_7$&$\omega_2$&0&$-\omega_2$&-1&0\\
         \hline
      $E_8$&$\omega_6$&0&$\omega_8$&0&0\\
         \hline
         $D_4$& $\omega_1+\omega_3+\omega_4$& $\left\{\begin{tabular}{c}
           $\omega_1+\omega_3+\omega_4$ \, \text {on the} $Exc$ \text{ line}\\
           0 \, \text{on the}\, $so$ \, \text{line}
           \end{tabular} \right. $ &
         $\left\{\begin{tabular}{c}
                    3 \, \text {on the} $Exc$ \text{ line}\\
                    0 \, \text{on the}\, $so$ \, \text{line}
                    \end{tabular} \right. $&0&0\\
         \hline
         \end{tabular}
         \label{tab:x2bex}
      \end{table}   
  
  $X_2(k,\gamma, \alpha,\beta)$ for the exceptional algebras are given in table \ref{tab:x2gex}.

 \begin{table}
 \caption{$X_2(k,\gamma,\alpha,\beta)$ for the exceptional algebras} \centering
     \begin{tabular}{|c|c|c|c|c|}
     \hline 
          $k$&1&2&3&$\geq4$\\
          \hline
       $G_2$&$3\omega_1$&$3\omega_1$&0&0\\
       \hline
       $F_4$&$\omega_2$&$\omega_2$&0&0\\
       \hline
       $E_6$&$\omega_3$&$\omega_3$&0&0\\
       \hline
       $E_7$&$\omega_2$&$\omega_2$&0&0\\
       \hline
    $E_8$&$\omega_6$&$\omega_6$&0&0\\
       \hline
       $D_4$& $\omega_1+\omega_3+\omega_4$& $\left\{\begin{tabular}{c}
           $\omega_1+\omega_3+\omega_4$ \, \text {on the} $Exc$ \text{ line}\\
           0 \, \text{on the}\, $so$ \, \text{line}
           \end{tabular} \right. $ &
         $\left\{\begin{tabular}{c}
                    0 \, \text {on the} $Exc$ \text{ line}\\
                    0 \, \text{on the}\, $so$ \, \text{line}\end{tabular}\right.$&0\\
                    \hline
       \end{tabular}
       \label{tab:x2gex}
 \end{table}

Again, when restricted to the exceptional line $\alpha=y,\beta=1-y,\gamma=2$ and in the limit $x \rightarrow 0$, 
 $X_2(2,\gamma)$ gives the following formula 

$$
\frac{5 (y-6) (y-4) (y+3) (y+5)}{(y-1)^2 y^2}
$$

which coincides with the dimensional formula for $X_2$ from \cite{Cohen}, and agrees with table \ref{tab:x2gex}.

 \begin{table}
 \caption{$X_2(k,\gamma,\beta,\alpha)$ for the classical algebras} \centering
     \begin{tabular}{|c|c|c|c|c|} 
     \hline
          $k$&1&2&3&$\geq4$\\
          \hline
       $A_1$&0&$-2\omega$ on the $sl$ line&0&0\\
          \hline
       $A_2$&$3\omega_1\oplus 3\omega_2$&$-(\omega_1+\omega_2)$&0&0\\
          \hline
       $A_n,n\geq3$&$(2\omega_1+\omega_{n-1})\oplus
       (\omega_2+2\omega_{n})$&$-(\omega_1+\omega_n)$&0&0\\
          \hline
       $B_2$&$\omega_1+2\omega_2$&0 \text{on the $so$ line}&0 \text{on the $so$ line}&0\\
          \hline
       $B_3$&$\omega_1+2\omega_3$&0&0&0\\
          \hline
       $B_n,n\geq4$&$\omega_1+\omega_3$&0&0&0\\
          \hline
       $C_n,n\geq3$&$2\omega_1+\omega_2$&0&0&0\\
          \hline
       $D_4$&$\omega_1+\omega_3+\omega_4$&0 \text{on the $so$ line}&0 \text{on the $so$ line}&0\\
          \hline
       $D_5$&$\omega_1+\omega_3$&0&0&0\\
          \hline
       $D_6$&$\omega_1+\omega_3$&0 \text{on the $so$ line}&0&0\\
          \hline                
       $D_n,n\geq7$&$\omega_1+\omega_3$&0&0&0\\
          \hline
    \end{tabular}
    \label{tab:x2gcl}
\end{table}
    
    For the classical algebras $X_2(k,\beta)$ is given in the table \ref{tab:x2bcl}. For small ranked algebras there shows up a complicated picture, so we present the stabilized answers for 
    sufficiently large ranks. The boundary depends on $k$, the larger $k$, the larger the boundary. At least the rank should be large enough to allow the existence of the fundamental 
    weights mentioned in the table. 
 \begin{table}
 \caption{$X_2(k,\beta,\alpha,\gamma)$ for the classical algebras for sufficiently large $n$ (depends on $k$)} \centering
     \begin{tabular}{|c|p{2.1cm}|p{2cm}|p{2cm}|p{2cm}|c|} 
     \hline
          $k$&1&2&3&4&$\geq5$\\
          \hline
       $A_n$&$ (2\omega_1+\omega_{n-1}) \oplus (2\omega_{n}+\omega_{2})$ &$ (2\omega_2+\omega_{n-3})\oplus (2\omega_{n-1}+\omega_{4})$&$(2\omega_3+\omega_{n-5})\oplus (2\omega_{n-2}+\omega_{6})$&$(2\omega_4+\omega_{n-7})\oplus (2\omega_{n-3}+\omega_{8})$&$\cdots$\\
       \hline
       $B_n$&$\omega_1+\omega_3$&0  & $0$&$0$&$0$\\
       \hline
       $C_n$&$2\omega_1+\omega_2$&$2\omega_2+\omega_4$&$2\omega_3+\omega_6$&$2\omega_4+\omega_8$&$\cdots$\\
       \hline
       $D_n$&$\omega_1+\omega_3$&0&$0$&$0$&$0$\\
       \hline
    \end{tabular}
    \label{tab:x2bcl}
    \end{table}

Now, we present the similar results for the more general formula – $X(x,k,n,\alpha,\beta,\gamma)$.
For the exceptional algebras the results are shown via tables \ref{tab:xknbe1} and \ref{tab:xknge1}.
\begin{table} 
	\caption{$X(x,k,n,\beta,\alpha,\gamma)$ for the exceptional algebras} \centering 
	\scalebox{0.8}{\begin{tabular}{|c|c|c|c|c|c|}
		\hline
		
		$k,n$&$G_2$&$F_4$&$E_6$&$E_7$&$E_8$\\
		\hline
		1,0&$3\omega_1$&$\omega_2$ &  $\omega_3$ & $\omega_2$&$\omega_6$ \\
		\hline
		1,1& $\omega_1+\omega_2$&$\omega_3+\omega_4$&{\begin{tabular}{c}
				$(\omega_1+\omega_2)$\\
				$\oplus(\omega_4+\omega_5)$
			\end{tabular}} & $\omega_6+\omega_7$ & $\omega_8$ \\
		\hline
		1,2& 0 & $\omega_1+\omega_4$ & $\omega_3$&0 & $-\omega_8$ \\
		\hline
		1,3& 0 & 0 & 0& E:$-2\omega_6$ & $-\omega_6$ \\
		\hline
		1,4& 0 & 0 & -1 & 0 & 0  \\
		\hline
		1,5& 0 & 0 & 0 & 0 & 1 \\
		\hline
		2,0& 0 & $3\omega_4$ & $3\omega_1 \oplus 3\omega_5$ & 0 & 0 \\
		\hline
		2,1& 0 & 0 & $-\omega_3$ & E:$-\omega_6-\omega_7$ & 0 \\
		\hline
		2,2& 0 & 0 & $-\omega_6$ & $-\omega_5$ & $\omega_6$ \\
		\hline
		2,3& 0 & 0 & 0 & 0 & $\omega_7$ \\
		\hline
		3,0& 0 & 0 &  $-\omega_1-\omega_5$ & E:$-\omega_2$ & $\omega_8$ \\
		\hline
		3,1& 0 & 0 & 0 & E:$-\omega_1$ & $\omega_1$ \\
		\hline
		4,0& 0 & 0 & 0 & -1 & 0 \\
		\hline
	\end{tabular}
	}
	\label{tab:xknbe1} \normalsize
\end{table}  

\begin{table} 
	\caption{$X(x,k,n,\gamma,\alpha,\beta)$ for the exceptional algebras} \centering
	\scalebox{0.8}{
	\begin{tabular}{|c|c|c|c|c|c|}
		\hline
		
		$k,n$&1,0&1,1&1,3&2,0&2,1\\
		\hline
		$G_2$&$3\omega_1$&$-3\omega_1$&1&$3\omega_1$&$\omega_2$\\
		\hline
		$F_4$&$\omega_2$ &$-\omega_2$&1&$\omega_2$&$\omega_1$\\
		\hline
		$E_6$& $\omega_3$&$-\omega_3$&1&$\omega_3$ &$\omega_6$  \\
		\hline
		$E_7$& $\omega_2$&$-\omega_2$&1&$\omega_2$&$\omega_1$ \\
		\hline
		$E_8$&$\omega_6$ &$-\omega_6$&1&$\omega_6$&$\omega_7$\\
		\hline
		
	\end{tabular}
	}
	\label{tab:xknge1}
\end{table}  

For the points associated with the classical algebras one has the results shown in tables \ref{tab:xknb} and \ref{tab:xkng1}.

As we see the situation is more complex in this case..
 In the table \ref{tab:xknb} we present the outputs of the $X(x,k,n,\beta,\alpha,\gamma)$ for "sufficiently large" rank of the corresponding algebra. 

One can prove the following 

{\bf Proposition 2.3}. 

{\it At the points in Vogel's plane, corresponding to the classical algebras with sufficiently large ranks, the function 
$X(x,k,n,\beta,\alpha,\gamma)$  
 is equal to the quantum dimension of representation of the corresponding algebra given in the table \ref{tab:xknb}.
 The ranges of the "sufficiently large" ranks ("Validity range") are presented in the tables.}

{\bf Remark 1.} The columns "Validity range/Regularity range" show the range of the rank $N$ where $X(x,k,n,\beta,\alpha,\gamma)$ yields the quantum dimension of the 
representation, given in the previous column, and the range of the rank $N$ where our formula is non-singular, respectively.
However, we do not claim that for the ranks less than the boundary, given in the "Regularity range",
 our formula is always singular. It is singular only for some of the ranks less than that boundary. We give some examples below. 

{\bf Remark 2.} For the ranks smaller than the boundary of the validity range we assume that $X(x,k,n,\beta,\alpha,\gamma)$ still yields quantum dimensions of some representations
 of the corresponding algebra. We do not prove that, and just present some information on the low ranks for the specific algebras. 

{\bf Remark 3.}Proposition 2.3 is proved by a direct case by case comparison of the output of our formula with the Weyl formula written for the corresponding 
highest weights. Calculations are similar to those implemented for the proof of  Proposition 2.2 given in the Appendix C.II, and we omit them. 

The possible singularities that may appear in the low rank domain will be studied in Section 3.

\begin{table} 
	\caption{$X(x,k,n,\beta,\alpha,\gamma)$ for the classical algebras} \centering
	\scalebox{0.8}{
	\begin{tabular}{|c|c|c|c|c|}
		\hline
		$k,n$&$1,n$& \vtop{\hbox{\strut Validity range} \hbox{\strut Regularity range}} &$k,n\,\, , \,\,k\geq 2$& \vtop{\hbox{\strut Validity range} \hbox{\strut Regularity range}}\\
		\hline
		$A_N$&  \vtop{\hbox{\strut  $(\omega_1+\omega_{1+n}+\omega_{N-1-n}) \oplus$}\hbox{\strut  $(\omega_{N}+\omega_{N-n}+\omega_{n+2})$}}    & \vtop{\hbox{\strut $N>2n+1$} \hbox{\strut $N>2n+1$}}  &  \vtop{\hbox{\strut  $(\omega_k+\omega_{k+n}+\omega_{N+1-2k-n})\oplus$}\hbox{\strut  $(\omega_{2k+n}+\omega_{N+1-k}+\omega_{N+1-k-n})$}}   & \vtop{\hbox{\strut $N>4k+2n-3$} \hbox{\strut $N>4k+2n-3$}}  \\
		\hline
		$B_N$&$\omega_1+\omega_{2n+3}$ & \vtop{\hbox{\strut $N>2n+3$} \hbox{\strut $N\geq2$}}  &0 &\vtop{\hbox{\strut $N\geq2$} \hbox{\strut $N\geq2$}}\\
		\hline
		$C_N$&$\omega_1+\omega_{n+1}+\omega_{n+2}$& \vtop{\hbox{\strut $N>n+1$} \hbox{\strut $N>n$}}   &$\omega_k+\omega_{k+n}+\omega_{2k+n}$&  \vtop{\hbox{\strut $N>2k+n-1$} \hbox{\strut $N>2k+n-2$}}  \\
		\hline
		$D_N$&$\omega_1+\omega_{2n+3}$& \vtop{\hbox{\strut $N>2n+4$} \hbox{\strut $N>2n+3$}}  & 0 &  \vtop{\hbox{\strut $N\geq4$} \hbox{\strut $N>4k+2n-1$}}   \\
		\hline
	\end{tabular}
	}
	\label{tab:xknb} \normalsize
\end{table}

\begin{table}[h]
	\caption{$X(x,k,n,\gamma,\alpha,\beta)$ for the classical algebras} \centering
	\begin{tabular}{|c|c|}
		\hline
		$k,n>0$&$1,2$\\
		\hline
		$A_N$&-1\\
			\hline
	\end{tabular}
	\label{tab:xkng1}
\end{table}

\section{Universal Casimir Eigenvalues on $(X_2)^k(g)^n$}

Let us now show, that the eigenvalues of the second Casimir operator on $(X_2)^k(g)^n$ representations can be written
in terms of Vogel's universal parameters.
The highest weights of the $X_2$ and $g$ are $k\lambda_{X_2}$ and $n\lambda_{g}$, correspondingly.
It is easy to check, that 
\begin{equation*}
 C_{k,n}= C_{k\lambda_{X_2}} + C_{n\lambda_{g}}
 + 2kn(\lambda_{X_2},\lambda_g)
 \end{equation*}
where $C_{k,n}$ is the Casimir eigenvalue on $(X_2)^k(g)^n$.
Substituting the corresponding highest weights (see Table \ref{tab:hw}) in the expression, written above for the Casimir eigenvalue, 
one obtains the expressions shown in the following table 4:
\begin{table}[h]
   \caption{Casimir Eigenvalues} \centering
   \scalebox{0.8}{
     \begin{tabular}{|c|c|c|c|c|}
     \hline
       
                    &$C_{k\lambda_{X_2}}$&$C_{n\lambda_{g}}$&$2kn(\lambda_{X_2},\lambda_g)$&$C_{k,n}$ \\
                                     \hline
               $A_N,N\geq3$&$6k^2+k(4N-2)$&$2n(n+N)$&$6kn$&$6k^2+k(4N-2)+2n(n+N)+6kn$\\
       \hline
               $B_N,N\geq4$&$6k^2+k(8N-10)$&$2n(n+2N-2)$&$6kn$&$6k^2+k(8N-10)+2n(n+2N-2)+6kn$\\
       \hline
        $C_N,N\geq3$&$5k^2+k(4N-1)$&$2n(n+N)$&$6kn$&$5k^2+k(4N-1)+2n(n+N)+6kn$\\
       \hline
              $D_N,N\geq5$&$6k^2+k(8N-14)$&$2n(n+2N-3)$&$6kn$&$6k^2+k(8N-14)+2n(n+2N-3)+6kn$\\
       \hline
        $G_2$&$6k^2+10k$&$2n(n+3)$&$6kn$&$6k^2+10k+2n(n+3)+6kn$\\
       \hline
       $F_4$&$6k^2+30k$&$2n(n+8)$&$6kn$&$6k^2+30k+2n(n+8)+6kn$\\
       \hline
       $E_6$&$6k^2+42k$&$2n(n+11)$&$6kn$&$6k^2+42k+2n(n+11)6kn$\\
       \hline
       $E_7$&$6k^2+66k$&$2n(n+17)$&$6kn$&$6k^2+66k+2n(n+17)+6kn$\\
       \hline
    $E_8$&$6k^2+114k$&$2n(n+29)$&$6kn$&$6k^2+114k+2n(n+29)+6kn$\\
       \hline
Universal Form&$ 3\alpha(k-k^2)+4tk   $&$ \alpha (n-n^2)+2tn  $& $-3\alpha kn$&$\alpha(3k-3k^2+n-n^2-3kn)+t(4k+2n)$\\

       \hline
       \end{tabular}
       }
       \label{tab:x2acl}
         \end{table}     
         \newline

One can check, that for each of these cases (except for the $C_N$)  the universal expression for the Casimir 
eigenvalues on the Cartan powers of $X_2$ and $g$ representations can be written through a linear function in terms of Vogel's
universal parameters:
\begin{multline}
C_{k,n}=
3\alpha(k-k^2)+4tk+\alpha (n-n^2)+2tn=\\
\alpha(3k-3k^2+n-n^2-3kn)+t(4k+2n)
 \end{multline}
which proves, that the Casimir eigenvalues on $(X_2)^k(g)^n$ representations are universal.

\subsection{Conformity Check}

Now we turn to the comparison of our universal expression with the values presented in \cite{Cohen}.
The representations on which the Casimir eigenvalues are to be compared are those defined
with the following highest weights: $2\lambda_{X_2}, \lambda_{X_2}+\lambda_g$ and $\lambda_{X_2}+2\lambda_g$.
So, we calculate $\gamma(H), \gamma(C), \gamma(G)$ (i.e. Casimirs in \cite{Cohen} notation) and compare them with
$C_{2,0}, C_{1,1}, C_{1,2}$, written in the corresponding scaling.
\newline
For the $k=2$ and $n=0$ case our formula in the corresponding scaling gives: 
$$C_{2,0}=\frac{3\alpha(k-k^2)+4tk}{2t}=\frac{-3\alpha+4t}{t}=\frac{6+4t}{t}.$$
For $k=1, n=1$ 
$$C_{1,1}=\frac{6t-3\alpha}{2t}=\frac{3(t-1)}{t}$$
Finally, for $k=1,n=2$
$$C_{1,2}=\frac{-8\alpha+8t}{2t}=\frac{8+4t}{t}$$

In the following table the corresponding Casimir eigenvalues calculated in \cite{Cohen} and those obtained by our formula are shown.\begin{table}[ht]
\caption{Conformity check of the Casimir Eigenvalues}     \label{tab:V1}
 \centering
 \scalebox{0.75}{
\begin{tabular}{|c|c|c|c|c|c|c|c|c|}
\hline  
&$a$&$\gamma(H)=4+6a$&$\gamma(C)=3+3a$&$\gamma(G)=4+8a$&$t$&$C_{2,0}=(6+4t)/t$&$C_{1,1}=3(t-1)/t$&$C_{1,2}=(8+4t)/t$\\
\hline  
$A_1$ &  $1/2$     & $7$ & $9/2$ & $8$ &  $2$     & $7$ & $9/2$ & $8$  \\
\hline
$A_2$ &   $1/3$    & $6$ & $4$ & $20/3$ &  $3$    & $6$ & $4$ & $20/3$ \\
\hline
$G_2$ & $ 1/4$    & $11/2$ & $15/4$ & $8$ & $4$    & $11/2$ & $15/4$ & $8$ \\
\hline
$D_4$ &   $1/6$    & $5$ &  $7/2$ & $16/3$ &  $6$    & $5$ & $7/2$ & $16/3$\\
\hline
$F_4$ & $1/9$    & $14/3$ & $10/3$ & $14/3$ & $9$    & $14/3$ & $10/3$ & $14/3$ \\
\hline
$E_6$ &  $1/12$    & $9/2$ & $13/4$ & $14/3$ & $12$    & $9/2$ & $13/4$ & $14/3$\\
\hline
$E_7$ & $1/18$    & $13/3$ &  $19/6$ & $40/9$ & $18$    & $13/3$ & $19/6$ & $40/9$  \\
\hline
$E_8$ & $1/30$    & $21/5$& $31/10$ & $64/15$ & $30$    & $21/5$ & $31/10$ & $64/15$\\
\hline  
\end{tabular} 
}
\end{table}

Thus, we see that the Casimir eigenvalues coincide.

 \subsection{Non-zero Universal Values of Casimir on Zero Representations} 
  
A notable quality of the $X_2(k,\alpha,\beta,\gamma)$ formula, presented above, is that for the
parameters, corresponding to the $C_N$ algebra it gives $0$ for any $k\geq2$, while we see that
the Casimir eigenvalues on those irreps are not 0.
 \newline
 A similar situation regarding $A_2$ algebra takes place. The universal decomposition of the symmetric square of the adjoint representation 
 writes as follows:
 
 $$S^2 \mathfrak{g}= 1 +Y_2(\alpha)+Y_2(\beta)+Y_2(\gamma) $$
 
 The $Y_2(\beta)$ for $A_2$ is $0$, whilst the Casimir eigenvalue on the same representation is
 $4t-2\beta$. 
 At first glance
it seems natural to expect, that the Casimir eigenvalues on that representations should be equal to 0, while we see,
that they are not.
If one thinks deeper, it is easy to understand, that the Casimir eigenvalue does not have to be equal to 0 on a zero-dimensional
representation.
Indeed, for the points close to the (-2,2,3) on the Vogel plane the Casimir operator acting on the symmetric 
square of the adjoint representation of $A_2$ has three eigenvalues, so in an appropriate basis, it has a block-diagonal form.
At (-2,2,3) point all that happens is that $Y_2(\gamma)$ becomes zero for that particular combination of parameters, and the 
corresponding block of the Casimir operator acts on a zero-dimensional vector subspace. Thus we do not see anything that dictates 
that block to be a zero-matrice at that particular point. 
\newline
After the discussion of this situation one concludes, that the universal description sheds a light on the fact, that
it is not just only reasonable, but turns out to be necessary to consider some non-zero eigenvalues of Casimir operators
on non-existing, i.e. zero-dimensional representations. 
Thus, it seems natural to believe, that the universal formulae "take care" of the "invisibility" of that sort of Casimirs.
In other words, we expect that in the universal formulae the Casimir eigenvalues appear in the product with 
the universal dimensions, or, more generally, with expressions, which are necessarily zero, if
the dimension is zero.
\newline
 In support of this idea we bring a formula, presented by Deligne in
 \cite{Del}:
 $$Tr(C_2, [R]V)=\frac{1}{n!}\sum_{\sigma}\chi(\sigma)m(\sigma)(dimV)^{n(\sigma)-1}Tr(C_2, V)$$
 where $V$ is a representation of the algebra, $R$ is a representation of the $S_n$ group, 
 $[R](V):=Hom_{S_n}(R, \otimes^nV)$, $\sigma$ is an element of $S_n$,
  $\chi(\sigma)$ is the character on that element, $m(\sigma)$ is the sum of the squares of the lengths of cycles 
  of $\sigma$, $n(\sigma)$ is the number of cycles of $\sigma$.
\newline
For the symmetric square of the adjoint representation, we rewrite this formula explicitly:
$$1\cdot C_2(\mathbb{1})+ \text{dim}Y_2(\alpha)C_2(Y_2(\alpha))+ \text{dim}Y_2(\beta)C_2(Y_2(\beta))+ \text{dim}Y_2(\gamma)C_2(Y_2(\gamma))=
(2+ \text{dim}g)\cdot  \text{dim}g C_2(g),$$
where $g$ is the adjoint representation.

Substituting the corresponding universal formulae, one can check, that for $A_2$ algebra this formula is true.

\subsection{Conformity With $sp(-2n)=so(2n)$ Duality}
In (\cite{MV}) R.Mkrtchyan and A.Veselov  have discussed the duality of higher-order Casimir operators for $SO(2n)$ and $Sp(2n)$ groups. Using the Perelomov and Popov 
(\cite{PP}) formula for the generating function for the Casimir spectra and parametrizing the 
Young diagrams in a different way (\cite{MV}), they have explicitly shown the $C_{Sp(2n)}(\lambda,z)=-C_{SO(-2n)}(\lambda',-z)$ duality for 
arbitrary Young diagrams.
\newline
Here we write the expressions for the corresponding eigenvalues of the second Casimir operator
($C_2$) for $so(2n)$ and $sp(2n)$ algebras, in the $A, B$ parametrization, used in \cite{MV}. 
\subsection*{$so(2n)$}
For $so(2n)$ the Casimir spectra write as follows

\begin{align*}
C_{so(2n)}(z,A,B)= \sum_{p=0}^{ \infty}C_{p_{so(2n)}} z^p=
\frac{(1- z n) (2-z (4 n-3))}{z (1-z (n-1)) (2-z (4 n-2))} \times \\
\prod _{i=0}^k \frac{1-z (-A_{k-i}+B_i+2 n-1)}{1-z (A_{k-i}-B_i)} \times
\prod _{i=1}^k \frac{1-z (A_{-i+k+1}-B_i)}{1-z (-A_{-i+k+1}+B_i+2 n-1)}
\end{align*}

After a proper expansion of $C_{so(2n)}(z,A,B)$ into series in the vicinity of the $z_0=0$ point, one can check, that the coefficient of
$z^2$, i.e. $C_{2_{so(2n)}}$ can be expressed as follows:

\begin{align*}
C_{2_{so(2n)}}(A,&B)=  \sum _{i=1}^k \big(4 n A_i (B_{-i+k+1}-B_{k-i})+2 A_i^2 (B_{k-i}-B_{-i+k+1})+ \\
 &+ 2 A_i (B_{k-i}-B_{-i+k+1})+2 B_i^2 (A_{-i+k+1}-A_{k-i}) \big)-4 n A_0 B_k+\\
 &+A_0^2 (2 B_k+4 B_0)+2 A_0 (B_k-B_0)-B_0^2 (2 A_k+4 A_0)-\\
 &-n (A_0-B_0)+2 n \left(A_0^2+B_0^2\right)+2 \left(B_0^3-A_0^3\right)+1/2(A_0-B_0).
\end{align*}

\subsection*{$sp(2n)$}
The Casimir spectra for this case is
\begin{align*}
C_{sp(2n)}(z, &A, B)= \sum_{p=0}^{\infty} C_{p_{sp(2n)}} z^p=\frac{(1- z n) (2-z (4 n+3))}{z (1-z (n+1)) (2-z (4 n+2))} \times  \\
& \prod _{i=0}^k \frac{1-z (B_{k-i}-A_i+2 n+1)}{1-z (-B_{k-i}+A_i)} \times \prod _{i=1}^k \frac{1-z (-B_{-i+k+1}+A_i)}{1-z (B_{-i+k+1}-A_i+2 n+1)}
\end{align*}
And for $C_{2sp(2n)}$ one has
\begin{align*}
C_{2_{sp(2n)}}(A,&B) =  -\sum _{i=1}^k \big(-4 n B_i (A _{-i+k+1}-A _{k-i})+2 A_i^2 (B _{-i+k+1}-B_{k-i})+\\
& 2 B_i^2 (A_{k-i}-A_{-i+k+1})++2 B_i (A_{k-i}-A_{-i+k+1})\big)-4n B_0 A_k+\\
&+A_0^2 (2 B_k+4B_0)-2 B_0 (A_k-A_0)-B_0^2 (2 A_k+4A_0)-\\
&-n (B_0-A_0)+2 n \left(A_0^2+B_0^2\right)+1/2(A_0-B_0)-2 \left(A_0^3-B_0^3\right).
\end{align*}

Therefore, we have obtained formulae for second Casimir eigenvalues on irreps of  $sp(2n)$ and $so(2n)$ algebras,
corresponding to any Young diagram (any $(A,B)$ set).
\newline
It can be checked, that
$$C_{2_{so(2n)}}(A,B)=-C_{2_{sp(-2n)}}(B,A)$$
i.e. the Casimir duality for the second Casimir holds for any Young diagram (for any $A,B$ set).
In particular, for $X_2$ one has the values, shown in the Table 4.
It can be observed, that $C_{2_{so(2n)}}=2C_{2_{sp(2n)}}=1/2C_{2_{so(2n)}}(A,B)$, which indicates the difference of the definition of the
Killing form in \cite{MV} \footnote{in \cite{MV} the Killing form is defined as $Tr(\hat{X^a}, \hat{X^b})$ in the fundamental representation, while our normalization (so called Cartan-Killing normalization)
corresponds to the Killing form, defined as $Tr(ad\hat{X^a}, ad\hat{X^b})$, i.e. in the adjoint representation. }.

\begin{table}
\caption{Comparison} 
 \centering
\begin{tabular}{|c|c|c|c|c|}
\hline  
$Algebra$ & $ Diagram$ & $A,B$ & $C_2(A,B)$ & $C_2$ \\
\hline  
$so(2n)$ & \scriptsize \ydiagram{2,1,1} & $A_1=B_1=1,A_2=3,B_2=2$  & $16n-16$ & $8n-8$ \\
\hline
$sp(2n)$ & \scriptsize  \ydiagram{3,1} & $A_1=B_1=1,A_2=2,B_2=3$   & $16n+16$ & $4n+4$ \\
\hline
\end{tabular}
\end{table}
 
 In \cite{AM} it has been shown, that when permuting the Vogel parameters
 corresponding to the $so(2n)$ algebra in this way: $(\alpha, \beta, \gamma)\to (\beta, \alpha, \gamma)$, the $X_2(k)$ formula
 gives dimensions for some representations of the $sp(2n)$ algebra. More precisely, that permutation specifies a correspondence
 between $\lambda^{so(2n)}=k(\omega_1+\omega_3)$ and $\lambda^{sp(2n)}=2\omega_k+\omega_{2k}$ representations.
 One can notice, that the Young diagrams, associated with these representations are conjugate with each other. Indeed, in $A,B$
 parametrization the associated sets are $$\lambda^{so(2n)} \leftrightarrow {A_0=B_0=0, A_1=1, B_1=k, A_2=3, B_2=2k},$$
 $$\lambda^{sp(2n)} \leftrightarrow {A_0=B_0=0, A_1=k, B_1=1, A_2=2k, B_2=3}.$$
 Therefore, it is reasonable to check the Casimir duality for these representations.
 Substituting the corresponding $(A,B)$ sets into the expressions for $C_2(A,B)$ written above, one gets
 $$C_{2_{so(2n)}}(A,B)=12k^2+k(16n-28),$$
 $$C_{2_{sp(2n)}}(B,A)=-12k^2+k(16n+28)=-(12k^2+k(16(-n)-28)=-C_{2_{so(2n)}}(A,B).$$
 So, the Casimir duality holds for representations, associated with the  \newline
 $X_2(k,-2,4,2n-4) \leftrightarrow X_2(k,4,-2,2n-4)$ transformation
 of the $X_2(k,\alpha,\beta,\gamma)$ universal formula \cite{AM}.\newline
 For the same representations in the Cartan-Killing normalization we have $$C_{2_{so(2n)}}=6k^2+k(8n-14),$$  $$C_{2_{sp(2n)}}=-3k^2+k(4n+7),$$
 i.e. $$C_{2_{so(2n)}}(\lambda)=-2C_{2_{sp(-2n)}}(\lambda'),$$ as expected.

\section{Appendix C.II}

\subsection*{Proof of the Propositions}

The proof is carried out case by case:  
for each set of the parameters 
$\alpha, \beta, \gamma$ from the Vogel's table  (except  $C_n$) we
compare the expression (\ref{main})  with the  quantum
dimension obtained by Weyl formula (\ref{W}) for the corresponding algebra.

\subsection{$A_{N-1}$}
 Substituting $\alpha=-2,\beta=2,\gamma=N$
in the 
$L$-terms, one gets
$$L_{31}=\sinh\left[\frac{x}{2}: \right.
\frac{  
N+3k+n-2}{ N-2 },$$

$$L_{32}=\sinh\left[\frac{x}{2}: \right.
\frac{ N+3k+2n-1}{N-1},$$

$$L_{21s1}=\sinh\left[\frac{x}{2}: \right.
\frac{  
  (N-2) \cdot  
\left(N-1 \right)\dots  
\left(N+2k+n-3  \right)}{1\cdot2\dots (2k+n)},$$

$$L_{21s2}=\sinh\left[\frac{x}{2}: \right.
\frac{  
\left(N-1\right)\cdot  
  N \dots  
\left(N+2k+n-2\right)}{ N/2\cdot (N/2+1) \dots (N/2+2k+n-1)},$$

$$L_{21s3}=\sinh\left[\frac{x}{2}: \right.
\frac{ N/2+2k+n-1}{N/2-1},$$

$$L_{10s1}=\cdot \sinh\left[\frac{x}{2}: \right.
\frac{ (N-2)\cdot
   (N-1)\cdot N\dots 
(N+k-3)  }{ 1
\cdot 2 \dots
k},$$

$$L_{10s2}=\cdot \sinh\left[\frac{x}{2}: \right.
\frac{ (N/2-1)\cdot
   N/2\cdot (N/2+1)\dots 
(N/2+k-2)  }
{(\alpha+\beta)
\cdot 1 \dots
(k-1)},$$

$$L_{10s3}=- \sinh\left[\frac{x}{2}: \right.
\frac{ (2\alpha+2\beta)\cdot
   1\cdot 2\dots 
(k-1) }{ (N/2-1)
\cdot N/2 \dots
(N/2+k-2)},$$

$$L_{11s1}=\sinh\left[\frac{x}{2}: \right.
\frac{  
\left(N/2-1\right)\cdot  
\left(N/2\right)\dots  
\left(N/2+k+n-2 \right)}
{ 2\cdot3\dots (k+n+1)},$$

$$L_{11s2}=\sinh\left[\frac{x}{2}: \right.
\frac{ N/2 \cdot
\left(N/2+1\right) \dots  
\left(N/2+k+n-1 \right)}
{ 1\cdot 2\cdot 3\dots (k+n)},$$

$$L_{11s3}=1/L_{11s2}$$

$$L_{01}=\sinh\left[\frac{x}{2}: \right. 
\frac{n+1}{1}$$

$$L_{c2}=\sinh\left[\frac{x}{2}: \right.
\frac{ (N/2+k+n-1) \cdot
\left(N/2+k+n)\right) \dots  
\left(N/2+2k+n-2 \right)}
{ (N+k+n-1)\cdot (N+k+n)\dots (N+2k+n-2)},$$

The product of all these terms gives 

\begin{multline*}
   X(x,k,n,-2,2,N+1)=\\
   =L_{31}\cdot L_{32}\cdot L_{21s1}\cdot L_{21s2}\cdot L_{21s3}\cdot L_{10s1}\cdot
L_{10s2}\cdot L_{10s3}\cdot L_{11s1}\cdot L_{11s2}\cdot L_{11s3}\cdot L_{01}\cdot L_{c2}=\\
2\cdot \sinh\left[\frac{x}{2}: \right.
\prod _{i=1}^k \frac{i+N-3}{i} \times \prod _{i=1}^{k+n} \frac{i+N-2}{i+1} \times \prod _{i=1}^{2 k+n} \frac{i+N-3}{i} \times \\ 
\frac{n+1}{1} \cdot \frac{N+3k+2n-1}{N-1} \cdot \frac{N+3k+n-2}{N-2}
\end{multline*}

which equals to the double of the expression of the Weyl formula, written for $\lambda=(2k+n)\omega_1+k\omega_{N-1}+n\omega_N$ highest weight representation of $A_N$ algebra, 
as expected.

\subsection{$B_N$}

 For this case we should substitute $\alpha=-2,\beta=4,\gamma=2N-3$, so 
 
$$L_{31}=\sinh\left[\frac{x}{2}: \right.
\frac{2N+3k+n-3}
{2N-3},$$

$$L_{32}=\sinh\left[\frac{x}{2}: \right.
\frac{2N+3k+2n-2}
{2N-2},$$

$$L_{21s1}=\sinh\left[\frac{x}{2}: \right.
\frac{  
  (2N-3) \cdot  
\left(2N-2 \right)\dots  
\left(2N+2k+n-4  \right)}{3\cdot4\dots (2+2k+n)},$$

$$L_{21s2}=\sinh\left[\frac{x}{2}: \right.
\frac{  
\left(2N-3\right)\cdot  
  (2N-2)\dots  
\left(2N+2k+n-4\right)}{ (N-1/2)\cdot (N+1/2) \dots (N+2k+n-3/2)},$$

$$L_{21s3}=\sinh\left[\frac{x}{2}: \right.
\frac{ N+2k+n-1/2}{N-1/2},$$

$$L_{10s1}=\cdot \sinh\left[\frac{x}{2}: \right.
\frac{ (2N-5)\cdot
   (2N-4)\dots 
(2N+k-6)  }{ 1
\cdot 2 \dots
k},$$

$$L_{10s2}=\cdot \sinh\left[\frac{x}{2}: \right.
\frac{ (N-3/2)\cdot
   (N-1/2)\dots 
(N+k-5/2)  }
{1\cdot 2 \dots
k},$$

$$L_{10s3}=\cdot \sinh\left[\frac{x}{2}: \right.
\frac{ 2\cdot
   3\cdot 4\dots 
(k+1) }{ (N-5/2)
\cdot (N-3/2) \dots
(N+k-7/2)},$$

$$L_{11s1}=\sinh\left[\frac{x}{2}: \right.
\frac{  
\left(N-1/2\right)\cdot  
\left(N+1/2\right)\dots  
\left(N+k+n-3/2 \right)}
{ 2\cdot3\dots (k+n+1)},$$

$$L_{11s2}=\sinh\left[\frac{x}{2}: \right.
\frac{ (N-1/2) \cdot
\left(N+1/2\right) \dots  
\left(N+k+n-3/2 \right)}
{ 2\cdot 3\cdot 4\dots (k+n+1)},$$

$$L_{11s3}=\sinh\left[\frac{x}{2}: \right.
\frac{ 3 \cdot
4 \dots  (k+n+2)}
{ (N-3/2)\cdot (N-1/2)\dots (N+k+n-5/2)},$$

$$L_{01}=\sinh\left[\frac{x}{2}: \right. 
\frac{n+1}{1}$$

$$L_{c2}=\sinh\left[\frac{x}{2}: \right.
\frac{ (N+k+n-1/2) \cdot
(N+k+n+1/2) \dots  
\left(N+2k+n-3/2 \right)}
{ (2N+k+n-4)\cdot (2N+k+n-3)\dots (2N+2k+n-5)},$$

So, the product of all  $L$-terms is:
\begin{multline*}
     X(x,k,n,-2,4,2N-3)=\\
     L_{31}\cdot L_{32}\cdot L_{21s1}\cdot L_{21s2}\cdot L_{21s3}\cdot L_{10s1}\cdot
L_{10s2}\cdot L_{10s3}\cdot L_{11s1}\cdot L_{11s2}\cdot L_{11s3}\cdot L_{01}\cdot L_{c2}=\\
\sinh\left[\frac{x}{2}: \right.
\prod _{i=1}^k \frac{i+2 N-6}{i}\times \prod _{i=1}^{2 k+n} \frac{i+2 N-4}{i+2} \times \prod _{i=1}^{k+n} \frac{i+2 N-5}{i+1} \times \\
\frac{n+1}{1} \cdot \frac{k+1}{1}  \cdot \frac{k+n+2}{2} \cdot \frac{N+2 k+n-1/2}{N-1/2}\cdot \frac{N + k + n  - 3/2}{N - 3/2} \cdot \\
\frac{N + k  - 5/2}{N-5/2} \cdot \frac{2N+3k+2n-2}{2N-2} \cdot \frac{2N+3k+n-3}{2N-3} \cdot \frac{2N+2k+n-4}{2N-4}.
 \end{multline*}

It coincides with the Weyl formula, written for $\lambda=k\omega_1+n\omega_2+k\omega_3$ highest weight representation of $B_N$ algebra.

\subsection{$C_N$}

 The Vogel parameters in this case are $\alpha=-2,\beta=1,\gamma=N+2$, and we notice, that for $k\geq2$ and for any $n$, the formula gives $0$, due to the contribution of $L_{10s3}$ term.
 So, we observe the $L_{?}$ terms for $k=1$ and any $n$.
 Thus, one has
 
$$L_{31}=\sinh\left[\frac{x}{2}: \right.
\frac{N+n+2}
{N-1},$$

$$L_{32}=\sinh\left[\frac{x}{2}: \right.
\frac{N+2n+3}
{N},$$

$$L_{21s1}=\sinh\left[\frac{x}{2}: \right.
\frac{  
  (N-1) \cdot  
N \dots  
\left(N+n \right)}{(-\alpha-2\beta)/2\cdot1\dots (n+1)},$$

$$L_{21s2}=\sinh\left[\frac{x}{2}: \right.
\frac{  
\left(N+1/2\right)\cdot  
  (N+3/2)\dots  
\left(N+n+3/2\right)}{ (N/2+1/2)\cdot (N/2+3/2) \dots (N/2+n+3/2)},$$

$$L_{21s3}=\sinh\left[\frac{x}{2}: \right.
\frac{ N/2+n+1}{N/2-1},$$

$$L_{10s1}=\cdot \sinh\left[\frac{x}{2}: \right.
\frac{ N  }{1},$$

$$L_{10s2}=\cdot \sinh\left[\frac{x}{2}: \right.
\frac{ N/2-1/2 }
{1/2},$$

$$L_{10s3}=\cdot \sinh\left[\frac{x}{2}: \right.
\frac{ 1 }{ N/2},$$

$$L_{11s1}=\sinh\left[\frac{x}{2}: \right.
\frac{  
\left(N/2-1\right)\cdot  
N/2\dots  
\left(N/2+n-1 \right)}
{ 2\cdot3\dots (n+2)},$$

$$L_{11s2}=\sinh\left[\frac{x}{2}: \right.
\frac{ (N/2+1/2) \cdot
\left(N/2+3/2\right) \dots  
\left(N/2+n+1/2 \right)}
{1/2\cdot 3/2\cdot 5/2\dots (n+1/2)},$$

$$L_{11s3}=\sinh\left[\frac{x}{2}: \right.
\frac{ (-\alpha-2\beta)/2 \cdot
1 \dots  n}
{ (N/2+1)\cdot (N/2+2)\dots (N/2+n+1)},$$

$$L_{01}=\sinh\left[\frac{x}{2}: \right. 
\frac{n+1}{1}$$

$$L_{c2}=\sinh\left[\frac{x}{2}: \right.
\frac{ N/2+n}
{ N+n+2},$$

And the product is
\begin{multline*}
     X(x,k,n,-2,4,2N-3)=\\
     L_{31}\cdot L_{32}\cdot L_{21s1}\cdot L_{21s2}\cdot L_{21s3}\cdot L_{10s1}\cdot
L_{10s2}\cdot L_{10s3}\cdot L_{11s1}\cdot L_{11s2}\cdot L_{11s3}\cdot L_{01}\cdot L_{c2}=\\
\sinh\left[\frac{x}{2}:
 \right. \prod _{i=1}^{n+2} \left( \frac{i+N-2}{i} \cdot \frac{i+N-1/2}{i-1/2} \right) \times 
 \frac{(N/2-1/2) \cdot(N+2n+3)}{(N-1) \cdot (N/2+n+3/2)} \times \frac{n+3/2}{1/2}
  \end{multline*}
which coincides with the Weyl formula, written for $\lambda=(2+2n)\omega_1+\omega_2$ highest weight representation of $C_N$ algebra.

\subsection{$D_N$}
 For this case
we substitute $\alpha=-2,\beta=4,\gamma=2N-4$, so $L$-terms become

$$L_{31}=\sinh\left[\frac{x}{2}: \right.
\frac{2N+3k+n-4}
{2N-4},$$

$$L_{32}=\sinh\left[\frac{x}{2}: \right.
\frac{2N+3k+2n-3}
{2N-3},$$

$$L_{21s1}=\sinh\left[\frac{x}{2}: \right.
\frac{  
  (2N-4) \cdot  
\left(2N-3 \right)\dots  
\left(2N+2k+n-5  \right)}{3\cdot4\dots (2+2k+n)},$$

$$L_{21s2}=\sinh\left[\frac{x}{2}: \right.
\frac{  
\left(2N-4\right)\cdot  
  (2N-3)\dots  
\left(2N+2k+n-5\right)}{ (N-1)\cdot N \dots (N-2+2k+n)},$$

$$L_{21s3}=\sinh\left[\frac{x}{2}: \right.
\frac{ N+2k+n-1}{N-1},$$

$$L_{10s1}=\cdot \sinh\left[\frac{x}{2}: \right.
\frac{ (2N-6)\cdot
   (2N-5)\dots 
(2N+k-7)  }{ 1
\cdot 2 \dots
k},$$

$$L_{10s2}=\cdot \sinh\left[\frac{x}{2}: \right.
\frac{ (N-2)\cdot
   (N-1)\dots 
(N+k-3)  }
{1\cdot 2 \dots
k},$$

$$L_{10s3}=\cdot \sinh\left[\frac{x}{2}: \right.
\frac{ 2\cdot
   3\cdot 4\dots 
(k+1) }{ (N-3)
\cdot (N-2) \dots
(N+k-4)},$$

$$L_{11s1}=\sinh\left[\frac{x}{2}: \right.
\frac{  
\left(N-1\right)\cdot  
N\dots  
\left(N+k+n-2 \right)}
{ 2\cdot3\dots (k+n+1)},$$

$$L_{11s2}=\sinh\left[\frac{x}{2}: \right.
\frac{ (N-1) \cdot
N \dots  
\left(N+k+n-2 \right)}
{ 2\cdot 3\cdot 4\dots (k+n+1)},$$

$$L_{11s3}=\sinh\left[\frac{x}{2}: \right.
\frac{ 3 \cdot
4 \dots  (k+n+2)}
{ (N-2)\cdot (N-1)\dots (N+k+n-3)},$$

$$L_{01}=\sinh\left[\frac{x}{2}: \right. 
\frac{n+1}{1}$$

$$L_{c2}=\sinh\left[\frac{x}{2}: \right.
\frac{ (N+k+n-1) \cdot
(N+k+n) \dots  
(N+2k+n-2)}
{ (2N+k+n-5)\cdot (2N+k+n-4)\dots (2N+2k+n-6)},$$

Overall, for $X(x,k,n,-2,4,2N-4)$ one gets
\begin{multline*}
     X(x,k,n,-2,4,2N-4)=\\
     L_{31}\cdot L_{32}\cdot L_{21s1}\cdot L_{21s2}\cdot L_{21s3}\cdot L_{10s1}\cdot
L_{10s2}\cdot L_{10s3}\cdot L_{11s1}\cdot L_{11s2}\cdot L_{11s3}\cdot L_{01}\cdot L_{c2}=\\
\sinh\left[\frac{x}{2}: \right.
\prod _{i=1}^{k+n} \frac{i+2 N-5}{i+1} \times \prod _{i=1}^{2 k+n} \frac{i+2 N-5}{i+2} \times \prod _{i=1}^k \frac{i+2 N-7}{i} \times \\
 \frac{n+1}{1} \cdot \frac{k+1}{1} \cdot \frac{k+n+2}{2} \cdot
 \cdot \frac{N + 2 k + n - 1}{N-1} \cdot  \frac{N +  k + n - 2}{N-2} \cdot \\ \frac{N +  k - 3}{N-3} \cdot 
 \frac{2 N + 3 k + 2n - 3}{2N-3}
\cdot \frac{2 N + 3 k + n - 4}{2N-4} \cdot \frac{2 N + 2 k + n - 5}{2 N + k + n -5}
 \end{multline*}

This coincides with the Weyl formula answer for $\lambda=k\omega_1+n\omega_2+k\omega_3$ highest weight representation.

\subsection{$G_2$}
For $G_2$ exceptional algebra Vogel's parameters 
take  values $\alpha=-2, \beta=10/3, \gamma=8/3$.
Substituting them in the $L$-terms, one has

$$L_{31}=\sinh\left[\frac{x}{2}: \right.
\frac{3k+n+2}{2},$$

$$L_{32}=\sinh\left[\frac{x}{2}: \right.\frac{3k+2n+3}{3}
,$$
$$L_{21s1}\times L_{21s2}=1,$$

$$L_{21s3}=\sinh\left[\frac{x}{2}: \right.\frac{2k+n+5/3}{5/3}
,$$

$$L_{10s1} \times L_{10s2}=1,$$

$$L_{10s3}=\sinh\left[\frac{x}{2}: \right.\frac{4/3\cdot7/3
\cdot10/3\dots(k+1/3)}
{1/3\cdot 4/3\dots\ (k-2/3)}=\sinh\left[\frac{x}{2}: \right.\frac{k+1/3}
{1/3},$$

$$L_{11s1} \times L_{11s2}=1,$$

$$L_{11s3}=\sinh\left[\frac{x}{2}: \right.\frac{7/3\cdot10/3
\dots(k+n+4/3)}
{4/3\cdot 7/3\dots\ (k+n+1/3)}=\sinh\left[\frac{x}{2}: \right.\frac{k+n+4/3}
{4/3},$$

$$L_{01}=\sinh\left[\frac{x}{2}: \right. 
\frac{n+1}{1}$$

$$L_{c2}=1$$

\begin{multline*}
 X(x,k,n,-2,10/3,8/3)=\\
 \sinh\left[\frac{x}{2}: \right.
  \frac{(n+1)(k+1/3)(k+n+4/3)(2k+n+5/3)(3k+n+2)(3k+2n+3)}
{1\cdot 1/3\cdot 4/3 \cdot 5/3 \cdot 2 \cdot 3 },
\end{multline*}

which coincides with the expression the Weyl 
formula (\ref{W}) gives for quantum dimension of $G_2$ algebra.

\subsection{$F_4$}
In this case we have $\alpha=-2, \beta=5, \gamma=6$

$$L_{31}=\sinh\left[\frac{x}{2}: \right.
\frac{3k+n+7}{7},$$

$$L_{32}=\sinh\left[\frac{x}{2}: \right.\frac{3k+2n+8}{8}
,$$

$$L_{21s1}=\sinh\left[\frac{x}{2}: \right.\frac{7\cdot8
\cdot9\dots(2k+n+6)}
{4\cdot 5\dots\ (2k+n+3)}=\sinh\left[\frac{x}{2}: \right.\frac{(2k+n+4)(2k+n+5)(2k+n+6)}
{4\cdot5\cdot6},$$

$$L_{21s2}=\sinh\left[\frac{x}{2}: \right.\frac{(2k+n+9/2)(2k+n+11/2)}{9/2\cdot 11/2}
,$$

$$L_{21s3}=\sinh\left[\frac{x}{2}: \right.\frac{2k+n+5}{5}
,$$

$$L_{10s1}=\sinh\left[\frac{x}{2}: \right.\frac{4\cdot5
\cdot6\dots(k+3)}
{1\cdot 2\dots\ k}=\sinh\left[\frac{x}{2}: \right.\frac{(k+1)(k+2)(k+3)}
{1\cdot2\cdot3},$$

$$L_{10s2}=\sinh\left[\frac{x}{2}: \right.\frac{7/2\cdot9/2
\cdot10/2\dots(k+5/2)}
{3/2\cdot 5/2\dots\ (k+1/2)}=\sinh\left[\frac{x}{2}: \right.\frac{(k+3/2)(k+5/2)}
{3/2\cdot5/2},$$

$$L_{10s3}=\sinh\left[\frac{x}{2}: \right.\frac{3\cdot4
\cdot5\dots(k+2)}
{2\cdot 3\dots\ (k+1)}=\sinh\left[\frac{x}{2}: \right.\frac{k+2}
{2},$$

$$L_{11s1}=\sinh\left[\frac{x}{2}: \right.\frac{5\cdot6
\cdot7\dots(k+n+4)}
{2\cdot 3\dots\ (k+n+1)}=\sinh\left[\frac{x}{2}: \right.\frac{(k+n+2)(k+n+3)(k+n+4)}
{2\cdot3\cdot4},$$

$$L_{11s2}=\sinh\left[\frac{x}{2}: \right.\frac{9/2\cdot11/2
\cdot13/2\dots(k+n+7/2)}
{5/2\cdot 7/2\dots\ (k+n+3/2)}=\sinh\left[\frac{x}{2}: \right.\frac{(k+n+5/2)(k+n+7/2)}
{5/2\cdot7/2},$$

$$L_{11s3}=\sinh\left[\frac{x}{2}: \right.\frac{4\cdot5
\dots(k+n+3)}
{3\cdot 4\dots\ (k+n+2)}=\sinh\left[\frac{x}{2}: \right.\frac{k+n+3}
{3},$$

$$L_{01}=\sinh\left[\frac{x}{2}: \right. 
\frac{n+1}{1}$$

$$L_{c2}=1$$

The product of all these terms

     \begin{multline*}
    X(x,k,n,-2,5,6) = \\
         \sinh\left[\frac{x}{2}: \right.
      \frac{(2 k + n+4)( 2 k + n+5)(2 k + n+6)}{4\cdot 5\cdot 6} \times \frac{(2k+n+9/2)(2k+n+11/2)}{9/2\cdot 11/2} \times \\
       \frac{(k+1) (k+2) (k+3)}{1\cdot 2\cdot 3} \times
       \frac{(k+3/2) (k+5/2)}{3/2 \cdot 5/2} \times \frac{(k+n+2) (k+n+3) (k+n+4)}{2\cdot 3\cdot 4} \times \\ \frac{(k+n+5/2) (k+n+7/2)}{5/2\cdot 7/2} \times
       \frac{n+1}{1} \cdot \frac{k+2}{2} \cdot \frac{k+n+3}{3}  \cdot \frac{2k+n+5}{5} \cdot \frac{3k+n+7}{7} \cdot \frac{3k+2n+8}{8} 
             \end{multline*}    
This immediately coincides with the  Weyl formula output for the  representations of $F_4$ 
algebra with highest weights $\lambda=k\omega_2+n\omega_1$.
 
 \subsection{$E_6$}
 For $E_6$ the Vogel parameters are $\alpha=-2, \beta=6, \gamma=8$.

$$L_{31}=\sinh\left[\frac{x}{2}: \right.
\frac{3k+n+10}{10},$$

$$L_{32}=\sinh\left[\frac{x}{2}: \right.\frac{3k+2n+11}{11}
,$$

$$L_{21s1}=\sinh\left[\frac{x}{2}: \right.\frac{10\cdot11
\dots(2k+n+9)}
{5\cdot 6\dots\ (2k+n+4)}=\sinh\left[\frac{x}{2}: \right.\frac{(2k+n+5)\dots(2k+n+9)}
{5\cdot6\dots9},$$

$$L_{21s2}=\sinh\left[\frac{x}{2}: \right.\frac{9\cdot10
\dots(2k+n+8)}
{6\cdot 7\dots\ (2k+n+5)}=\sinh\left[\frac{x}{2}: \right.\frac{(2k+n+6)(2k+n+7)(2k+n+9)}
{6\cdot7\cdot8}
,$$

$$L_{21s3}=\sinh\left[\frac{x}{2}: \right.\frac{2k+n+7}{7}
,$$

$$L_{10s1}=\sinh\left[\frac{x}{2}: \right.\frac{6\cdot7
\dots(k+5)}
{1\cdot 2\dots\ k}=\sinh\left[\frac{x}{2}: \right.\frac{(k+1)\dots(k+5)}
{1\cdot2\dots5},$$

$$L_{10s2}=\sinh\left[\frac{x}{2}: \right.\frac{5\cdot6
\dots(k+4)}
{2\cdot 3\dots\ (k+1)}=\sinh\left[\frac{x}{2}: \right.\frac{(k+2)(k+3)(k+4)}
{2\cdot3\cdot4},$$

$$L_{10s3}=\sinh\left[\frac{x}{2}: \right.\frac{4\cdot5
\dots(k+3)}
{3\cdot 4\dots\ (k+2)}=\sinh\left[\frac{x}{2}: \right.\frac{k+3}
{3},$$

$$L_{11s1}=\sinh\left[\frac{x}{2}: \right.\frac{7\cdot8
\dots(k+n+6)}
{2\cdot 3\dots\ (k+n+1)}=\sinh\left[\frac{x}{2}: \right.\frac{(k+n+2)\dots(k+n+6)}
{2\cdot3\dots6},$$

$$L_{11s2}=\sinh\left[\frac{x}{2}: \right.\frac{6\cdot7
\dots(k+n+5)}
{3\cdot 4\dots\ (k+n+2)}=\sinh\left[\frac{x}{2}: \right.\frac{(k+n+3)(k+n+4)(k+n+5)}
{3\cdot4\cdot5},$$

$$L_{11s3}=\sinh\left[\frac{x}{2}: \right.\frac{k+n+4}
{4},$$

$$L_{01}=\sinh\left[\frac{x}{2}: \right. 
\frac{n+1}{1}$$

$$L_{c2}=1$$
The product of all these terms is
  \begin{multline*}
    X(x,k,n,-2,6,8) = \\
         \sinh\left[\frac{x}{2}: \right.
         \prod _{i=5}^9 \frac{i+2 k+n}{i} \times \prod _{i=6}^8 \frac{i+2 k+n}{i} \times \prod _{i=1}^5 \frac{i+k}{i} \times \prod _{i=2}^4 \frac{i+k}{i} \times
         \prod _{i=2}^6 \frac{i+k+n}{i} \times \\ \prod _{i=3}^5 \frac{i+k+n}{i} \times  
         \frac{n+1}{1}  \cdot \frac{k+3}{3} \cdot \frac{k+n+4}{4} \cdot \frac{2k+n+7}{7}  \cdot \frac{3k+n+10}{10} \cdot \frac{3k+2n+11}{11}
         \end{multline*}

     which coincides with the quantum dimension (\ref{W}) of the $\lambda=k\omega_3+n\omega_6$ irrep.
     
     \subsection{$E_7$}
 For $E_7$ Vogel's parameters are $\alpha=-2, \beta=8, \gamma=12$.
$$L_{31}=\sinh\left[\frac{x}{2}: \right.
\frac{3k+n+16}{16},$$

$$L_{32}=\sinh\left[\frac{x}{2}: \right.\frac{3k+2n+17}{17}
,$$

$$L_{21s1}=\sinh\left[\frac{x}{2}: \right.\frac{16\cdot17
\dots(2k+n+15)}
{7\cdot 8\dots\ (2k+n+6)}=\sinh\left[\frac{x}{2}: \right.\frac{(2k+n+7)\dots(2k+n+15)}
{7\cdot8\dots15},$$

$$L_{21s2}=\sinh\left[\frac{x}{2}: \right.\frac{14\cdot15
\dots(2k+n+13)}
{9\cdot 10\dots\ (2k+n+8)}=\sinh\left[\frac{x}{2}: \right.\frac{(2k+n+9)\dots(2k+n+13)}
{9\cdot10\dots13}
,$$

$$L_{21s3}=\sinh\left[\frac{x}{2}: \right.\frac{2k+n+11}{11}
,$$

$$L_{10s1}=\sinh\left[\frac{x}{2}: \right.\frac{10\cdot11
\dots(k+9)}
{1\cdot 2\dots\ k}=\sinh\left[\frac{x}{2}: \right.\frac{(k+1)\dots(k+9)}
{1\cdot2\dots9},$$

$$L_{10s2}=\sinh\left[\frac{x}{2}: \right.\frac{8\cdot9
\dots(k+7)}
{3\cdot 4\dots\ (k+2)}=\sinh\left[\frac{x}{2}: \right.\frac{(k+3)\dots(k+7)}
{3\cdot4\dots7},$$

$$L_{10s3}=\sinh\left[\frac{x}{2}: \right.\frac{6\cdot7
\dots(k+5)}
{5\cdot 6\dots\ (k+4)}=\sinh\left[\frac{x}{2}: \right.\frac{k+5}
{5},$$

$$L_{11s1}=\sinh\left[\frac{x}{2}: \right.\frac{11\cdot12
\dots(k+n+10)}
{2\cdot 3\dots\ (k+n+1)}=\sinh\left[\frac{x}{2}: \right.\frac{(k+n+2)\dots(k+n+10)}
{2\cdot3\dots10},$$

$$L_{11s2}=\sinh\left[\frac{x}{2}: \right.\frac{9\cdot10
\dots(k+n+8)}
{4\cdot 5\dots\ (k+n+3)}=\sinh\left[\frac{x}{2}: \right.\frac{(k+n+4)\dots(k+n+8)}
{4\dots8},$$

$$L_{11s3}=\sinh\left[\frac{x}{2}: \right.\frac{k+n+6}
{6},$$

$$L_{01}=\sinh\left[\frac{x}{2}: \right. 
\frac{n+1}{1}$$

$$L_{c2}=1$$
The product of all these terms gives
  \begin{multline*}
    X(x,k,n,-2,8,12) = \\
         \sinh\left[\frac{x}{2}: \right.
         \prod _{i=7}^{15} \frac{i+2 k+n}{i} \times \prod _{i=9}^{13} \frac{i+2 k+n}{i} \times \prod _{i=1}^9 \frac{i+k}{i} \times \prod _{i=3}^7 \frac{i+k}{i} \times
         \prod _{i=2}^{10} \frac{i+k+n}{i} \times \\ \prod _{i=4}^8 \frac{i+k+n}{i} \times 
         \frac{n+1}{1} \cdot \frac{k+5}{5} \cdot \frac{k+n+6}{6}  \cdot \frac{2k+n+11}{11} \cdot \frac{3k+2n+17}{17} \cdot \frac{3k+n+16}{16}
\end{multline*}

     which coincides with quantum dimension of the $\lambda=k\omega_2+n\omega_1$ irrep.
     
     \subsection{$E_8$}
 For $E_8$ the Vogel parameters are $\alpha=-2, \beta=12, \gamma=20$.
$$L_{31}=\sinh\left[\frac{x}{2}: \right.
\frac{3k+n+28}{28},$$

$$L_{32}=\sinh\left[\frac{x}{2}: \right.\frac{3k+2n+29}{29}
,$$

$$L_{21s1}=\sinh\left[\frac{x}{2}: \right.\frac{28\cdot29
\dots(2k+n+27)}
{11\cdot 12\dots\ (2k+n+10)}=\sinh\left[\frac{x}{2}: \right.\frac{(2k+n+11)\dots(2k+n+27)}
{11\cdot12\dots27},$$

$$L_{21s2}=\sinh\left[\frac{x}{2}: \right.\frac{24\cdot25
\dots(2k+n+23)}
{15\cdot 16\dots\ (2k+n+14)}=\sinh\left[\frac{x}{2}: \right.\frac{(2k+n+15)\dots(2k+n+23)}
{15\cdot16\dots23}
,$$

$$L_{21s3}=\sinh\left[\frac{x}{2}: \right.\frac{2k+n+19}{19}
,$$

$$L_{10s1}=\sinh\left[\frac{x}{2}: \right.\frac{18\cdot19
\dots(k+17)}
{1\cdot 2\dots\ k}=\sinh\left[\frac{x}{2}: \right.\frac{(k+1)\dots(k+17)}
{1\cdot2\dots17},$$

$$L_{10s2}=\sinh\left[\frac{x}{2}: \right.\frac{14\cdot15
\dots(k+13)}
{5\cdot 6\dots\ (k+4)}=\sinh\left[\frac{x}{2}: \right.\frac{(k+5)\dots(k+13)}
{5\cdot6\dots13},$$

$$L_{10s3}=\sinh\left[\frac{x}{2}: \right.\frac{10\cdot11
\dots(k+9)}
{9\cdot 10\dots\ (k+8)}=\sinh\left[\frac{x}{2}: \right.\frac{k+9}
{9},$$

$$L_{11s1}=\sinh\left[\frac{x}{2}: \right.\frac{19\cdot20
\dots(k+n+18)}
{2\cdot 3\dots\ (k+n+1)}=\sinh\left[\frac{x}{2}: \right.\frac{(k+n+2)\dots(k+n+18)}
{2\cdot3\dots18},$$

$$L_{11s2}=\sinh\left[\frac{x}{2}: \right.\frac{15\cdot16
\dots(k+n+14)}
{6\cdot 7\dots\ (k+n+5)}=\sinh\left[\frac{x}{2}: \right.\frac{(k+n+6)\dots(k+n+14)}
{6\cdot7\dots14},$$

$$L_{11s3}=\sinh\left[\frac{x}{2}: \right.\frac{k+n+10}
{10},$$

$$L_{01}=\sinh\left[\frac{x}{2}: \right. 
\frac{n+1}{1}$$

$$L_{c2}=1$$
     
       \begin{multline*}
    X(x,k,n,-2,8,12) = \\
         \sinh\left[\frac{x}{2}: \right.
         \prod _{i=11}^{27} \frac{i+2 k+n}{i} \times \prod _{i=15}^{23} \frac{i+2 k+n}{i} \times \prod _{i=1}^{17} \frac{i+k}{i} \times \prod _{i=5}^{13} \frac{i+k}{i} \times
         \prod _{i=2}^{18} \frac{i+k+n}{i} \times \\ \prod _{i=6}^{14} \frac{i+k+n}{i} \times 
         \frac{n+1}{1} \cdot \frac{k+9}{9} \cdot \frac{k+n+10}{10}  \cdot \frac{2k+n+19}{19} \cdot \frac{3k+n+28}{28} \cdot \frac{3k+2n+29}{29} 
     \end{multline*}    
      coinciding with direct calculation by (\ref{W}) carried out for $\lambda=k\omega_6+n\omega_7$ irrep.

\newpage
\chapter{On singularities of universal formulae. Revelation and proof of the {\it linear resolvability} property}
It has been observed that the known universal formulae show quite an interesting behavior when considering them at the permuted coordinates of the initial special 
points in Vogel's plane, which correspond to the simple Lie algebras. Namely, in case they are not singular at a given permuted point,
 they (usually) yield some reasonable outputs, which naturally correspond to some other 
representations of the algebra, associated with the permuted coordinates. 
In this chapter, we will show that the quantum dimension $X(x,k,n,\alpha,\beta,\gamma)$, derived in the previous chapter, has a feature which allows
obtaining finite answers at its singular points, associated with those from Vogel's table.
Below we present the formal definition of that feature and call it {\it linear resolvability}. Then we show that all universal formulae known so far, including the newly derived
$X(x,k,n,\alpha,\beta,\gamma)$,
are linearly resolvable at the points from Vogel's table.
\newpage
\section{Definition of linear resolvability for universal formulae and the main statement}

In this section, we give the definition of linear resolvability LR and set out the method, which has been used for the proof of the main results stated in Proposition 3.1 and Proposition 3.2.

{\bf Definition. }
{\it
 A multivariable function is said to be LR at its singular point, if it yields finite output when approaching that point through all (regular, by definition) but a finite number of (irregular) lines.}

Note, that all known universal (quantum) dimension formulae, particularly (\ref{main}) are ratios of 
a special form, where both the numerator and denominator decompose into products of a finite number of 
(sines of) linear functions of parameters $(\alpha,\beta,\gamma)$, so that at their singular points some of the factors of the denominator are necessarily zeroing.
 
This feature allows us to prove the following

{\bf Lemma.}
{\it A universal formula is LR at its singular point iff the number of zeroing factors 
 in the denominator is less or equal to those in the numerator at that point. }

\begin{proof}
Suppose for a universal formula $F$ the number of zeroing terms in the numerator and denominator is $n$ and $d$ respectively.
If $d>n$ then approaching the singular point through lines, other than those given by the equations coinciding with any of the zeroing factors in the numerator, the formula 
obviously yields an infinite output. As the number of such choices is infinite, then $F$ is not LR.

Now suppose $d \leq n$. If we approach the singular points through all but the line given by the equation coinciding with any of the $d$ factors, we will
 necessarily get a finite output for $F$, which means that it is LR. It is clear, that as long as $d<n$, $F$ yields zero when restricted at any regular line. 
\end{proof}

{\bf Remark 1.}
Obviously, when considered a universal formula on any regular line, both $n$ and $d$ do not change.
 It means that the complete examination of LR can be made by observing the function on a single regular line.
 
{\bf Remark 2.} 
All irregular lines for a given universal formula are exactly determined by each of the factors in its denominator; if there is, 
say a $ c_1\alpha+c_2 \beta+c_3\gamma$ factor in the denominator of the universal formula, it cannot yield a finite output, when 
restricted on the associated  $c_1\alpha+c_2 \beta+c_3\gamma=0$ line, meaning, that each of the factor of the denominator determines 
an irregular line of the corresponding function. 

{\bf Remark 3.} 
Based on the previous remark, one can easily check, that any of the $sl, so, exc$ lines (see Table \ref{tab:V2}) is regular for \ref{main} formula.

The {\bf Lemma} is of essential importance for the proof of the following

{\bf Proposition 3.1}

{\it  At the points from the Vogel's table the function $X(x,k,n,\beta,\alpha,\gamma)$ (\ref{main}) and and functions, obtained from it
 by all possible permutations of the corresponding 
parameters $(\alpha, \beta, \gamma)$, are LR 
for any set $(k,n)$ with integer non-negative numbers $k,n$. }

The proof is carried out by case by case (for each algebra and for each permutation) examination of the structure of (\ref{main}), restricting it
 on the corresponding line
and tracking all possible zero factors appearing both in the numerator and the denominator. In fact, the procedure of the proof automatically highlights all
possible singular points.

Finally, we propose a conjecture:

{\bf Conjecture.}

{\it The values of functions $X$, calculated at the singular points by restricting the functions to the corresponding $sl, so, sp$ or 
$exc$ lines,
 are equal to the quantum dimensions of some representations
 of the corresponding algebra.
 Particularly, if a singular point belongs to two distinguished lines simultaneously, the same statement is true for each of the obtained values.}

This conjecture has been tested in a number of cases. 

\section{Proof of the linear resolvability of $X(x,k,n,\beta,\alpha,\gamma)$}
Since the \ref{main} function is symmetric w.r.t. the two last arguments, there are only two relevant permutations to be examined: $X(x,k,n,\beta,\alpha,\gamma)$ and 
$X(x,k,n,\gamma,\alpha,\beta)$.
\subsection{Exceptional algebras}
At the points, corresponding to the exceptional algebras, the behavior of $X(x,k,n,\beta,\alpha,\gamma)$ and 
$X(x,k,n,\gamma,\alpha,\beta)$ functions is shown in the tables \ref{tab:xknbe} and \ref{tab:xknge}. Namely, they yield 
 quantum dimensions of representations with highest weights given in these tables,
 provided that in marked cases ("E:") the singularities are linearly resolved on the exceptional line Exc (see table \ref{tab:V2}).
For the values of the parameters, exceeding the corresponding numbers of rows/columns, they yield $0$. 
\begin{table} 
	\caption{$X(x,k,n,\beta,\alpha,\gamma)$ for the exceptional algebras}
	 \centering
	  \scalebox{0.8}{
	\begin{tabular}{|c|c|c|c|c|c|}
		\hline
		
		$k,n$&$G_2$&$F_4$&$E_6$&$E_7$&$E_8$\\
		\hline
		1,0&$3\omega_1$&$\omega_2$ &  $\omega_3$ & $\omega_2$&$\omega_6$ \\
		\hline
		1,1& $\omega_1+\omega_2$&$\omega_3+\omega_4$&{\begin{tabular}{c}
				$(\omega_1+\omega_2)$\\
				$\oplus(\omega_4+\omega_5)$
			\end{tabular}} & $\omega_6+\omega_7$ & $\omega_8$ \\
		\hline
		1,2& 0 & $\omega_1+\omega_4$ & $\omega_3$&0 & $-\omega_8$ \\
		\hline
		1,3& 0 & 0 & 0& E:$-2\omega_6$ & $-\omega_6$ \\
		\hline
		1,4& 0 & 0 & -1 & 0 & 0  \\
		\hline
		1,5& 0 & 0 & 0 & 0 & 1 \\
		\hline
		2,0& 0 & $3\omega_4$ & $3\omega_1 \oplus 3\omega_5$ & 0 & 0 \\
		\hline
		2,1& 0 & 0 & $-\omega_3$ & E:$-\omega_6-\omega_7$ & 0 \\
		\hline
		2,2& 0 & 0 & $-\omega_6$ & $-\omega_5$ & $\omega_6$ \\
		\hline
		2,3& 0 & 0 & 0 & 0 & $\omega_7$ \\
		\hline
		3,0& 0 & 0 &  $-\omega_1-\omega_5$ & E:$-\omega_2$ & $\omega_8$ \\
		\hline
		3,1& 0 & 0 & 0 & E:$-\omega_1$ & $\omega_1$ \\
		\hline
		4,0& 0 & 0 & 0 & -1 & 0 \\
		\hline
	\end{tabular}
	}
	\label{tab:xknbe} \normalsize
\end{table}

\begin{table} 
	\caption{$X(x,k,n,\gamma,\alpha,\beta)$ for the exceptional algebras} 
	 \centering
	 \scalebox{0.9}{
	\begin{tabular}{|c|c|c|c|c|c|}
		\hline
		
		$k,n$&1,0&1,1&1,3&2,0&2,1\\
		\hline
		$G_2$&$3\omega_1$&$-3\omega_1$&1&$3\omega_1$&$\omega_2$\\
		\hline
		$F_4$&$\omega_2$ &$-\omega_2$&1&$\omega_2$&$\omega_1$\\
		\hline
		$E_6$& $\omega_3$&$-\omega_3$&1&$\omega_3$ &$\omega_6$  \\
		\hline
		$E_7$& $\omega_2$&$-\omega_2$&1&$\omega_2$&$\omega_1$ \\
		\hline
		$E_8$&$\omega_6$ &$-\omega_6$&1&$\omega_6$&$\omega_7$\\
		\hline
		
	\end{tabular}
	}
	\label{tab:xknge} \normalsize
\end{table}  

Thus we see, that for the exceptional algebras, all singularities can be resolved, moreover, their resolution on the exceptional line yield some quantum dimensions of representations of
the corresponding algebra, affirming the statement of the Conjecture of the previous section. 

\subsection{Classical algebras}
{\it $X(x,k,n,\gamma,\alpha,\gamma)$}
The direct substitution of Vogel's parameters, corresponding to the classical algebras yield the following result: 
at the points, corresponding to the classical algebras, the $X(x,k,n,\gamma,\alpha,\beta)$ is always zero, except
when  $(k,n)=(k,0)$, (see Table \ref{tab:x2gcl}), and  $(k,n,\gamma,\alpha,\beta)=(1,2,N+1,-2,2)$ i.e. for $k=1,n=2$ and the algebra $A_N$, only. 
For the latter case $X(x,k,n,\gamma,\alpha,\beta)=-1$, (see Table \ref{tab:xkng}).

\FloatBarrier
 \begin{table}
 \caption{$X(x,k,0,\gamma)$ for the classical algebras}
  \centering
     \begin{tabular}{|c|c|c|c|c|}
     \hline
          $k$&1&2&3&$\geq4$\\
          \hline
       $A_1$&0&$-2\omega$ on the $sl$ line&0&0\\
          \hline
       $A_2$&$3\omega_1\oplus 3\omega_2$&$-(\omega_1+\omega_2)$&0&0\\
          \hline
       $A_N,N\geq3$&$(2\omega_1+\omega_{N-1})\oplus
       (\omega_2+2\omega_{N})$&$-(\omega_1+\omega_N)$&0&0\\
          \hline
       $B_2$&$\omega_1+2\omega_2$&0 \text{on the $so$ line}&0 \text{on the $so$ line}&0\\
          \hline
       $B_3$&$\omega_1+2\omega_3$&0&0&0\\
          \hline
       $B_N,N\geq4$&$\omega_1+\omega_3$&0&0&0\\
          \hline
       $C_N,N\geq3$&$2\omega_1+\omega_2$&0&0&0\\
          \hline
       $D_4$&$\omega_1+\omega_3+\omega_4$&0 \text{on the $so$ line}&0 \text{on the $so$ line}&0\\
          \hline
       $D_5$&$\omega_1+\omega_3$&0&0&0\\
          \hline
       $D_6$&$\omega_1+\omega_3$&0 \text{on the $so$ line}&0&0\\
          \hline                
       $D_N,N\geq7$&$\omega_1+\omega_3$&0&0&0\\
          \hline
    \end{tabular}
    \label{tab:x2gcl}
\end{table}
    \FloatBarrier

\begin{table}[h]
	\caption{$X(x,k,n,\gamma,\alpha,\beta)$ for the classical algebras}
	 \centering
	\begin{tabular}{|c|c|}
		\hline
		$k,n>0$&$1,2$\\
		\hline
		$A_N$&-1\\
			\hline
	\end{tabular}
	\label{tab:xkng}
\end{table}

{\it $X(x,k,n,\beta,\alpha,\gamma)$}

For $A_N$ algebra, i.e. the universal parameters $(\alpha,\beta,\gamma)=(-2,2,N+1)$, the function $X(x,k,n,\beta,\alpha,\gamma)$ is equal to

\begin{multline} \label{aNfinal}
X(x,k,n,2,-2,N+1)=\\
 L_{31}\cdot L_{32}\cdot L_{21s1}\cdot L_{21s2}\cdot L_{21s3}\cdot L_{10s1}\cdot
L_{10s2}\cdot L_{10s3}\cdot L_{11s1}\cdot L_{11s2}\cdot L_{11s3}\cdot L_{01}\cdot L_{c2}=\\
2\times
\sinh\left[\frac{x}{2}: \right.
\frac{(n+1)}
{1\cdot2\cdot3\cdot \dots (k+n+1) }
\times (N+2)(N+1)\dots(N-(k+n-3)) \\
\frac{(N+3)\cdot(N+2)\cdot(N+1)\cdot N\cdot(N-1)\dots(N-(2k+n-4))
}{1\cdot2 \cdot \dots (2k+n)}
\times \\
\frac{(N+3)\cdot (N+2)\cdot\dots(N-(k-4))}{1\cdot2 \cdot \dots k}\times \frac{(N-(3k+n-3))\cdot(N-(3k+2n-2))}{(N+3)\cdot(N+2)}
\end{multline}

It obviously is non-singular.

For $B_N$ algebra, i.e. $(\alpha,\beta,\gamma)=(4,-2,2N-3)$, $X(x,k,n,\beta,\alpha,\gamma)$ has no zeroing terms in the denominator for $N\in Z_+$, so that it also is non-singular.

For $C_N$ algebra, i.e. for the parameters $(\alpha,\beta,\gamma)=(-2,1,N+2)$, the function  $X(x,k,n,\beta,\alpha,\gamma)$ is equal to

\begin{multline} \label{cnfinal}
X(x,k,n,1,-2,N+2)=\\
 L_{31}\cdot L_{32}\cdot L_{21s1}\cdot L_{21s2}\cdot L_{21s3}\cdot L_{10s1}\cdot
L_{10s2}\cdot L_{10s3}\cdot L_{11s1}\cdot L_{11s2}\cdot L_{11s3}\cdot L_{01}\cdot L_{c2}=\\
\sinh\left[\frac{x}{4}: \right.
\frac{(n+1)(k+1)(k+n+2)(N-k+3)(N-k-n+2)(N-2k-n+1)}
{1^2\cdot2\cdot(N+1)(N+2)(N+3)}
\times \\
\frac{(2N-3k-n+4)(2N-3k-2n+3)}{(2N+4)(2N+3) } \times
\frac{(2N+4)\cdot(2N+3)\cdot\dots(2N-2k-n+5)}{3\cdot4\cdot \dots (2k+n+2)}
\times \\
\frac{(2N+4)\cdot(2N+3)\cdot \dots (2N-k-n+6)\times(2N-2k-n+5)}{2\cdot3\cdot4\dots(k+n+1)} \\
\times 
\frac{(2N+6)\cdot(2N+5)\dots(2N-k+7)}{1\cdot2\cdot3\dots k}
\end{multline}

It also is non-singular.

For $D_N$ algebra, i.e. for the parameters $(\alpha,\beta,\gamma)=(-2,1,2N-4)$, and for  $k=1$, the function $X(x,1,n,\beta,\alpha,\gamma)$ is equal to

\begin{multline}  \label{dnfinal}
X(x,1,n,4,-2,2N-4)=\\
L_{31}\cdot L_{32}\cdot L_{21s1}\cdot L_{21s2}\cdot L_{21s3}\cdot L_{10s1}\cdot
L_{10s2}\cdot L_{10s3}\cdot L_{11s1}\cdot L_{11s2}\cdot L_{11s3}\cdot L_{01}\cdot L_{c2}=\\
\sinh\left[x: \right.
\frac{(N-2n-3)(N/2+1/2)}
{(N/2-n-3/2) \cdot 1/2\cdot(N+1)\cdot (n+2)}
\times \\
\frac{(N+1) \cdot N \cdot (N-1) \dots (N-n)}{1\cdot 2\cdot 3\dots (n+1)}
\times \\
\frac{(N-1/2)\cdot(N-3/2) \dots(N-n-3/2)}{1/2 \cdot 3/2 \dots (n+1/2)}
\end{multline}

It can easily be seen, that the number of zero terms in the numerator is not less than those in the denominator, which according to the Proposition 3.1 means
that the corresponding function is linearly resolvable.

At last, the remaining case of $D_N$ algebra at $k>1$, the function  $X(x,k,n,\beta,\alpha,\gamma)$ is identically zero due to the  $2\alpha+\beta$ term
in the numerator:

\begin{multline} \label{dnzero}
\sinh\left[x: \right.
\frac{(2\alpha+\beta)\cdot N/2 }
{(N/2-k-n)(k+n)(k+n+1)}
\times \\
\frac{(N-1/2) \cdot (N-3/2) \dots (N-2k-n+1/2)}{1/2\cdot 3/2\dots (k+n-1/2) \times (N/2-k-n-1/2)\cdot (N/2-k-n-3/2) \dots (N/2-2k-n+1/2)}
\times \\
\frac{(N+1)\cdot N\dots(N-k-n)}{2 \cdot 3 \dots (2k+n-1)}\times \frac{(N/2-k-n)\dots (N/2-2k-n+2)}{(N-2k-n+1)(N-2k-n)} \times \\
\frac{1}{(k-1)\cdot k}\times \frac{N\cdot (N-1) \dots (N-k+1)}{(-1/2) \cdot 1/2 \dots (k-3/2)} \times \\
\frac{(N/2+1/2)\cdot (N/2-1/2)\cdot (N/2-k+3/2)}{N/2\cdot (N/2-1)\dots (N/2-k+1)} \times \\
\frac{(N-3k-n+1)\cdot(N-3k-2n)\cdot (N/2-2k-n+1) \cdot (n+1)}{1\cdot (N+1)\cdot N }
\end{multline} 

After a careful inspection, we see that the number of zero terms in the denominator again is not greater than those in the numerator for any natural value
of the rank $N$, which means that the whole function $X(x,k,n,\beta,\alpha,\gamma)$ is linearly resolvable. 

Overall, we proved the Proposition 3.1 by case by case inspection of the main formula (\ref{main}).

\section{Permutation of the parameters, corresponding to $so(8)$ algebra}

Below we present some interesting results regarding the $so(8)$ algebra. 
It belongs both to the orthogonal and the 
exceptional lines and its Dynkin diagram has the largest symmetry group, $S_3$. However, as it is shown below, that symmetry group reveals itself when we consider 
$so(8)$ algebra as a member of the exceptional family, i.e. resolve the singularities on the exceptional line. If we consider it as a member of the orthogonal algebras 
our formulae reveal only the $Z_2$ symmetry. 
 So, the expectation that our formula for $X(x,k,n,\alpha,\beta,\gamma)$ yields reasonable results when restricted on either of these lines is totally met.
 In the table \ref{tab:SO8} we present the results of permutation and restriction on each of the mentioned lines. 

E.g. the number (minus) $2$ in $(k,n)=(1,3)$ case on the exceptional line is the dimension of the standard representation of $S_3$ group. 
The number $3$ in $(3,0)$ case also can be interpreted as a (reducible) representation of $S_3$ group, $so$ part is represented trivially. 
The weight $\omega_{1}$, for $(1,2)$ case, is invariant w.r.t. the $Z_2$ group of automorphism of the orthogonal algebras, that is why it appears alone when resolved on the orthogonal line.

\begin{table} 
	\caption{Permutations of the parameters for $so(8)$ algebra} 
	 \centering
	\begin{tabular}{|c|c|c|c|c|c|c|c|c|}
		\hline
		
		Line&$k,n \, :$&1,0&1,1&1,2&1,3&2,0&2,1&3,0\\
		\hline
		Exc &$X(x,k,n,\beta,\alpha,\gamma)$&$\lambda_{X_2}$&$\lambda_{X_2}\oplus \lambda_{X_2}$&$\omega_1\oplus\omega_3\oplus\omega_4$&-2&$\lambda_{X_2}$&$\lambda_{ad}$&3\\
		\hline
		Exc & $X(x,k,n,\gamma,\alpha,\beta)$&$\lambda_{X_2}$ &$-\lambda_{X_2}$&0&1&$\lambda_{X_2}$&$\lambda_{ad}$&0\\
		\hline
		SO &$X(x,k,n,\beta,\alpha,\gamma)$& $\lambda_{X_2}$&$\lambda_{X_2}$&$\omega_{1}$&-1&0&0&0  \\
		\hline
		SO&$X(x,k,n,\gamma,\alpha,\beta)$& $\lambda_{X_2}$&0&0&0&0&0&0\\
				\hline		
	\end{tabular}
	\label{tab:SO8} 
\end{table}  

\section{On LR of all known universal quantum dimensions}
After revelation that $X(x,k,n,\alpha,\beta,\gamma)$ is LR, we tested all known universal quantum dimension formulae on this property, \cite{Zresolv}.
Namely, we tested the following series of universal (quantum) dimensions $(\mathfrak{g})^kY_2^n(\beta)$ (\cite{LM1,M16QD}):

\begin{multline}\label{Z}
Z(x,k,l,\alpha,\beta,\gamma)= \\
    \sinh\left[\frac{x}{4}: \right. \prod_{i=1}^{k+l} \frac{(\alpha(3-i)+2\gamma)\cdot(\alpha(4-i)+\beta+2\gamma)\cdot(\alpha(3-i)+2\beta+\gamma)}{(\alpha(1-i)+2\beta)\cdot(-\alpha i+\beta)\cdot(\alpha(1-i)+\gamma)} \times \\
    \prod_{i=1}^{k} \frac{\alpha(i-1)-2\beta}{(\alpha i)} \times  \prod_{i=1}^{k+2l} \frac{\alpha(4-i)+2\beta+2\gamma}{\alpha(3-i)+2\gamma} \times \\
    \prod_{i=1}^{l} \frac{(\alpha(3-i)-\beta+2\gamma)\cdot(\alpha(3-i)+\beta+\gamma)\cdot(\alpha(4-i)+2\gamma)}{(\alpha(1-i)-\beta+\gamma)\cdot(\alpha(1-i)+\beta)\cdot(-\alpha i)} \times \\
    \frac{(\alpha(3-2k-2l)+2\beta+2\gamma)\cdot(\alpha(3-2l)+2\gamma)\cdot(\alpha(3-k-2l)+\beta+2\gamma)\cdot(-\alpha k+\beta)}{ \beta \cdot (3\alpha+2\beta+2\gamma)\cdot(3\alpha+2\gamma)
    \cdot(3\alpha+\beta+2\gamma)} 
       \end{multline} 
and proved that it also has the feature of LR.
The proof is carried out by case by case (for each algebra and for each permutation) examination of the structure of (\ref{Z}),
 restricting it on the corresponding line
and tracking all possible zero factors appearing both in the numerator and the denominator. In fact, the procedure of the proof automatically highlights all
possible singular points.
Particularly, it turns out that there is an infinite number or series of singular points, (see Appendix C.III). However, the patterns, governing 
the appearance of them is pretty complicated, so we do not classify them in the scope of this work.

Finally, joining this result with the one in Proposition 3.1 we claim our ultimate result:

 {\bf Proposition 3.2}
 
{\it  At the points from Vogel's table all universal quantum dimensions known so far, and functions, obtained from them
 by all possible permutations of the corresponding 
parameters $(\alpha, \beta, \gamma)$, are LR. }

\section{LR beyond Vogel's table}

 Besides the points corresponding to the simple Lie algebras, there are other notable ones in Vogel's plane. These points
  have been revealed in \cite{M16} (see also \cite{AM-dubna}. Some of them were studied earlier in \cite{W3,LM2}) using the requirement that the 
universal quantum dimension of the adjoint representation (\ref{cad}) be a regular function of $x$ in the finite complex plane.
 In other words, the quantum dimension, associated with these points, rewrites as 
 a finite sum of exponents. 
  \begin{eqnarray}  \label{cad}
 f(x) &=& -\frac{\text{sinh}\left(\frac{\gamma+2\beta+2\alpha}{4}x\right)}{\text{sinh}\left(\frac{\alpha}{4}x\right)}
 \frac{\text{sinh}\left(\frac{2\gamma+\beta+2\alpha}{4}x\right)}{\text{sinh}\left(\frac{\beta}{4}x\right)}
 \frac{\text{sinh}\left(\frac{2\gamma+2\beta+\alpha}{4}x\right)}{\text{sinh}\left(\frac{\gamma}{4}x\right)}
 \end{eqnarray}
 These points are listed in Tables (\ref{isV}) and (\ref{YV}), along with the points, which correspond to the exceptional simple
  Lie algebras. 
Note the $E_{7\frac{1}{2}},  X_1$ and  $X_2$ points there, which belong to the physical region of Vogel's plane. They were 
suggested to have the following interpretations:  $E_{7\frac{1}{2}}$, with dimension $190$ and rank $8$, is proven  to be the
 semidirect product of $e_7$ and $H_{56}$ - (56+1)-dimensional Heisenberg algebra \cite{W3, LM2} 
 $X_1$, with dimension 156 and rank 8, is proposed to be $\mathfrak {so}_{14}\rtimes H_{64}$ semidirect product, and $X_2$ is proposed
 to be the $\mathfrak {so}_{12}\rtimes H_{32}$ semidirect product \cite{W3,LM2,M16}.

Examining the behavior of $Z(x,k,l,\alpha,\beta,\gamma)$ (\ref{Z}) and $X(x,k,n,\alpha,\beta,\gamma)$ (\ref{main}) functions at these points we present the following:

{\bf Proposition 3.3}
{\it At the $X_1, X_2,E_{7\frac{1}{2}}$ points in the Vogel's plane, both $Z(x,k,l)$ and $X(x,k,n)$ functions are LR.}

The proof is straightforward.

The remaining 48 points, corresponding to the so-called Y-objects, are given in Table 2. Dimensions of their "adjoint representation",
 i.e. values of $f(x)$ at the associated points, when $x \rightarrow 0$, are negative\footnote{Note some irregularity in the notations: 
there is an object  $Y_6^{\prime}$, which stands out from the remaining ones ($Y_i, i=1,2,...,47$) in its notation. The reason is that in \cite{M16}  two different solutions of 
Diophantine equations were accidentally denoted by the same notation $Y_6$, and here, trying to have minimal changes in notations, we denote one of them as $Y_6^{\prime}$.}.

We tested both $Z(x,k,l,\alpha,\beta,\gamma)$ (\ref{Z}) and $X(x,k,l,\alpha,\beta,\gamma)$ (\ref{main}) \cite{X2kng} formulae on LR at those points and obtained the following result:

{\bf Proposition 3.4 }
{\it At the points $Y_2, Y_6, Y_{32}$ from Table 2 both $Z(x,k,l,\alpha,\beta,\gamma)$ and $X(x,k,l,\alpha,\beta,\gamma)$ formulae
are regular. At all other points from the same table, it is  possible to choose a $(k,l)$ pair, for which either $Z(x,k,l,\alpha,\beta,\gamma)$ or $X(x,k,n,\alpha,\beta,\gamma)$ is singular and 
not LR for some permutation of the Vogel's parameters.}

\begin{proof}
The desired result follows from the direct substitution of the corresponding sets of parameters $(\alpha,\beta,\gamma)$ (with all possible permutations) into the denominators of 
 $Z(x,k,l,\alpha,\beta,\gamma)$ and $X(x,k,l,\alpha,\beta,\gamma)$.
 \end{proof}

{\bf Remark.}
A notable fact is that when taking the $x\rightarrow 0$ limit, both
 $Z(x,k,l,\alpha,\beta,\gamma)$ and $X(x,k,n,\alpha,\beta,\gamma)$ formulae yield integer-valued outputs at the 
$Y_2$ and $Y_{32}$ points, pointing out a remarkable similarity of those "unknown" objects with
the simple Lie algebras.

We see that among so-called Y-objects, with corresponding points belonging to the non-physical region of the 
Vogel's plane, there are another three points at which all universal 
quantum dimensions are regular. Furthermore,
we observe that two of them behave like real existing algebras, in the sense that
the universal dimension formulae yield integer-valued output at those points.

This results prompt a number of natural questions. For example it would be interesting to find out what is the underlying
reason for universal quantum dimensions possessing the LR feature. 
Also, it is intriguing where is the remarkable
 property of $Y_2$ and $Y_{32}$ points 
inducing integer-valued outputs from universal 
dimension formulae rooted in.

\begin{table}[ht]
\caption{Isolated solutions in the physical region of Vogel's plane.}
\centering
{\begin{tabular}{@{}|c|c|c|c|@{}} 
\hline
$\alpha \beta \gamma$ & Dim & Rank & Notation \\ 
\hline
-6	-10	1	&	248	&	8	&	$	E_8	$		\\
-8	1	-5	&	190	&	8	&	$	E_{7\frac{1}{2}}	$	\\
-4	1	-7	&	156	&	8	&	$	X_1	$	\\
-6	-4	1	&	133	&	7	&	$	E_7	$	\\
1	-3	-5	&	99	&	7	&	$	X_2	$	\\
-3	-4	1	&	78	&	6	&	$	E_6	$	\\
-6	2	-5	&	52	&	4	&	$	F_4	$	\\
3	-5	-4	&	14	&	2	&	$	G_2	$	\\
\hline
\end{tabular}}\label{isV}
\end{table}

\begin{table}[ht]
\caption{Isolated solutions in the non-physical region of Vogel's plane.}
\centering
{\begin{tabular}{@{}|c|c|c|c|@{}}
\hline
$\alpha \beta \gamma$ & Dim & Rank & Notation \\ 
\hline
1	1	1	&	-125	&	-19	&	$	Y_1	$      \\
10	8	7	&	-129	&	-1	&	$	Y_2	$	\\
6	4	5	&	-130	&	-4	&	$	Y_3	$	\\
2	2	3	&	-132	&	-10	&	$	Y_4	$	\\
5	7	8	&	-132	&	-2	&	$	Y_5	$	\\
5	8	6	&	-132	&	-2	&	$	Y_6	$	\\
4      5       3      &       -133&       -2     &      $      Y_6^{'} $       \\
4	7	5	&	 -135&	-3	&	$	Y_7	$       \\
7	6	4	&	-135	&	-3	&	$	Y_8	$	\\
2	4	3	&	-140	&	-8	&	$	Y_9	$	\\
2	1	2	&	-144	&	-14	&	$	Y_{10}$	\\
2	1	1	&	-147	&	-17	&	$	Y_{11}	$	\\
7	3	4	&	-150	&	-4	&	$	Y_{12}	$	\\
2	4	5	&	-153	&	-7	&	$	Y_{13}	$	\\
5	3	2	&	-153	&	-7	&	$	Y_{14}	$	\\
1	2	3	&	-165	&	-13	&	$	Y_{15}	$	\\
2	6	5	&	-168	&	-6	&	$	Y_{16}	$	\\
6	2	7	&	-184	&	-6	&	$	Y_{17}	$	\\
4	5	13	&	-186	&	-2	&	$	Y_{18}	$	\\
3	10	4	&	-186	&	-4	&	$	Y_{19}	$	\\
3	7	2	&	-187	&	-7	&	$	Y_{20}	$	\\
1	1	3	&	-189	&	-17	&	$	Y_{21}	$	\\
11	5	3	&	-189	&	-3	&	$	Y_{22}	$	\\
4	1	3	&	-195	&	-11	&	$	Y_{23}	$	\\
2	1	4	&	-195	&	-13	&	$	Y_{24}	$	\\
3	11	4	&	-200	&	-4	&	$	Y_{25}	$	\\
2	3	8	&	-207	&	-7	&	$	Y_{26}	$	\\
2	5	9	&	-207	&	-5	&	$	Y_{27}	$	\\
3	1	5	&	-221	&	-11	&	$	Y_{28}	$	\\ 
1	4	5	&	-228	&	-10	&	$	Y_{29}	$	\\
2	1	5	&	-231	&	-13	&	$	Y_{30}	$	\\
4	1	1	&	-242	&	-18	&	$	Y_{31}	$	\\
6	5	22	&	-244	&	-2	&	$	Y_{32}	$	\\
18	4	5	&	-245	&	-3	&	$	Y_{33}	$	\\
14	4	3	&	-247	&	-5	&	$	Y_{34}	$	\\
10	2	3	&	-252	&	-8	&	$	Y_{35}	$	\\
1	4	6	&	-252	&	-10	&	$	Y_{36}	$	\\
3	5	16	&	-258	&	-4	&	$	Y_{37}	$	\\
6	1	2	&	-272	&	-14	&	$	Y_{38}	$	\\
1	3	7	&	-285	&	-11	&	$	Y_{39}	$	\\
1	5	7	&	-285	&	-9	&	$	Y_{40}	$	\\
14	2	5	&	-296	&	-6	&	$	Y_{41}	$	\\
6	8	1	&	-319	&	-9	&	$	Y_{42}	$	\\
1	3	8	&	-322	&	-12	&	$	Y_{43}	$	\\
4	1	9	&	-342	&	-10	&	$	Y_{44}	$	\\
10	1	4	&	-377	&	-11	&	$	Y_{45}	$	\\
12	1	5	&	-434	&	-10	&	$	Y_{46}	$	\\
1	6	14	&	-492	&	-10	&	$	Y_{47}	$       \\  
\hline
\end{tabular}}\label{YV}
\end{table}

\section{Appendix C.III: Proof of LR of $Z(x,k,l)$ formula}

The procedure of the proof is carried out in the following way: first, we take the main formula (\ref{Z}), and for each of the point $(\alpha, \beta,\gamma)$ 
from Table \ref{tab:V2}
in the Vogel's plane, including those obtained by another 5 permutations of the parameters, 
 examine its expression in the neighborhood of the point in question, restricting it on the corresponding distinguished line (Table \ref{tab:V2}) beforehand.
 Then, we trace the number of zeroing factors in both its numerator ($n$) and denominator ($d$) at the corresponding points.
 Based on the Lemma 1, the proof of LR is, in fact, equivalent to the checking of
  the realization of the $n\geq d$ inequality in each of the possible cases, namely,
 for every possible non-negative integer-valued set $(k,l)$ for each of the permutations of the corresponding Vogel's parameters.
 
 Since the implementation of this procedure is quite repetitive, we find it reasonable to
 present the explicit calculations for several key cases only, which are sufficient to outline the essence of the proof.
 They are presented in the following section.

\paragraph{Classical algebras.}

\subsection{$A_N$}
\subsubsection{$\alpha,\gamma,\beta$}
Here we examine the $Z(x,k,l,\alpha,\gamma,\beta)$
for the parameters, corresponding to the $A_N$ algebra, by presenting the corresponding formulae, which are obtained by
every possible choice of the set $k,l$:

\subsubsection{$l=0,k=1$}
\begin{equation*}
Z(x,1,0,-2,N+1,2)=\sinh\left[\frac{x}{4}: \right.\frac{(2N)\cdot(2N+4)}{2^2}
\end{equation*}

\subsubsection{$l=0,k>1$}
\begin{equation*}
Z(x,k,0,-2,N+1,2)=\sinh\left[\frac{x}{4}: \right.\frac{(2N)\cdot(2N+4k)}{(2k)^2}\times \frac{(2N+2)^2\cdot(2N+4)^2\dots(2N+2k-2)^2}
{2^2\cdot4^2\dots (2k-2)^2}
\end{equation*}

\subsubsection{$l=1,k=0$}
\begin{equation*}
Z(x,0,1,-2,N+1,2)=\sinh\left[\frac{x}{4}: \right.\frac{(2N)\cdot(2N+4)}{2^2}
\end{equation*}

\subsubsection{$l=1,k\geq1$}
\begin{equation*}
Z(x,k,1,-2,N+1,2)=\sinh\left[\frac{x}{4}: \right.\frac{(2N)\cdot(2N+4k+4)}{(2k+2)^2}\times \frac{(2N+2)^2\cdot(2N+4)^2\dots(2N+2k)^2}
{2^2\cdot4^2\dots (2k)^2}
\end{equation*}
Obviously, each of the functions written above is regular for any $N\in \mathbb{N}$.

Let's move on to the remaining $(k,l)$ sets:

\subsubsection{$l=2,k=0$}
\begin{multline*}
Z(x,0,2,-2,N+1,2)=\\
\sinh\left[\frac{x}{4}: \right.
\frac{(2\alpha+2\beta)\cdot(N-1)\cdot N \cdot (N+1)^2\cdot (N+7) \cdot(2 N+2)\cdot (2 N+6)\cdot (2 N+8)}{2^2\cdot4^3\cdot(N-3)\cdot (N+3)^2\cdot (N+5)}
\end{multline*}
We see, that there is a $2\alpha+2\beta$ factor, which is zeroing at any point of the 
$2\alpha+2\beta=0$ line, so that one can easily determine, that the possible number of zeroing factors in the numerator is
always greater or equal to those in the denominator, namely $d\leq n$, which means, that the initial function is LR.

\subsubsection{$l=2,k\geq1$}
\begin{multline*}
Z(x,k,2,-2,N+1,2)=\\
\sinh\left[\frac{x}{4}: \right.\frac{(2\alpha+2\beta)\cdot6\cdot(2N)\cdot(N+1)\cdot(N-1)\cdot(2N+4k+8)}
{2\cdot4\cdot(2k+4)\cdot(N+2k+3)\cdot(N+2k+5)\cdot(N+3)}\times \\ 
\frac{(N+2k+7)\cdot(N+2k+1)\cdot(2N+2k+2)\cdot(2N+2k+6)}{(N-3)\cdot(2k+2)\cdot(2k+4)\cdot(2k+6)}\times\\
\frac{(2N+2)^2\cdot(2N+4)^2\dots(2N+2k)^2}{2^2\cdot\ 4^2\dots (2k)^2}
\end{multline*}

Proof of the LR of this function is similar to that of the previous one.
Notice, that for each of the integer $k\geq1$, the corresponding function has a singularity (linear resolvable, of course), when $N=3$.
This particular case is interesting in the sense, that it explicitly demonstrates, that the set of singularities of the function (\ref{Z}) is basically infinite.

\subsubsection{$l\geq3,k=0$}
\begin{multline*}
Z(x,0,l,-2,N+1,2)=\\
\sinh\left[\frac{x}{4}: \right.\frac{(2\alpha+2\beta)\cdot(2N)\cdot(N+1)^2\cdot(N-1)\cdot(2N+4l)}
{(2l-2)\cdot(2l)^2}\times \\
\frac{(N+4l-1)\cdot(4l-2)}{(N-2l+1)\cdot(2N+2l)\cdot(N+2l-1)^2\cdot(N+2l+1)} \times \\
\frac{(2N+2)\dots(2N+4l-2)}{2\cdot\ 4\dots (4l-2)}
\end{multline*}

\subsubsection{$l\geq3,k\geq1$}
\begin{multline*}
Z(x,k,l,-2,N+1,2)=\\
\sinh\left[\frac{x}{4}: \right.\frac{(2\alpha+2\beta)\cdot(N+1)\cdot(N-1)\cdot(2N+4k+4l)}
{(2l-2)\cdot(2l)\cdot(2k+2l)\cdot(N+2l-1)}\times \\
\frac{(N+2k+4l-1)\cdot(N+2k+1)\cdot(4l-2)}{(N-2l+1)\cdot(2N+2k+2l)\cdot(N+2k+2l-1)\cdot(N+2k+2l+1)} \times \\
\frac{(2N+2)\cdot(2N+4)\dots(2N+2k)}{2\cdot\ 4\dots 2k} \times 
\frac{2N\cdot(2N+2)\dots(2N+2k+4l-2)}{2\cdot\ 4\dots (2k+4l-2)}
\end{multline*}

The same reasoning, which proves the LR, holds for the latter two cases.

Thus, we proved the LR of the $Z(x,k,l,\alpha,\gamma,\beta)$ function at any $N \in \mathbb{N}$ point lying on the $sl$ line.

\subsection{$B_N$}
\subsubsection{$\beta,\gamma,\alpha$}
Let's prove, that the $Z(x,k,l,4,2N-3,-2)$ function is LR for any non-negative integer set $(k,l)$.
To prove the LR of $Z(x,k,l,\beta,\gamma,\alpha)$  at the $(4,2N-3,-2)$ points, where $N \in \mathbb{N}$, we examine it on the $so$ line in the following cases:
\subsubsection{$l=1,k=0$}
In this case, $Z$ writes as follows
\begin{equation*}
Z(x,0,1,4,2N-3,-2)=\sinh\left[\frac{x}{4}: \right.\frac{(4N)(4N-2)(2N+3)}{2\cdot 4\cdot (2N-1)}
\end{equation*}
it is regular on the $so$ line for any integer $N$.

One can easily determine, that the following 4 functions are also regular for any integer $N$.

\subsubsection{$l=1,k\geq1$}
\begin{multline}
Z(x,k,1,4,2N-3,-2)=\\
=
\sinh\left[\frac{x}{4}: \right.
\frac{(2\alpha+\beta)\cdot4\dots (4k-4)}{6\cdot 10\cdot 14\dots (4k+2)} \times \\
\frac{(2N+5)\cdot(2N+1)\dots (2N-4k+5)}{(2N-7)\cdot(2N-11)\dots (2N-4k-7)} \times 
\frac{4N\cdot(4N-4)\dots (4N-4k)}{(4N-6)\cdot (4N-10)\dots (4N-4k-6)}
\times \\
\frac{(4N-6)^2\cdot(4N-10)^2\dots(4N-4k-2)^2}{4^2\cdot8^2\cdot \dots (4k)^2}\times \\
\frac{(4N-2)\cdot(2N-4k-3)^2\cdot(4N-8k-6)\cdot(2N-7)\cdot(2N+3)}{2\cdot4\cdot(2N-3)^2\cdot(2N+5)\cdot(1-2N)} .
\end{multline}

\subsubsection{$l=2,k=0$}
\begin{equation*}
Z(x,0,2,4,2N-3,-2)=\\
=
-\sinh\left[\frac{x}{4}: \right. \frac{(4N)\cdot(4N-4)\cdot(4N-2)\cdot(2N+1)\cdot(4N-14)}{2\cdot4\cdot6\cdot8\cdot(2N-7)}
\end{equation*}

\subsubsection{$l=2,k\geq1$}
\begin{multline}
Z(x,k,2,4,2N-3,-2)=\\
=
-\sinh\left[\frac{x}{4}: \right.
\frac{4\dots (4k)}{10\cdot 14\dots (4k+6)} \times \\
\frac{(2N+5)\cdot(2N+1)\dots (2N-4k+1)}{(2N-7)\cdot (2N-11)\dots (2N-4k-11)} \times 
\frac{4N\cdot(4N-4)\dots (4N-4k-4)}{(4N-6)\cdot (4N-10)\dots (4N-4k-10)}
\times \\
\frac{(4N-6)\cdot(4N-10)\dots(4N-4k-2)}{4\cdot8\cdot \dots 4k}\times \\
\frac{(4N+2)\cdot (4N-2)\cdot\dots(4N-4k-10)}{2\cdot 6\cdot 4\cdot 8  \dots (4k+8)}\times \\
\frac{(2N-4k-3)\cdot(2N-4k-11)\cdot(4N-8k-14)}{(2N-3)\cdot(2N+5)\cdot (4N+2)} .
\end{multline}

\subsubsection{$l\geq3,k=0$}

\begin{multline}
Z(x,0,l,4,2N-3,-2)=\\
=
\sinh\left[\frac{x}{4}: \right.
\frac{(4N)\cdot(4N-4)\dots(4N+4-4l)}{2\cdot6 \cdot \dots(4l-2)}\times \\
\frac{(2N+5)\cdot (2N+1)\cdot\dots(2N+9-4l)}{(2N-7)\cdot (2N-11) \dots (2N-4l-3)}\times \\
\frac{(4N-2-4(l+1))\dots(4N-2-4(2l-2))}{(4(l-1))\cdot(4l)\dots (4(2l-2))} \times \\ 
\frac{4\cdot (2\alpha+\beta)}{(4l-8)\cdot(4l-4)\cdot 4l} \times \frac{(2N+1)\cdot(2N+5)\dots (2N+4l-11)}{(2N+7)\cdot(2N+11)\dots (2N+4l-5)} \times  \\ 
\frac{(2N-5)\cdot(2N-9)\dots (2N+7-4l)}{(2N-11)\cdot(2N-15)\dots (2N+1-4l)} \times \\
\frac{(4N-2)\cdot(2N-4k-3)\cdot(2N-8l+5)\cdot (8l-8) \cdot 
	(4N-8l+2)}{(2N-3)\cdot(2N+5)}.
\end{multline}

\subsubsection{$l\geq3,k\geq1$}

\begin{multline}
Z(x,k,l,4,2N-3,-2)=\\
=
\sinh\left[\frac{x}{4}: \right.
\frac{(4N)\cdot(4N-4)\dots(4N+4-4k-4l)}{2\cdot6 \cdot \dots(4k+4l-2)}\times \\
\frac{(2N+5)\cdot (2N+1)\cdot\dots(2N+9-4k-4l)}{(2N-7)\cdot (2N-11) \dots (2N-4k-4l-3)}\times \\
\frac{(4N-2-4(k+l+1))\dots(4N-2-4(k+2l-2))}{4\cdot 8\dots 4k} \times \frac{4\cdot (2\alpha+\beta)}{(4l-8)\cdot(4l-4)\cdot 4l}
\times \\ \frac{(2N+1)\cdot(2N+5)\dots (2N+4l-11)}
{(2N+7)\cdot(2N+11)\dots (2N+4l-5)} \times  \frac{(4N+2)\cdot(4N-2)\dots (4N-2-4k)}
{(4(k+l-1))\cdot(4(k+l))\dots (4(k+2l-2))} \times \\
\frac{(2N-5)\cdot(2N-9)\dots (2N+7-4l)}{(2N-11)\cdot(2N-15)\dots (2N+1-4l)} 
\times \\ \frac{(2N-4k-3)\cdot(2N-4k-8l+5)\cdot (8l-8) \cdot 
	(4N-8k-8l+2)}{(2N-3)\cdot(2N+5)\cdot(4N+2)}.
\end{multline}

\subsection{$C_N$}
\subsubsection{$\beta,\alpha,\gamma$}
Let's examine the following functions:

\subsubsection{$l=0,k\geq1$}
\begin{multline}
Z(x,k,0,1,-2,N+2)=\\
=
\sinh\left[\frac{x}{4}: \right.
\frac{(2N+5)\cdot(2N+4)\dots(2N+6-k)}{3\cdot4 \cdot \dots(k+2)}\times \\
\frac{(2N+6)\cdot (2N+5)\cdot\dots(2N+7-k)}{4\cdot 5 \dots (k+3)}\times \frac{(k+1)\cdot(k+2)^2\cdot(k+3)}
{1\cdot2^2\cdot 3} \times \\ \frac{(N-k+2)\cdot(N-k+1)\cdot(2N+6-k)\cdot(2N+5-k)\cdot(2N+4-k)}
{(2N+3)\cdot(2N+4)\cdot(2N+5)^2\cdot (2N+6)\cdot(2N+7)}  \times \\
\frac{(2N+5-k)\cdot(2N+7)\cdot(2N+3-2k)}{(N+1)\cdot(N+2)} 
\end{multline}

\subsubsection{$l\geq1,k=0$}
\begin{multline}
Z(x,0,l,1,-2,N+2)=\\
=
\sinh\left[\frac{x}{4}: \right.
\frac{(2N+5)\cdot(2N+4)\dots(2N+6-l)}{3\cdot4 \cdot \dots(l+2)}\times \\
\frac{(2N+6)\cdot (2N+5)\cdot\dots(2N+7-l)}{4\cdot 5 \dots (l+3)}\times \\
\frac{(2N+7)\dots(2N+8-l)}{1\cdot 2\dots l} \times \frac{(2N+8)\cdot(2N+7)\dots(2N+9-l)}{2\cdot3\dots (l+1)}
\times \\ \frac{(N-l+2)\cdot(N-l+1)\cdot(2N+6-2l)\cdot(2N+5-2l)\cdot(2N+4-2l)}
{(2N+3)\cdot(2N+4)\cdot(2N+5)^2\cdot (2N+6)\cdot(2N+7)} \times \\
\frac{(N+4-l)\cdot(N+3-l)\cdot(2N+5-2l)\cdot(2N+7-2l)\cdot(2N+3-2l)}{(N+1)\cdot(N+2)\cdot(N+3)\cdot(N+4)} 
\end{multline}

\subsubsection{$l\geq1,k\geq1$}
The $Z(x,k,l,1,-2,N+2)$ rewrites as follows:
\begin{multline}
Z(x,k,l,1,-2,N+2)=\\
=
\sinh\left[\frac{x}{4}: \right.
\frac{(2N+5)\cdot(2N+4)\dots(2N+6-k-l)}{3\cdot4 \cdot \dots(k+l+2)}\times \\
\frac{(2N+6)\cdot (2N+5)\cdot\dots(2N+7-k-l)}{4\cdot 5 \dots (k+l+3)}\times \\
\frac{(2N+7)\dots(2N+8-l)}{1\cdot 2\dots l} \times \frac{(2N+8)\cdot(2N+7)\dots(2N+9-l)}{2\cdot3\dots (l+1)}
\times \frac{(k+1)\cdot(k+2)^2\cdot(k+3)}
{1\cdot2^2\cdot 3} \times \\ \frac{(N-k-l+2)\cdot(N-k-l+1)\cdot(2N+6-k-2l)\cdot(2N+5-k-2l)\cdot(2N+4-k-2l)}
{(2N+3)\cdot(2N+4)\cdot(2N+5)^2\cdot (2N+6)\cdot(2N+7)} \times \\
\frac{(N+4-l)\cdot(N+3-l)\cdot(2N+5-k-2l)\cdot(2N+7-2l)\cdot(2N+3-2k-2l)}{(N+1)\cdot(N+2)\cdot(N+3)\cdot(N+4)} 
\end{multline}

As we see, in all three cases the $Z(x,k,l,1,-2,N+2)$ function is regular for any integer valued set $(k,l)$ and $N \in \mathbb{N}$.

\subsection{$D_N$}
\subsection{$\gamma,\beta,\alpha$}
\subsubsection{$l+k>4$ and $l,k\geq1$}
In the following expression $T=2N-4$, $T\geq 4$, since $N\geq4$:
\begin{multline}
Z(x,k,l,T,4,-2)=\\
=
\sinh\left[\frac{x}{4}: \right.
\frac{(4+T)}{(4-3T)\cdot(4-4T)} \times \frac{(4+2T)\cdot(4+3T)\dots(4+(k+l-3)T)}{(4-5T)\cdot(4-6T)\dots(4-(k+l)T)} \times \\
\frac{(2T)\cdot T \cdot (2\alpha+\beta)}{8\cdot(8-T)\cdot(8-2T)\cdot(8-3T)} \times \frac{T\cdot(2T)\dots((k+l-4)T)}{(8-4T)\cdot(8-5T)\dots(8-(k+l-1)T)} \times \\
\frac{(6+2T)\cdot(6+T)\cdot6\cdot(6-T)}{2\cdot(2+T)\cdot(2+2T)\cdot(2+3T)} \times \frac{(6-2T)\cdot(6-3T)\dots(6-(k+l-3)T)}{(2+4T)\cdot(2+5T)\dots(2+(k+l-1)T)} \times \\
\frac{8\cdot(8-T)\dots(8-(k-1)T)}{T\cdot(2T)\dots(kT)} \times \\
\frac{(4-2T)\cdot(4-3T)\cdot(4-4T)\dots(4-(k+2l-4)T)}
{(4+3T)\cdot(4+4T)\cdot(4+5T)\dots(4+(k+2l-3)T)} \times \\
\frac{(8-2T)\cdot(8-T)\cdot8\dots(8+(l-3)T)}{4\cdot(4-T)\cdot(4-2T)\dots (4-(l-1)T)}\times \\ \frac{(2+2T)\cdot(2+T)\cdot2\dots(2-(l-3)T)}{6\cdot(6+T)\cdot(6+2T)\dots
	(6+(l-1)T)} \times \\
\frac{(4-3T)\cdot(4-2T)\dots(4+(l-4)T)}{T\cdot(2T)\dots(lT)}\times \\ \frac{(4+(3-2k-2l)T)\cdot(4+(2l-3)T)\cdot((2l+k-3)T)\cdot(kT-4)}{(4-2T)\cdot(4-3T)}
\end{multline}

A careful inspection of the above-written formula shows, that for any integer-valued $T\geq 4$ the number of zeroing factors
in the denominator is not greater than those in the numerator: $d\leq n$, which proves the LR of it.

In the following cases, proofs are either evident or repeat those for the previous case.

\subsubsection{$l+k=4$ and $k, l\geq1$}
In this case we have the following function:
\begin{multline}
Z(x,k,l,T,4,-2)=\\
=
\sinh\left[\frac{x}{4}: \right.
\frac{4\cdot(4+T)}{(4-3T)\cdot(4-4T)}\times \frac{(3T)\cdot(2T)\cdot T \cdot (2\alpha+\beta)}{8\cdot(8-T)\cdot(8-2T)\cdot(8-3T)} \times \\
\frac{(6+2T)\cdot(6+T)\cdot6\cdot(6-T)}{2\cdot(2+T)\cdot(2+2T)\cdot(2+3T)} \times \frac{8\cdot(8-T)\dots(8-(3-l)T)}{T\cdot(2T)\dots((4-l)T)} \times \\
\frac{(4+3T)\cdot(4+2T)\dots(4-lT)}{(4-2T)\cdot(4-T)\dots(4+(l+1)T)} \times  \frac{(8-2T)\cdot(8-T)\cdot8\dots(8+(l-3)T)}{4\cdot(4-T)\cdot(4-2T)\dots (4-(l-1)T)}\times \\ 
\frac{(2+2T)\cdot(2+T)\cdot2\dots(2-(l-3)T)}{6\cdot(6+T)\cdot(6+2T)\dots(6+(l-1)T)} \times \\
\frac{(4-3T)\cdot(4-2T)\dots(4+(l-4)T)}{T\cdot(2T)\dots(lT)}\times \\
\frac{(4-5T)\cdot(4+(2l-3)T)\cdot((l+1)T)\cdot(4-(4-l)T)}{4\cdot(3T)\cdot(3T-4)\cdot(3T+4)}
\end{multline}
\subsubsection{$l+k=4$ and $k=0$}
\begin{multline}
Z(x,0,4,T,4,-2)=\\=
\frac{(2\alpha+\beta)\cdot4\cdot (5T)\cdot(2-T)\cdot(6-T)\cdot(8+T)\cdot(4-5T)\cdot(4+T)}{(3T)\cdot(4T)\cdot(8-3T)\cdot(2+3T)\cdot(4+4T)\cdot(4-3T)\cdot(4+3T)
	\cdot(6+3T)}
\end{multline}

\subsubsection{$l+k=4$ and $l=0$}
\begin{multline}
Z(x,4,0,T,4,-2)=\\
=
-\sinh\left[\frac{x}{4}: \right.
\frac{(2\alpha+\beta)\cdot6\cdot (T)\cdot(4+T)\cdot(6+2T)\cdot(6+T)\cdot(6-T)\cdot(4+2T)\cdot(4-5T)}{2\cdot(3T)\cdot(4T)\cdot(4-3T)\cdot(2+T)\cdot(2+2T)\cdot(2+3T)\cdot(4-2T)\cdot(4-T)}
\end{multline}

\subsubsection{$l+k=3$ and $l\geq1$}
\begin{multline}
Z(x,k,l,T,4,-2)=\\
=
\sinh\left[\frac{x}{4}: \right.
\frac{(2T)\cdot T}{8\cdot(8-T)\cdot(8-2T)} \times \\
\frac{(6+2T)\cdot(6+T)\cdot6}{2\cdot(2+T)\cdot(2+2T)} \times \frac{8\cdot(8-T)\dots(8-(2-l)T)}{T\cdot(2T)\dots((3-l)T)} \times \\
\frac{(4+3T)\cdot(4+2T)\dots(4-(l-1)T)}{(4-2T)\cdot(4-T)\dots(4+lT)} \times  \frac{(8-2T)\cdot(8-T)\cdot8\dots(8+(l-3)T)}{4\cdot(4-T)\cdot(4-2T)\dots (4-(l-1)T)}\times \\ 
\frac{(2+2T)\cdot(2+T)\cdot2\dots(2-(l-3)T)}{6\cdot(6+T)\cdot(6+2T)\dots(6+(l-1)T)} \times
\frac{(4-3T)\cdot(4-2T)\dots(4+(l-4)T)}{T\cdot(2T)\dots(lT)}\times \\
\frac{4\cdot(4+(2l-3)T)\cdot(lT)\cdot(4-(3-l)T)}{(3T-4)^2\cdot(3T+4)}
\end{multline}

\subsubsection{$l+k=3$ and $k=0$}
\begin{equation}
Z(x,0,3,T,4,-2)=-1
\end{equation}

\subsubsection{$l+k=3$ and $l=0$}
\begin{multline}
Z(x,3,0,T,4,-2)=\\
=
\sinh\left[\frac{x}{4}: \right.
\frac{(2\alpha+\beta)\cdot6\cdot (6+T)\cdot(6+2T)\cdot(4+T)\cdot(4+2T)\cdot(4-3T)}{2\cdot4\cdot(3T)\cdot(2+T)\cdot(2+2T)\cdot(4-T)\cdot(4-2T)}
\end{multline}

\subsubsection{$l+k=2$ and $k,l>0$}
\begin{multline}
Z(x,1,1,T,4,-2)=\\
=
-\sinh\left[\frac{x}{4}: \right.
\frac{(2\alpha+\beta)\cdot(2T)\cdot(6+2T)\cdot(6+T)\cdot(4+2T)\cdot(4+T)^2\cdot(8-2T)\cdot(2+2T)\cdot(4-T)}{2\cdot4^3\cdot6\cdot(T)^2(8-T)\cdot(2+T)\cdot(4-2T)
} 
\end{multline}

\subsubsection{$l+k=2$ and $k=0$}
\begin{multline}
Z(x,0,2,T,4,-2)=
\sinh\left[\frac{x}{4}: \right.
=\frac{(4+T)\cdot(2+2T)\cdot(4+2T)\cdot(6+2T)\cdot(8-2T)}{2\cdot4\cdot6\cdot8\cdot(4-T)}
\end{multline}

\subsubsection{$l+k=2$ and $l=0$}
\begin{equation}
Z(x,2,0,T,4,-2)=
\sinh\left[\frac{x}{4}: \right.
\frac{(6+T)\cdot(6+2T)\cdot (4+2T)}{2\cdot4\cdot(2+T)}
\end{equation}

\subsubsection{$l+k=1$ and $l=0$}
\begin{equation}
Z(x,1,0,T,4,-2)=
\sinh\left[\frac{x}{4}: \right.
\frac{(2T)\cdot(6+2T)\cdot (4+T)}{2\cdot4\cdot(T)}
\end{equation}

\subsubsection{$l+k=1$ and $k=0$}
\begin{multline}
Z(x,0,1,T,4,-2)=\\
=
\sinh\left[\frac{x}{4}: \right.
\frac{(2+2T)\cdot(4+2T)\cdot(6+2T)\cdot (8-2T)\cdot(4+T)}{2\cdot4\cdot6\cdot8\cdot(4-T)}
\end{multline}

{\bf Exceptional algebras:} 

For the exceptional algebras, the procedure is technically similar to that for the classical algebras, so we omit its detailed presentation.

\newpage

\chapter{On the problem of uniqueness of universal formulae for simple Lie algebras. 
Geometrical {\it configurations of points and lines} from the perspective of the uniqueness of universal formulae}
In this chapter, we describe how Vogel's universal description of Lie algebras makes it possible to connect two distinct areas of mathematics – the theory of Lie algebras and
geometrical {\it configurations of points and lines}. 

Firstly, we formulate the problem of the uniqueness of universal dimension formulae and introduce the notion of a {\it non-uniqueness factor}.

Then, we present a geometrical formulation of the uniqueness problem and show that it brings us to a completely new area of 
mathematics – the theory of configurations of points and lines.
Finally, we employ the geometrical formulation by deriving an explicit expression for a four-by-four non-uniqueness factor, making use of a known $(16_3, 12_4)$ configuration.

\newpage

\section{The problem of uniqueness of universal dimensions}
The emergence of the uniqueness problem of universal formulae for simple Lie algebras was motivated by the notice that the variance of intricacies of known formulae
is quite big.
For example, take a look at the following dimension formulae:
$$ dim X_2=\frac{(2 \alpha +\beta +\gamma ) (\alpha +2 \beta +\gamma ) (2 \alpha +2 \beta +\gamma ) (\alpha +\beta +2 \gamma ) (2 \alpha +\beta +2 \gamma ) (\alpha +2 (\beta +\gamma ))}
{\alpha ^2 \beta ^2 \gamma ^2} $$
which is the dimension of the $X_2$ representation, and this one \cite{LM1}

\begin{multline}
dim((\mathfrak{g})^2 (Y_2(\beta))^2)=\\
-\frac{(\alpha +\gamma ) (2 \gamma -\alpha ) (\beta +\gamma ) (2 \beta +\gamma ) (\beta +2 \gamma ) (-\alpha +\beta +\gamma ) (\alpha +\beta +\gamma )^2}
{\alpha ^4 \beta ^2 \gamma} \times \\
\frac{ (2 \alpha +\beta +\gamma ) (-\alpha +2 \beta +\gamma ) (\alpha +2 \beta +\gamma ) (2 \alpha +2 \beta +\gamma )(\alpha -\beta +2 \gamma ) (2 \alpha -\beta +2 \gamma )  }
{  (\beta -4 \alpha ) (\beta -3 \alpha ) (\beta -\alpha )^3 (2 \beta -3 \alpha ) (\gamma -3 \alpha ) } \times \\
 \frac{(-3 \alpha +\beta +2 \gamma ) 
(\alpha +\beta +2 \gamma ) (2 \alpha +\beta +2 \gamma ) (-5 \alpha +2 \beta +2 \gamma ) (-\alpha +2 \beta +2 \gamma ) (\alpha +2 \beta +2 \gamma )}
	{(\gamma -2 \alpha ) (\gamma -\alpha )^2 (2 \gamma -3 \alpha ) (\gamma -\beta ) (\alpha +\beta -\gamma )}
\end{multline}
which gives the dimensions of Cartan product of the squares of $Y_2(\beta)$ and the adjoint representations
 \footnote{
 they appear in the following universal
 decomposition of the symmetric square of the adjoint representation:
  \begin{eqnarray}
	S^2 \mathfrak{g} = \mathds{1} \oplus Y_2(\alpha) \oplus Y_2(\beta) \oplus Y_2(\gamma)
\end{eqnarray}
 }

As we see, the first formula writes much simpler than the latter.
So, a natural question arises: for a given universal formula can we find a more “simple-looking” one, with the same 
features of universal nature? Or, generally, are the known universal formulae unique?

Note, that all known universal formulae  possess a specific structure: they are rational functions, where both the numerator and denominator decompose into products of the same number  of 
linear factors of universal parameters:
\begin{eqnarray}\label{Qx}
	F=\prod_{i=1}^{k}\frac{n_i\alpha+x_i\beta+y_i\gamma}{m_i\alpha+z_i\beta+t_i\gamma}
\end{eqnarray}

 Now let $F_1$ and $F_2$ be two universal formulae, which are rational functions, where both the numerator and denominator decompose into products of 
 the same finite number of linear factors of Vogel's parameters,
 and yield the same outputs at the points from Table \ref{tabV}, which correspond to the simple Lie algebras.
 
 \begin{table}[ht] 
	\caption{Vogel's parameters for simple Lie algebras and the distinguished lines}
	\centering
	\begin{tabular}{|r|r|r|r|r|r|} 
		\hline Algebra/Parameters & $\alpha$ &$\beta$  &$\gamma$  & $t$ & Line \\ 
		\hline  $sl(N)$  & -2 & 2 & $N$ & $N$ & $\alpha+\beta=0$ \\ 
		\hline $so(N)$ & -2  & 4 & $N-4$ & $N-2$ & $ 2\alpha+\beta=0$ \\ 
		\hline $sp(2N)$ & -2  & 1 & $N+2$ & $N+1$ & $ \alpha+2\beta=0$ \\ 
		\hline $exc(n)$ & $-2$ & $2n+4$  & $n+4$ & $3n+6$ & $\gamma=2(\alpha+\beta)$\\ 
		\hline 
	\end{tabular}
	            \\ {On the $exc$ line $n=-2/3,0,1,2,4,8$ for $G_2, so(8), F_4, E_6, E_7,E_8 $, 
		respectively.}
		 \label{tabV}
\end{table} 

Then their ratio $Q$, which has the same structure, is obviously equal to $1$ at those points.
\begin{eqnarray}
	Q=\frac{F_1}{F_2}
\end{eqnarray}
We call such functions $Q$ {\it non-uniqueness factors}.
In fact, the problem of uniqueness of dimension formulae formulates as the search for possible non-uniqueness factors $Q$.

Note that the points from Vogel's table occupy the following distinguished lines \cite{LM1}:
\begin{eqnarray} \label{lines}
sl:	\alpha+\beta=0, \\ \label{sl}
so:	2\alpha+\beta=0, \\ \label{so}
sp:	\alpha+2\beta=0,\\
exc:	\gamma-2(\alpha+\beta)=0, \label{exc}
\end{eqnarray}
on which the linear, orthogonal, symplectic and the exceptional algebras are situated, 
respectively. 
Based on this fact, we add an additional requirement to the problem, namely,
 we search for a $Q$, which is equal to 1 not only at the points, associated
with the simple Lie algebras, but also on the entire distinguished lines.

In \cite{Uniq21} (see Appendix C.IV) we have derived the following general expression for such non-uniqueness factors, equivalent to $1$ on each of the four distinguished lines:

\begin{eqnarray}
	Q=\prod_{i=1}^{k}\frac{n_i\alpha+x_i\beta+y_i\gamma}{ k_i n_{s(i)}\alpha+ x_i\beta+y_{i}\gamma}=\prod_{i=1}^{k}\frac{ n_i\alpha+x_i\beta+y_i\gamma}{ c_i n_{p(i)}\alpha+ x_i\beta+y_{i}\gamma} \label{8} \\ 
	x_i=c_ix_{p(i)} \label{9} \\
	y_i=k_iy_{s(i)} \label{10} \\    
	k_i n_{s(i)}=c_i n_{p(i)} \label{11} \\  
	y_i=r_i y_{v(i)} \label{12} \\
	c_i n_{p(i)}+3x_i= r_i(n_{v(i)}+3x_{v(i)}) \label{13} \\ 
	c_1c_2...c_k=1 \label{14} \\ 
	k_1k_2...k_k=1\label{15} \\   
	r_1 r_2...r_k=1 \label{16}
\end{eqnarray}

for some permutations $s(i), p(i), v(i), \, i=1,2...k$. Note, that this expression is written after the following change of coordinates was made:

$$\alpha^{\prime}=\alpha+\beta,  \,\, \beta^{\prime}= 2\alpha+\beta  \,\, \gamma^{\prime} = \gamma-2(\alpha+\beta) $$

so that in the primed coordinates the equations of the distinguished lines $sl,so,exc$ will simply be 
$$\alpha^{\prime}=0, \,\, \beta^{\prime}=0, \,\, \gamma^{\prime}=0$$ 
respectively, and thus the one for the $sp$ line will be $3\alpha^{\prime}-\beta^{\prime}=0$.
The prime mark in (\ref{8}) is dropped for convenience.

In order to write a non-uniqueness factor explicitly, one has to choose an appropriate set of the $s(i), p(i), v(i), \, i=1,2...k$ permutations, then solve the presented equations, using it.
It is easy to see, that the choice of the set of permutations is quite wide: there are generally $k!$ variations for each of them.
Remarkably, there is a rather smart option of the derivation of an explicit expression for some concrete $k$, which opens up after interpreting the non-uniqueness factors geometrically.
In the next section we outline the geometrical rephrasing of the problem, then derive a four-by-four (at $k=4$) non-uniqueness factor, invoking a specific $(16_3 12_4)$ 
geometrical configuration \cite{GB}.

\section{Rephrazing the problem of uniquiness}
In this section, we provide a geometric point of view to the problem
of uniqueness, setting up its connection with a classical problem of 
the so-called {\it configurations}, namely {\it configurations of points and lines}. 

\subsection{Geometric representation of universal formulae}
First, observe, that each of the linear factors in the expression of $Q$ corresponds to a line in the projective Vogel's plane. Indeed, to each of the factors $x\alpha+y\beta+z\gamma$ 
one can put in correspondence the line equation $x\alpha+y\beta+z\gamma=0$.

Thus, for any given expression for universal (quantum) dimension, with say $k$ multipliers, we can draw a unique picture in the Vogel's plane, consisting of $k$ lines, 
corresponding to the 
linear factors in the numerator, which will be referred to as {\it red lines} for convenience, and $k$ {\it green} lines for those in the denominator.
In addition, we can draw a number of {\it black} lines, 
corresponding to the distinguished $sl, so, exc$, lines as well as those, associated to the permuted coordinates - such as the $sp$ line.  

\begin{figure}
 \centering
\begin{tikzpicture} \tkzDefPoints{0/0/F, -2/3/A, 1/0/B, 2/0/C, 4/4/D, 4/0/E, 0/4/G, 0/-4/K, 4/-3/L, 4/-4/M, 1/4/N, 1/-4/J, 2/4/H, 2/-4/P}
\tkzDrawLines[color=green](K,G E,F)
\tkzDrawLines[color=red](E,A L,A G,M)
\tkzDrawLines(N,J H,P D,M K,D)
	\tkzLabelLine[pos=-0.22, left](G,K){$g_2$}
		\tkzLabelLine[pos=-0.2, left](F,E){$g_3$}
		\tkzLabelLine[pos=-0.18, left](A,E){$r_2$}
		\tkzLabelLine[pos=-0.2, right](A,L){$r_1$}
		\tkzLabelLine[pos=-0.15, left](G,M){$r_3$}
		\tkzLabelLine[pos=-0.2, left](N,J){$sp$}
		\tkzLabelLine[pos=-0.2, left](H,P){$sl$}
		\tkzLabelLine[pos=-0.2, left](D,M){$so$}
		\tkzLabelLine[pos=-0.2, right](D,K){$exc$}
\end{tikzpicture}
\caption{The "sketch" of the dimension formula of the adjoint.} \label{fig:Conf1}
\end{figure}
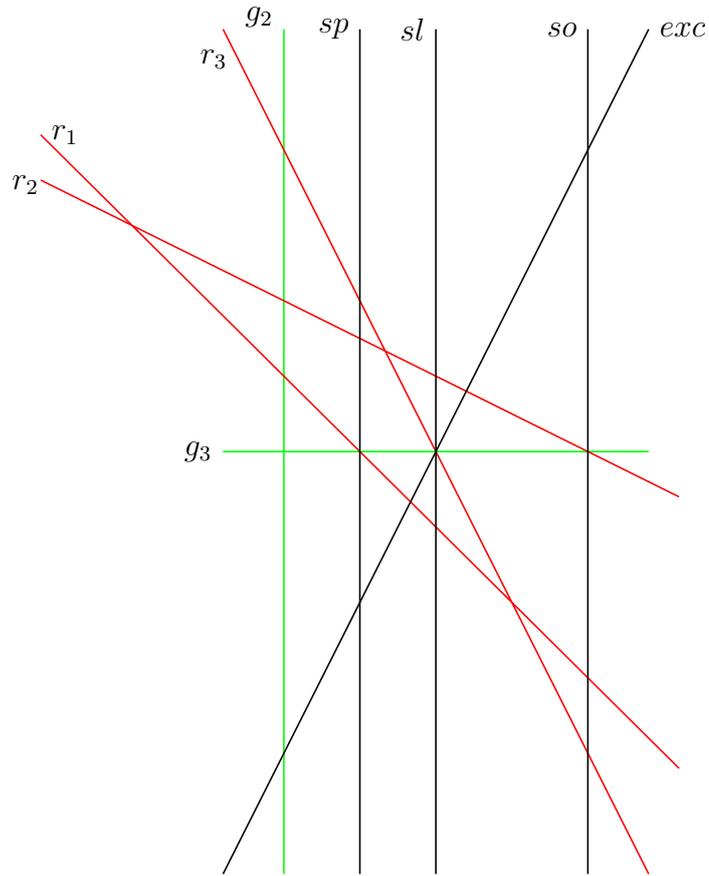

One can see the corresponding picture\footnote{The labeling of the lines in the following figures is meant to identify the corresponding colors
 they are given.
For example, in Figure \ref{fig:Conf1}, $r_2$ identifies the line, associated to the second factor in the numerator of (\ref{ad}), and $g_3$ - 
to the third factor in the corresponding denominator.} for the simplest universal formula, namely the dimension of the 
adjoint representation (\ref{ad}), in Figure \ref{fig:Conf1}.

Let's consider the picture, associated with a non-uniqueness factor $Q$.
 It turns out that each of the black lines must contain $k$ points, at which a green and a red line intersect. 

Indeed, this statement exactly rephrases the cancellation mechanism, necessary for the non-uniqueness factor $Q$ to be $1$ at the distinguished lines:
 when restricting $Q$ to a black line, each of the factors from the numerator is proportional to some factor from the denominator.
This means that these two factors are zeroing simultaneously, meaning, that residing on a black and, say, a red line at once, we necessarily reside on a green line too. 

It is easy to notice, that this corresponding picture also contains information about the choice of the permutations $s(i), p(i), ...$, -
the intersection points of three different-colored lines obviously define the pairs of canceling factors, when restricting the function to each of the distinguished black lines. 

Thus, the picture of $k$ black, $k$ red, and $k$ green lines, corresponding to a non-uniqueness factor $Q$, has the following characteristic feature:
 on each of the black lines, there are $k$ points at which a red and a green line intersect. 
 Note that besides these points of intersection of three differently colored 
 lines, there may be some other intersection points, which however will not be 
 of interest for us. 

\subsection{Configurations}

Let's introduce the following standard definitions \cite{GKF,GB}:

{\bf Definition 1.}

{\it We say a line is incident with a point, (equivalently, a point is incident with a line) if it passes through it (equivalently, if it lies on it).}

{\bf Definition 2.}

{\it A { \bf configuration} $(p_{\gamma} l_{\pi})$ is a set of $p$ points and $l$ lines, such that every point is incident with precisely $\gamma$ of these lines and
every line is incident with precisely $\pi$ of these points.}

{\bf Remark 1.} Notice, that the total number of incidences, on one hand,
 is equal to $p \gamma$, and is $l \pi$, on the other hand, so that from 
Definition 2 it 
follows, that $p \gamma= l \pi$.

{\bf Remark 2.} If $p=l, \gamma=\pi$, the configuration is denoted by $(p_\gamma)$.

We see that the picture of $k$ black, $k$ red, and $k$ green lines, possessing the feature described in the previous subsection,
 turns into a configuration iff the
number of black, red, and green lines coincide and is equal to
 $k$.
Obviously, the corresponding configuration will be $(k^2_3,3k_k)$. 

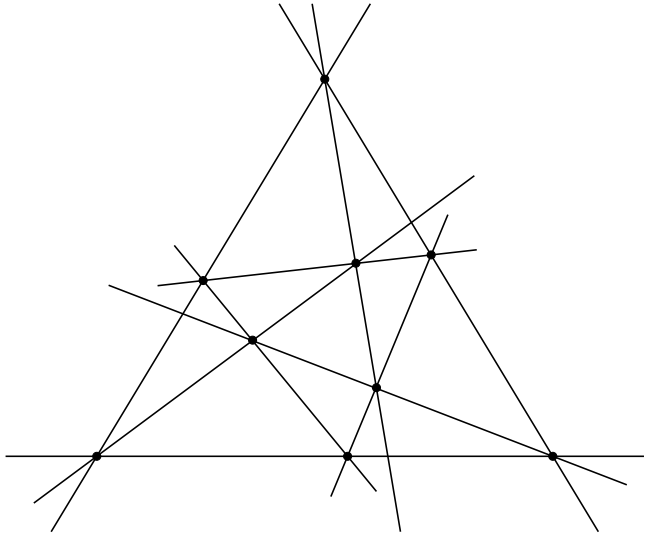
\begin{figure}
 \centering
	\begin{tikzpicture} \tkzDefPoints{0/0/A, 6/0/B, 3/5/C, 
			1.4/2.33/D, 4.4/2.67/E, 3.3/0/F, 3.68/0.91/G, 3.41/2.56/H,
			2.05/1.54/I, 4.14/3.1/J, 1.13/1.89/K, 3.83/0/L}
\tkzDrawLines(A,B C,B A,C F,D D,E E,F A,J C,L B,K)
\tkzDrawPoints(A,B,C,D,E,F,G,H,I) 
			\end{tikzpicture}
	\caption{The $(9_3)_2$ configuration, which cannot be "colored" in order to be corresponded to some $Q$} \label{fig:Conf3}
\end{figure}

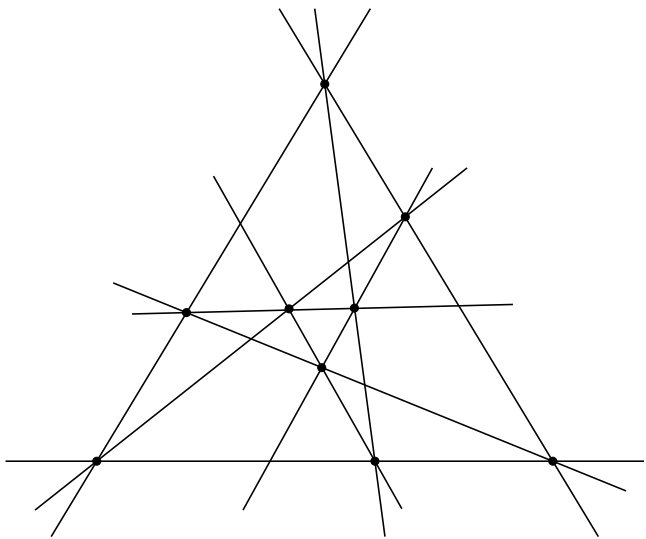
\begin{figure}
 \centering
\begin{tikzpicture} \tkzDefPoints{0/0/A, 6/0/B, 3/5/C, 3.66/0/D, 1.18/1.97/E, 4.06/3.24/F, 3.39/2.03/G, 2.96/1.24/H, 2.53/2.02/I, 1.89/3.15/J, 4.76/2.06/K, 2.28/0/L}
\tkzDrawLines(A,B C,B A,C A,F C,D B,E J,D E,K F,L)
 \tkzDrawPoints(A,B,C,D,E,F,H,I,G)
	\end{tikzpicture}
\caption{The $(9_3)_3$ uncolorable configuration} \label{fig:Conf4}
\end{figure}

However, if we have a configuration $(k^2_3,3k_k)$, it doesn't mean that we can definitely construct a corresponding $Q$. 
The possible obstacle is that one would not be able to attribute the black, red, and green colors to its $3k$ lines such that at each of the points, 
belonging to the configuration, three lines of different colors meet. Such configuration are presented in Figures \ref{fig:Conf3} and \ref{fig:Conf4}.

For any given configuration $(p_{\gamma} l_{\pi})$ one can construct a so-called {\it configuration table}: we label the points and lines of that configuration, 
then for each of the lines allocate a column, consisting of the labels of the 
points, which are incident with the corresponding line. 
Characteristic properties of a configuration table are the following: the label for each of the points occurs in exactly $\gamma$ columns, different columns do not
contain two similar labels of points, and each column contains exactly $\pi$ labels. 
Two configuration tables are identical, if they coincide after some relabeling of points and lines, and/or rearranging the points in a given column.  

So, "possible" configurations of a given type $(p_{\gamma} l_{\pi})$ can be considered simply as different configuration tables of that type. 

Further, a configuration table is called {\it realizable} if one can construct a geometrical picture of lines and points corresponding to it.
Not all tables are realizable. 

\section{The Pappus-Brianchon-Pascal configuration}

Let us take the (\ref{8})-(\ref{16}) general solution for a four-by-four non-uniqueness factor $Q$ and relax the (\ref{12}), (\ref{13}), and (\ref{16}) conditions.
We will get a solution for a three-by-three non-uniqueness factor, which is equivalent to $1$ on three basic lines – $sl, so$, and $exc$.
It can be shown that one can get the following non-trivial $Q$ for this case (see Appendix C.IV, (\ref{Q33})):
\begin{eqnarray}\label{Q33}
	\frac{(\alpha+\beta x+\gamma y) (\alpha c_1 c_2 +\beta c_2 x+\gamma y) (\alpha c_1+\beta c_1 c_2 x+\gamma y)}{(\alpha c_1+\beta x+\gamma y) (\alpha+\beta c_2 x+\gamma y) (\alpha c_1 c_2 +\beta c_1 c_2 x+\gamma y)}
\end{eqnarray}

A relevant configuration happens to be corresponding to this solution.
It is the configuration $(9_3,9_3)_1$, which is usually referred as $(9_3)_1$ \cite{GKF,GB}, since the terms in the standard notation 
coincide. This configuration is also known as the Pappus (Pappus of Alexandria) or Pappus-Brianchon-Pascal configuration, 
which is presented in Figure \ref{fig:Conf2}.

The index in the notation $(9_3)_1$ is to indicate the fact that there are several  $(9_3)$ configurations, so that it is used  to distinguish these.
Possible values of the index, i.e. the number of different configurations $(9_3)$ is $3$, equivalently, there are three different configuration tables for $(9_3)$ configuration. 
Each of these $3$ tables happens to be realizable. However, only one of them, presented in Figure \ref{fig:Conf2}, $(9_3)_1$ from \cite{GKF}, can be colored in the way we need. 
For example, for the configuration $(9_3)_2$ (see Figure \ref{fig:Conf3}) 
it is impossible to distinguish $3$ black lines, since for any two lines of the configuration there is always a third one, which intersects with one of them at some point,
 belonging to the configuration. This violates the requirement that at each point of the configuration three lines of different
 colors intersect. The same reasoning holds for the remaining third, i.e. the  $(9_3)_3$ configuration, see Figure \ref{fig:Conf4}.
So, we have the following
 
 {\bf Proposition 4.1}
{\it The non-uniqueness factor (\ref{Q33}) at $k=3$ is in one-to-one correspondence with
 the Pappus-Brianchon-Pascal $(9_3)_1$ configuration.}
 
 \begin{proof}
 The proof is obvious, since there is no other
  $(9_3)$ configuration which can be colored in the needed way. 
\end{proof}

\begin{figure}
 \centering
\begin{tikzpicture} \tkzDefPoints{3.67/2.98/A, 5/3/B, 9.78/2.945/C, 
 1.145/5.64/H, 2.28/8.275/I, 3.025/9.99/J}
 \tkzInterLL(C,A)(J,H)
\tkzGetPoint{O}
\tkzDrawPoint(O)
\tkzDrawLine(C,O)
\tkzDrawLine(J,O)
\tkzLabelPoint[below right](O){$(0,0)$}
\tkzDrawLines[color=red](H,B I,A J,C)
\tkzDrawLines[color=green](A,H C,I J,B)
\tkzLabelPoint[below left](A){$(-\frac{1}{c_2x},0)$}
\tkzLabelPoint[below left](C){$(-c_1/ x,0)$}
\tkzLabelPoint[left](I){$(0,-\frac{c_1}{y})$}
\tkzLabelPoint[below left](H){$(0,-\frac{1}{y})$}
\tkzLabelPoint[above right](B){$(-1/x,0)$}
\tkzLabelPoint[right](J){$(0,-\frac{1}{c_1 c_2 y})$}
  \tkzDrawPoints(A,B,C,H,I,J)
                 \tkzLabelLine[pos=0.3, right](H,B){$r_1$}
		\tkzLabelLine[pos=0.5, right](J,C){$r_2$}
		\tkzLabelLine[pos=0.45, right](I,A){$r_3$}
		\tkzLabelLine[pos=0.4, right](I,C){$g_1$}
		\tkzLabelLine[pos=0.5, left](H,A){$g_2$}
		\tkzLabelLine[pos=0.6, right](J,B){$g_3$}
		\tkzLabelLine[pos=-0.15, above](C,O){$exc: \ \gamma=0$}
		\tkzLabelLine[pos=-0.2, right](J,O){$so: \ \beta=0$}

\end{tikzpicture}
\caption{The Pappus-Brianchon-Pascal $(9_3)_1$configuration after a projective transformation} \label{fig:Conf5}
\end{figure}
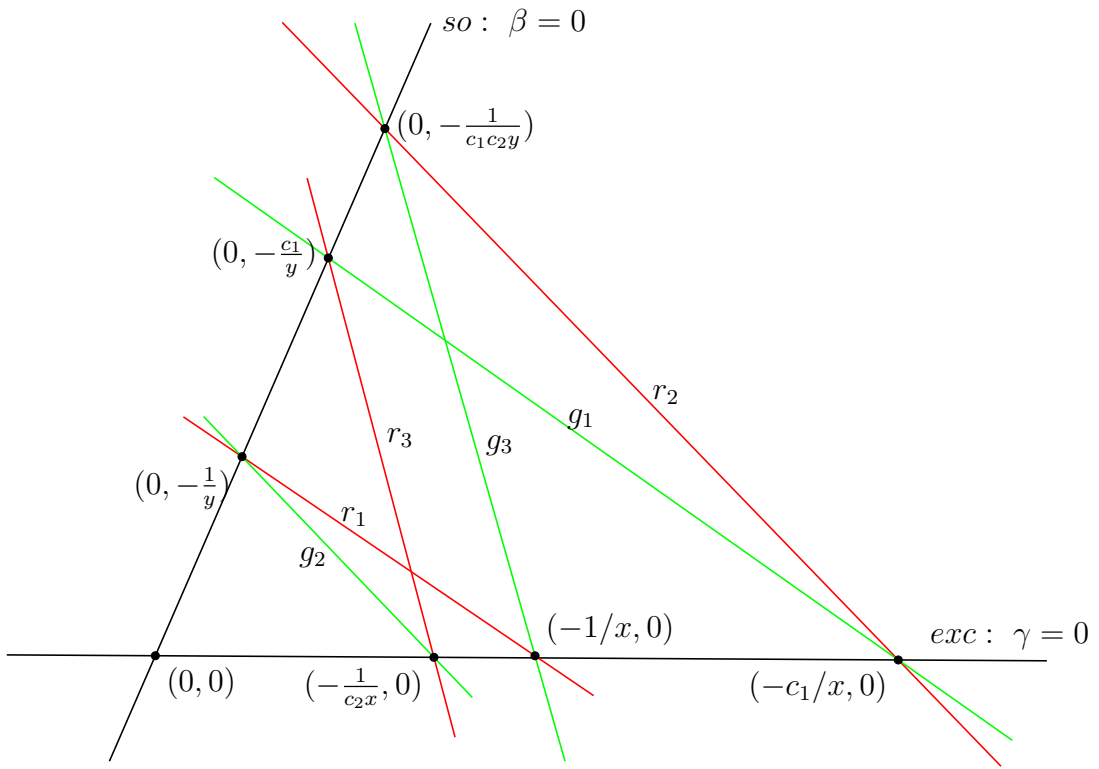

The picture, associated to the non-uniqueness factor (\ref{Q33}) is given in Figure \ref{fig:Conf5}, which is the $(9_3)_1$
 configuration after a projective transformation, which takes the $\alpha=0$ line to infinity.\footnote{One has to take into account that as the equations of the three distinguished lines are $\alpha=0, \beta=0$ and $\gamma=0$,
one of them unavoidably will be the ideal line of the projective plane, i.e. the line in the infinity (we choose $\alpha=0$). }
 
 Finally, the geometrical roles of the free parameters $c_1,c_2,x,y$ in the 
 (\ref{Q33}) expression of $Q$ are easily observed in the same Figure 
 \ref{fig:Conf5}, where the 
 associated coordinates of the points of the configuration are shown explicitly.  
 \begin{figure}
 \centering
\begin{tikzpicture} \tkzDefPoints{0/0/A,2/0/B, 5/0/C, 0.5/4/D, 2.5/4.6/E, 5.5/5.5/F}
\tkzDrawLine(A,C)
\tkzDrawLine(D,F)
\tkzDrawLines[color=green](A,E B,F)
\tkzDrawLines[color=red](C,E B,D)
\tkzDrawLines[color=red](A,F)
\tkzDrawLines[color=green](C,D)
\tkzInterLL(A,F)(C,D)
\tkzGetPoint{I}
\tkzInterLL(A,E)(B,D)
\tkzGetPoint{H}
\tkzInterLL(B,F)(C,E)
\tkzGetPoint{J}
\tkzDrawLine(H,J)
\tkzDrawPoint(H)
\tkzDrawPoint(J)
\tkzDrawPoint(I)
 \tkzDrawPoints(A,B,C,D,E,F) 
                 \tkzLabelLine[pos=-0.2, right](D,B){$r_1$}
		\tkzLabelLine[pos=-0.2, right](E,C){$r_2$}
		\tkzLabelLine[pos=-0.15, right](F,A){$r_3$}
		\tkzLabelLine[pos=-0.15, left](D,C){$g_1$}
		\tkzLabelLine[pos=-0.15, right](E,A){$g_2$}
		\tkzLabelLine[pos=-0.2, left](F,B){$g_3$}
		\tkzLabelLine[pos=-0.38, right](D,F){$sl: \ \alpha=0$}
		\tkzLabelLine[pos=-0.2, left](H,J){$so: \ \beta=0$}
		\tkzLabelLine[pos=-0.68, right](A,C){$exc: \  \gamma=0$}
\end{tikzpicture}
\caption{The Pappus-Brianchon-Pascal, or $(9_3)_1$ configuration} \label{fig:Conf2}
\end{figure}
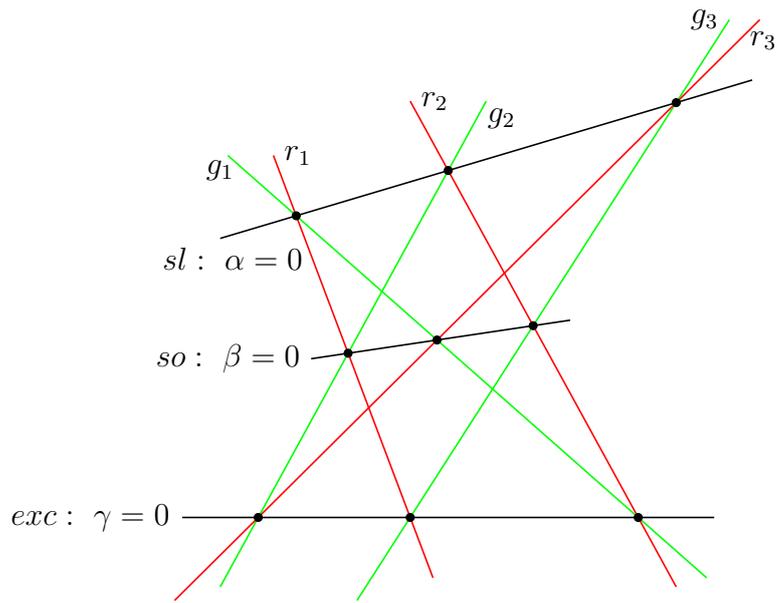

\section{A four-by-four non-uniqueness factor and a known $(16_3, 12_4)$ configuration}

\begin{figure}
	\begin{tikzpicture} \tkzDefPoints{-0.5/2/A, 1.5/2/B, -1.5/0.5/C, 
			2.5/0.5/D, 3.5/-1/E, 4.5/-2.5/F, -2.5/-1/G, -3.5/-2.5/H,
			0.5/1.5/I, 4.9/-0.7/L, -1.5/-0.7/M, 2.5/-0.7/N, 0.5/-0.1/O,
			0.5/1.1/J, -3.9/-0.7/K, 0.5/-1.645/P}
		\tkzDrawLines(A,H I,P B,F K,L)
		\tkzDrawLines[color=red](A,F B,K D,G L,H)
		\tkzDrawLines[color=green](A,L C,E K,F B,H)
		\tkzLabelLine[pos=-0.15, left](A,L){$g_1$}
		\tkzLabelLine[pos=-0.2, left](B,H){$g_2$}
		\tkzLabelLine[pos=-0.2, left](K,F){$g_4$}
		\tkzLabelLine[pos=1.18, left](E,C){$g_3$}
		\tkzLabelLine[pos=-0.2, right](B,K){$r_1$}
		\tkzLabelLine[pos=-0.2, right](A,F){$r_2$}
		\tkzLabelLine[pos=-0.18, right](D,G){$r_3$}
		\tkzLabelLine[pos=-0.25, left](L,H){$r_4$}
		\tkzLabelLine[pos=-0.2, below](K,L){$ \gamma=0$}
		\tkzLabelLine[pos=-0.2, below](H,A){$ \beta=3\alpha$}
		\tkzLabelLine[pos=-0.2, below](P,I){$\alpha=0$}
		\tkzLabelLine[pos=-0.25, left](F,B){$ \beta=0$}
				\tkzDrawPoints(A,B,C,D,E,F,G,H,I,L,M,N,O,J,K,P)
			\end{tikzpicture}
	\caption{A $(16_3 12_4)$ configuration} \label{fig:Conf16}
\end{figure}
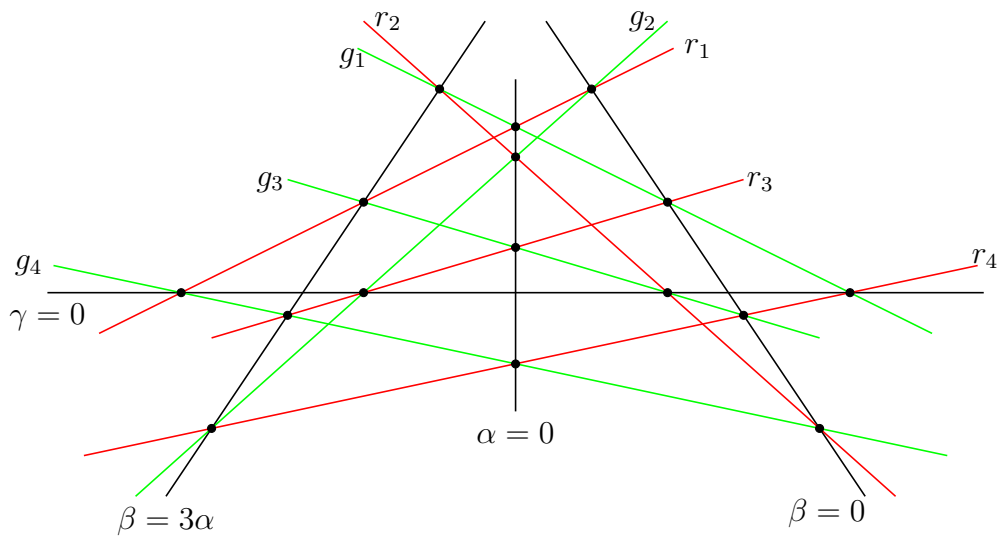

If we take the lines $sl, so, sp, exc$ as black lines and search for a
 non-uniqueness factor $Q$, which is equal to $1$ on each of these lines, we will happen to be dealing with the configuration  
$(16_3 12_4)$ ($k=4$). 
One of its realizations, taken from \cite{GB}, is
 presented in Figure \ref{fig:Conf16}.

{\bf Proposition 4.2}
{\it The configuration $(16_3 12_4)$ presented in Figure \ref{fig:Conf16} corresponds to the following non-uniqueness factor $Q$ for universal dimensions:
\begin{multline}\label{QQ}
	Q=
	\frac{(3 k k_1k_2 n \alpha+ (-k k_1 k_2 n-k_1 n) \beta + 3 c_2 k y \gamma) 
	(3 n \alpha +n (-k k_2-1) \beta +3 k k_2 y\gamma)}
	{(-3 k_1 n\alpha +(k k_1 k_2 n+k_1 n)\beta -3 c_2 k y \gamma)
	(-3 c_2 n\alpha+ (c_2 n+k_1 k_2 n)\beta-3 c_2 k k_2 y\gamma)
	} \times \\
	\frac{(3 c_2 n \alpha+(-c_2 n-k_1 k_2 n) \beta+3 c_2 y\gamma)
	 (-3 k_1 k_2 n \alpha+(c_2 n+k_1 k_2 n)\beta -3 c_2 k k_2 y\gamma)}
	 {(3 k_1 k_2 n\alpha +(-c_2 n-k_1 k_2 n)\beta +3 c_2 y\gamma )
	(3 k k_2 n \alpha +(-k k_2 n-n)\beta +3 k k_2 y \gamma)}
		\end{multline}
with $y=y_1,  \,\, n=n_1,  \,\, k=c_1 k_4,$ free parameters and which is equal to $1$ on each of the $sl, so, sp, exc$ distinguished lines in Vogel's plane. }
\begin{proof}
As it is seen, the configuration in Figure \ref{fig:Conf16} can be colored in a needed way, so that it corresponds to some non-uniqueness factor $Q$ at $k=4$.
Tracing the pairs of green and red lines, intersecting at each of the black lines, we 
almost immediately define the set of three permutations $s(i), p(i), v(i), \, i=1,2...k$:
$$s(1)=3, s(2)=1, s(3)=4, s(4)=2$$
$$p(1)=4, p(2)=3, p(3)=2, p(4)=1$$
$$v(1)=2, v(2)=4, v(3)=1, v(4)=3$$
Then, equations (\ref{9})-(\ref{12}) and (\ref{14})-(\ref{16}) yield:
\begin{eqnarray}
	x_3= \frac{x_2}{c_2}, \,\, x_4= \frac{x_1}{c_1}  \\
	y_2=k_2 y_1, \,\, y_3=\frac{y_1}{k_1}, \,\, y_4=k_2 k_4 y_1  \\
	n_2=\frac{n_1}{c_1 k_4}, \,\, n_3=\frac{k_2 n_1}{c_2},  \,\, n_4=\frac{k_1 k_2 n_1}{c_1 c_2} \\
	r_1=1/k_2, \,\, r_2=1/k_4, \,\, r_3=1/k_1, \,\, r_4=1/k_3 \\
	c_4=\frac{1}{c_1}, \,\, c_3=\frac{1}{c_2}
\end{eqnarray}

The remaining equation (\ref{13}) yields:
\begin{eqnarray}
	x_1=- \frac{k_1 k_2+c_2}{3 c_2} n_1,  \,\, x_2= -\frac{n_1}{c_1 k_4}-k_2 n_1 
\end{eqnarray}
So that we get the non-trivial solution (\ref{QQ}) for $Q$:
\end{proof}

\section{The $(144_3 36_{12})$ configuration and a symmetric non-uniqueness factor $Q$}

An immediate problem, arising after the previous investigation, is the derivation of
a symmetric non-uniqueness factor $Q$, which would be equivalent to $1$ on all the
$12$ lines, obtained by the basic lines after all possible permutations of the coordinates.
The search of such a $Q$ appears to be one of a realizable $(144_3 36_{12})$ configuration in the scope of the geometrical approach.
Unfortunately, this configuration has not been studied yet, so the existence
 of a symmetric $Q$ remains an open question.

\newpage

\section{Appendix C.IV}

Below we present the derivation of the general non-uniqueness factor $Q$ (i.e. (\ref{9})-(\ref{16})), which is equivalent to $1$ on $sl, so, exc$, and $sp$ lines in Vogel's plane.

To simplify calculations, we make the following change of coordinates :

\begin{tabular}{c|c}
\centering

$\alpha^{\prime}=\alpha+\beta$	& $\alpha=-\alpha^{\prime}+\beta^{\prime}$ \\
	\hline
$\beta^{\prime}= 2\alpha+\beta$	& $\beta=2\alpha^{\prime}-\beta^{\prime}$ \\
	\hline
	$\gamma^{\prime} = \gamma-2(\alpha+\beta) $	& $\gamma=2\alpha^{\prime}+\gamma^{\prime}$ \\
	
\end{tabular}

so that in the primed coordinates the equations of the basic lines $sl,so,exc$ will simply be 
$$\alpha^{\prime}=0, \beta^{\prime}=0,\gamma^{\prime}=0.$$ 
 And, consequently, the equation of the $sp$ line will take the following form:
 $$3\alpha^{\prime}-\beta^{\prime}=0$$

We drop the prime mark below.

Now let us take a universal dimension in its most general form

\begin{eqnarray}\label{Qx}
	Q=\prod_{i=1}^{k}\frac{n_i\alpha+x_i\beta+y_i\gamma}{m_i\alpha+z_i\beta+t_i\gamma}
\end{eqnarray}

and 
consider its values on the three lines $\alpha=0, \beta=0, \gamma=0.$

We require $Q\equiv1$ at $\alpha=0$. 
Then 

\begin{eqnarray}\label{Qx2}
	1\equiv\prod_{i=1}^{k}\frac{x_i\beta+y_i\gamma}{z_i\beta+t_i\gamma}
\end{eqnarray}

and one deduces, that $z_i=l_ix_{q(i)}, t_i=l_iy_{q(i)}$, with some permutation $q(i), i=1,...k$, and non-zero multipliers $l_i$ with $l_1l_2...l_k=1$. 

Substituting these relations into (\ref{Qx}), one has 

\begin{eqnarray}\label{Qx22}
	Q=\prod_{i=1}^{k}\frac{n_i\alpha+x_i\beta+y_i\gamma}{m_i\alpha+l_ix_{q(i)}\beta+l_iy_{q(i)}\gamma}
\end{eqnarray}

Absorbing the $1/l_i$ into $m_i$, renumbering $m_i\rightarrow m_{q(i)}$ and changing the order of the multipliers in the denominator, we rewrite $Q$ as:

\begin{eqnarray}\label{Qx23}
	Q=\prod_{i=1}^{k}\frac{n_i\alpha+x_i\beta+y_i\gamma}{m_i\alpha+x_{i}\beta+y_{i}\gamma}
\end{eqnarray}

Now let $Q\equiv1$ at $\beta=0$:
\begin{eqnarray}\label{Qx24}
	1\equiv\prod_{i=1}^{k}\frac{n_i\alpha+y_i\gamma}{m_i\alpha+y_{i}\gamma}
\end{eqnarray}

Then one must have $y_i=k_iy_{s(i)}, m_i=k_in_{s(i)}$, with some permutation $s(i)$ and with $k_1k_2...k_k=1$, so that $Q$ accepts the form: 

\begin{eqnarray}\label{Qx25}
	Q=\prod_{i=1}^{k}\frac{n_i\alpha+x_i\beta+y_i\gamma}{k_i n_{s(i)}\alpha+x_{i}\beta+k_iy_{s(i)}\gamma}=
	\prod_{i=1}^{k}\frac{n_i\alpha+x_i\beta+y_i\gamma}{k_i n_{s(i)}\alpha+x_{i}\beta+y_i\gamma}
\end{eqnarray}

Next we require $Q\equiv1$ at $\gamma=0$:

\begin{eqnarray}\label{Qx26}
	1\equiv \prod_{i=1}^{k}\frac{n_i\alpha+x_i\beta}{ k_i n_{s(i)}\alpha+ x_i \beta}
\end{eqnarray}

Again, from this relation we infer  

\begin{eqnarray}
	x_i=c_ix_{p(i)} \\
	k_i n_{s(i)}=c_i n_{p(i)}
\end{eqnarray}

for some permutation $p(i)$ and $c_i$ with $c_1c_2...c_k=1$.

So, altogether we have the following expression for $Q$ with the restrictions on its parameters:

\begin{eqnarray}\label{Qx27}
	Q=\prod_{i=1}^{k}\frac{n_i\alpha+x_i\beta+y_i\gamma}{ k_i n_{s(i)}\alpha+ x_i\beta+y_{i}\gamma}=\prod_{i=1}^{k}\frac{ n_i\alpha+x_i\beta+y_i\gamma}{ c_i n_{p(i)}\alpha+ x_i\beta+y_{i}\gamma} \\    \label{3lineBeq}
	x_i=c_ix_{p(i)} \\ 
	y_i=k_iy_{s(i)} \\
	k_i n_{s(i)}=c_i n_{p(i)} \\
	c_1c_2...c_k=1 \\  \label{3lineEeq}
	k_1k_2...k_k=1 
\end{eqnarray}
for some permutations $s(i), p(i)$. Note that after having solved these equations, one must check the $Q$ on absence of any cancellation in it. 

It is easy to show, that there is not a non-trivial solution if $k=1,2$. 
For $k=3$ one can show that the existence of a non-trivial solution requires that the permutations $s(i),p(i)$ do not have fixed points and do not coincide,
 i.e. $s(i)=i+1, p(i)=i+2 \,\, (mod \,\,3)$, or vice versa. One can also show that $n_i \neq 0$, so that one can factor them out, 
 or effectively put $n_i=1$, so that $k_i=c_i$, $y_3=c_3 y_1, y_2=c_2 c_3 y_1, x_2=c_2 x_1, x_3=c_2 c_3 x_1$. Denoting $x_1=x, y_1=y$, we get the final expression of $Q$:

\begin{eqnarray} \label{Q33}
	\frac{(\alpha+\beta x+\gamma y) (\alpha c_1 c_2 +\beta c_2 x+\gamma y) (\alpha c_1+\beta c_1 c_2 x+\gamma y)}{(\alpha c_1+\beta x+\gamma y) (\alpha+\beta c_2 x+\gamma y) (\alpha c_1 c_2 +\beta c_1 c_2 x+\gamma y)}
\end{eqnarray}

Finally, we require $Q\equiv1$ when $3\alpha-\beta=0$:

\begin{eqnarray}
	1\equiv \prod_{i=1}^{k}\frac{\alpha(n_i+3x_i)+y_i\gamma}{\alpha(c_i n_{p(i)}+3 c_i x_i)+y_i\gamma}
\end{eqnarray}

which leads to
\begin{eqnarray}
	c_i n_{p(i)}+3x_i= r_i(n_{v(i)}+3x_{v(i)}) \\
	y_i=r_i y_{v(i)} \\
	\prod_{i=1}^k r_i=1 \\
	i=1,2,...,k
\end{eqnarray}
for some permutation $v(i)$. 

So, altogether we have

{\bf Proposition A.C.IV.} {\it The general expression for a non-uniqueness factor $Q$ for universal dimensions, which is equal to $1$ on each of the
 $\alpha=0 (sl), \beta=0 (so)$, $\gamma=0 (exc)$, and $3\alpha-\beta=0 (sp)$ lines in Vogel's
plane, writes as follows:

\begin{eqnarray}\label{Qx27b}
	Q=\prod_{i=1}^{k}\frac{n_i\alpha+x_i\beta+y_i\gamma}{ k_i n_{s(i)}\alpha+ x_i\beta+y_{i}\gamma}=\prod_{i=1}^{k}\frac{ n_i\alpha+x_i\beta+y_i\gamma}{ c_i n_{p(i)}\alpha+ x_i\beta+y_{i}\gamma}  
\end{eqnarray}
	
with parameters, satisfying the following equations

\begin{eqnarray}	\label{4lineBeq}
 	x_i=c_ix_{p(i)} \\
	y_i=k_iy_{s(i)} \\   \label{end1} 
	k_i n_{s(i)}=c_i n_{p(i)} \\  \label{4-th-eq}
	y_i=r_i y_{v(i)} \\           \label{end2}
	c_i n_{p(i)}+3x_i= r_i(n_{v(i)}+3x_{v(i)}) \\
	c_1c_2...c_k=1 \\
	k_1k_2...k_k=1 \\   \label{4lineEeq}
	r_1 r_2...r_k=1 
\end{eqnarray}

for some permutations $s(i), p(i), v(i), \, i=1,2...k$.}

{\bf Remark.} As follows from the example above, one can get a trivial ($Q=1$) or a non-trivial non-uniqueness factor $Q$ depending on the particular choice of permutations.

\chapter{Vogel's universality and dualities in the physical theories. The universal-type partition functions of the refined Chern-Simons theories with arbitrary
gauge groups}

In this chapter, we generalize the universal partition function of the Chern-Simons theory on $S^3$ to the {\it refined} case, and present its explicit expression for an arbitrary gauge group.

Using this form of the partition function we show that the previously known Krefl-Schwartz representation of the partition function of the
refined Chern-Simons on $S^3$ can be generalized to all simply laced algebras.

Then, for all non-simply laced gauge algebras, we derive similar representations of that partition function, which makes it possible to transform it into a product of 
multiple sine functions aiming at the further establishment of duality with the refined topological strings.
\newpage
\section{Universal partition function of the Chern-Simons on $S^3$}
The partition function of Chern-Simons (CS) theory on a three-dimensional sphere $S^3$, first calculated in \cite{W1} (see below (\ref{Zk})), is presented
in a universal form in \cite{MV, M13}, which means that alternative to the pure Lie algebra data - roots, invariant scalar product, etc., it is now
expressed in terms of the so-called Vogel's universal parameters $\alpha, \beta, \gamma$ \cite{V,V0}, (see Vogel's table \ref{tab:V1}).
The advantage of this representation is that it is very convenient for the further transformation of the abovementioned partition function into the Gopakumar-Vafa partition function of topological strings, 
as shown in \cite{M13,SGV} for CS theory with the classical groups. In the recent work, \cite{RM20} this transformation has been
 extended to the CS with the exceptional groups, meaning that the partition function of CS on $S^3$ with an exceptional gauge group 
 has been presented in the form of a partition function of a specific refined topological string. This should be considered as a step towards the establishment of the duality of the corresponding theories.
 The fact that all exceptional algebras (actually all algebras in $E_8$ row of the Freudenthal magic square) belong to a line in Vogel's plane – the so-called Deligne's line, is exploited in that work. 
Deligne \cite{Del} suggested that all the points on that line make up the so-called series of Lie algebras, which was partially confirmed in \cite{Cohen}.

The main features of the presentation of the partition function discovered in \cite{MV,M13} have been extended to include the partition function of the refined CS theory on a $3d$
 sphere\footnote{We will omit to mention $S^3$ from now on, 
since we do not consider theories on other manifolds in this paper} for $A_n$ and $D_n$ algebras in \cite{KS}. It has also been shown to be very convenient for the derivation of the partition
 function of the dual refined topological strings in \cite{KM}. 
In the same work the non-perturbative corrections to the partition function of topological strings, derived from the universal CS partition function \cite{M13} 
(with $A_n$ gauge algebra), have been shown to coincide with those derived in \cite{HMMO13,H15} directly in the topological string theory framework,
 thus extending the CS/topological strings duality to the non-perturbative domain. 

The natural development of these investigations would be the extension of the universal-type representation of the refined CS theories with $A_n$ and $D_n$ algebras to the remaining 
algebras: the simply-laced $E_n$ and the non-simply laced classical ($B_n, C_n$) and exceptional ($F_4, G_2$) algebras with the final aim of setting up a connection of the corresponding 
refined CS theories with some (refined) topological strings. 

In the given thesis the first step has been taken. Here for the first time we present universal-type representations of the partition function of the refined CS theory  
with each of the remaining gauge groups. 

Below we present a new representation of the partition function of the refined CS theory for {\it all} simple Lie algebras. 
It is based on a new Lie-algebraic identity for the determinant of the symmetrized Cartan matrix (the refined version of that in the \cite{KP}) and generalizes a feature of the non-refined 
theory, exploited in \cite{M13} earlier, which states that the partition function is equal to $1$ when the coupling of CS is $0$. 

Then, we rewrite this partition function in a "universal" form, which means that instead of the roots and other standard characteristics of a gauge algebra it
is now expressed in terms of Vogel's parameters. Simultaneously, the range of the refinement parameter is extended to include non-integer values, too. 

\section{{\it Refined} CS theory on $S^3$}
The partition function of CS theory on $S^3$ sphere was given in Witten's seminal paper \cite{W1} as the $S_{00}$ element of the $S$ matrix of modular transformations. 
For an arbitrary gauge group, it is (see, e.g. \cite{DiF, MV})

 \begin{eqnarray} \label{Zk}
 	Z(k)= Vol(Q^{\vee})^{-1} (k+h^{\vee})^{-\frac{r}{2}} \prod_{\alpha_+} 2\sin \pi \frac{(\alpha,\rho)}{k+h^{\vee}}
 \end{eqnarray}
Here the so-called minimal normalization of the invariant scalar product $(,)$ in the root space is used, which implies that the square of the long roots equals $2$. 
Other notations are: $Vol(Q^{\vee})$ is the volume of the fundamental domain of the coroot lattice $Q^\vee$, the integer $k$ is the CS coupling constant, 
$h^\vee$ is the dual Coxeter number of the algebra, $r$ is the rank of the algebra,  
the product is taken over all positive roots $\alpha_+$. 

 $Vol(Q^{\vee})$ is equal to the square root of the determinant of the matrix of scalar products of the simple co-roots. For the simply laced algebras, in the minimal normalization,
  it is equal to the square root of the determinant of the Cartan matrix, accordingly:   

\begin{eqnarray}
	Vol(Q^{\vee})= (\det (\alpha_i^\vee,\alpha_j^\vee))^{1/2} \\
	\alpha_i^\vee=\alpha_i \frac{2}{(\alpha_i,\alpha_i)}, \,\,
	i=1,...,r
\end{eqnarray}

The same formula for the partition function, rewritten in an arbitrary normalization of the scalar product \cite{MV}, is

\begin{eqnarray} \label{Zkap}
	Z(\kappa)= Vol(Q^{\vee})^{-1} (\delta)^{-\frac{r}{2}} \prod_{\alpha_+} 2\sin \pi \frac{(\alpha,\rho)}{\delta}
\end{eqnarray}

where $k$ is now replaced by $\kappa$, $h^\vee$ by $t$, and  $\delta=\kappa+t$. 
In this form the r.h.s. is invariant w.r.t. the simultaneous rescaling of the scalar product, $\kappa$, and $t$ (and hence $\delta$). In the minimal normalization they accept their usual values in (\ref{Zk}).

In \cite{M13} it was noticed that from this formula for the partition function one can derive an interesting closed expression for $Vol(Q^{\vee})$, 
which agrees with that in the Kac-Peterson's paper \cite{KP}, (see eq. (4.32.2)), provided

 \begin{eqnarray} \label{Z0}
  Z(0)=1
  \end{eqnarray}

This equality is completely natural from the physical point of view. Indeed, the CS theory is based on the
unitary integrable representations of affine Kac-Moody algebras. At a given $k$ there is a finite number of such representations, and at $k=0$ there is no any non-trivial one. 

So, from (\ref{Zkap}) and (\ref{Z0}) we have

\begin{eqnarray} \label{voly=1}
	 Vol(Q^{\vee})= t^{-\frac{r}{2}} \prod_{\alpha_+} 2\sin \pi \frac{(\alpha,\rho)}{t}
\end{eqnarray}
 which, as mentioned, agrees with \cite{KP}. Below we generalize this equation by inclusion of a refinement parameter.

The generalization of the usual CS to the refined CS theory is given in \cite{AS11, AS12a, AS12}. 
It is based on Macdonald's deformation of e.g. the Shur polynomials, and other  "deformed" formulae, given in \cite{Mac1, Mac2, Mac3}. 
In a nutshell, Macdonald's deformation yields the deformed $S$ and $T$ matrices of the modular transformations, and since these matrices define all observables in CS theory,
one can naturally consider the "deformed" or the refined versions of all observables, i.e. the link/manifold invariants.

Particularly, the partition function of the refined CS theory on $S^3$ is given \cite{AS11} by the $S_{00}$ element of the refined $S$-matrix.
In \cite{AS11} instead of an orthonormal basis an orhogonal is sometimes used. We will use the orthonormal one only (as in \cite{KS}), so that
 there is no difference between e.g. $S_{00}$ and $S_0^0$.

We suggest the following expression for $S_{00}$ for the refined CS theory:

 \begin{eqnarray}\label{refCS}
	Z(\kappa,y)= Vol(Q^{\vee})^{-1} \delta^{-\frac{r}{2}} \prod_{m=0}^{y-1} \prod_{\alpha_+} 2\sin \pi \frac{y(\alpha,\rho)-m (\alpha,\alpha)/2}{\delta}
\end{eqnarray}

We assume that now $\delta=\kappa+y t$, $y$ is the refinement parameter, which we consider to be a positive integer at this stage. 

Although we could not find the $Z(\kappa,y)$ in this exact form in literature, the expression (\ref{refCS}) complies with the known formulae in different limits, 
e.g. at $y=1$ it yields the corresponding formula for the non-refined case (\ref{Zkap}). It also coincides with the corresponding formulae for the refined CS theory 
in \cite{AS11,AS12,KS} for $A_n, D_n$ algebras. 
The coefficient $(\alpha,\alpha)/2$ in front of the summation parameter $m$ coincides with that in the constant term formulae in \cite{Cher1,Cher2}.
 Actually, for non-simply laced algebras one can introduce two refinement parameters, one for each length of the roots, see e.g. \cite{Cher1,Cher2}. 
 However, we did not try to introduce a second parameter (and also are not aware of the physical interpretation of it), so below we consider them to be coinciding,
  so that we always have one refinement parameter.

The latter expression of the partition function is supported by the key feature of (\ref{refCS}): at $\kappa=0$ the equality $Z(0,y)=1$ holds, 
which is ensured by the following generalization of the formula (\ref{voly=1}) for the same object  $Vol(Q^{\vee})$:

\begin{eqnarray} \label{vol}
	 Vol(Q^{\vee})=  (ty)^{-\frac{r}{2}}  \prod_{m=0}^{y-1} \prod_{\alpha_+} 2\sin \pi \frac{y(\alpha,\rho)-m (\alpha,\alpha)/2}{ty}
\end{eqnarray}

For $A_{n}$ algebras this equality can be easily proved with the use of the following well-known identity, valid at an arbitrary positive integer $N$:
\begin{eqnarray}
	N=\prod_{k=1}^{N-1}2 \sin{\pi\frac{k}{N}}
\end{eqnarray}
Similarly it can be checked for all the remaining root systems. 

Next, with (\ref{vol}) taken into account, we obtain the following
expression of the partition function:

\begin{eqnarray}
	Z(\kappa,y)= \left(\frac{ty}{\delta}\right)^{\frac{r}{2}} \prod_{m=0}^{y-1} \prod_{\alpha_+} \frac{\sin \pi \frac{y(\alpha,\rho)-m (\alpha,\alpha)/2}{\delta}}{\sin \pi \frac{y(\alpha,\rho)-m (\alpha,\alpha)/2}{ty}}
\end{eqnarray}

which explicitly satisfies $Z(0,y)=1$, since $\delta=t y$ at $\kappa=0$.

\section{Integral representation of partition function for the refined CS theories}
In order to write the integral representation of the partition function presented above, we apply the transformation introduced in \cite{M13}. 
We transform each of the sines into a pair of Gamma-functions by the following 
well-known identity

\begin{eqnarray}
	\frac{\sin \pi z}{\pi z} = \frac{1}{\Gamma(1+z) \Gamma(1-z)}
\end{eqnarray}

and make use of the integral representation of (the logarithm of) the $\Gamma$ function:

\begin{eqnarray}
	\ln\Gamma(1+z)=\int_{0}^{\infty}dx \frac{e^{-zx}+z(1-e^{-x})-1}{x(e^{x}-1)}
\end{eqnarray}

Let us rewrite the partition function in the following form:

\begin{eqnarray}
Z(\kappa,y)=\left(\frac{ty}{\delta}\right)^{y\frac{dim-r}{2}+\frac{r}{2}}	\prod_{m=0}^{y-1} \prod_{\alpha_+} \frac{\sin \pi \frac{y(\alpha,\rho)-m (\alpha,\alpha)/2}{\delta}}{\pi \frac{y(\alpha,\rho)-m (\alpha,\alpha)/2}{\delta}} \times \\
 \prod_{m=0}^{y-1} \prod_{\alpha_+}\frac{\pi \frac{y(\alpha,\rho)-m (\alpha,\alpha)/2}{ty}}{\sin \pi \frac{y(\alpha,\rho)-m (\alpha,\alpha)/2}{ty}} \equiv\\
 \left(\frac{ty}{\delta}\right)^{y\frac{dim-r}{2}+\frac{r}{2}} Z_1 Z_2
\end{eqnarray}

and apply the abovementioned transformation to the first couple of products of sines (then similarly to the second couple of products):

\begin{eqnarray}
\ln Z_1=	\ln \left( \prod_{m=0}^{y-1} \prod_{\alpha_+} \frac{\sin \pi \frac{y(\alpha,\rho)-m (\alpha,\alpha)/2}{\delta}}{\pi \frac{y(\alpha,\rho)-m (\alpha,\alpha)/2}{\delta}} \right)= \\
	-\int_{0}^{\infty} \frac{dx}{x(e^{x}-1)} \sum_{m=0}^{y-1}\sum_{\alpha_+}\left( e^{x\frac{y(\alpha,\rho) -m (\alpha,\alpha)/2}{\delta}}+e^{-x\frac{y(\alpha,\rho) -m (\alpha,\alpha)/2}{\delta}}-2\right)
\end{eqnarray}

Let us introduce the following function for any simple Lie algebra $X$ of the rank $r$:

\begin{eqnarray}\label{FX2}
		F_X(x,y)=    r+	\sum_{m=0}^{y-1}	\sum_{\alpha_{+}} \left( e^{x(y(\alpha,\rho) -m (\alpha,\alpha)/2)}+e^{-x(y(\alpha,\rho) -m (\alpha,\alpha)/2)}\right)
\end{eqnarray}

Then 

\begin{eqnarray}
	\sum_{m=0}^{y-1}	\sum_{\alpha_+}\left( e^{x(y(\alpha,\rho) -m (\alpha,\alpha)/2)}+e^{-x(y(\alpha,\rho) -m (\alpha,\alpha)/2)}-2\right)= \\
	F_X(x,y) - r-y(dim-r)
\end{eqnarray}
and  $\ln Z_1$ becomes

\begin{eqnarray}
	\ln Z_1= -\int_{0}^{\infty} \frac{dx}{x(e^{x}-1)} \left( F_X\left( \frac{x}{\delta},y\right) - r-y(dim-r)\right)
\end{eqnarray}

A similar transformation applies to $\ln Z_2$:

\begin{eqnarray}
	\ln Z_2= \int_{0}^{\infty} \frac{dx}{x(e^{x}-1)} \left( F_X\left( \frac{x}{ty},y\right) - r-y(dim-r)\right)
\end{eqnarray}

and $\ln Z$ takes the form

\begin{eqnarray} \label{Zintg}
	\ln Z= \\
	\frac{1}{2} (y(dim-r)+r)  \ln \left(\frac{ty}{\delta}\right)  +\int_{0}^{\infty} \frac{dx}{x(e^{x}-1)} \left( F_X\left( \frac{x}{t y},y\right) - F_X\left( \frac{x}{\delta},y\right) \right)
\end{eqnarray}

Finally, one can further transform this formula into an expression, similar to the one derived in \cite{KM} for the non-refined theories. 

Let us make the $x\rightarrow ty x/\delta$ rescaling  in $\ln Z_2$, so that
\begin{eqnarray}
	\ln Z_2=\int_0^\infty \frac{dx}{x(e^{ty x/\delta}-1)} \left(F_X\left(\frac{x}{\delta},y\right)- r-y(dim-r)\right)\,.
\end{eqnarray}

Using the relation
\begin{eqnarray} \label{cotmcotId}
	\frac{1}{e^{b x}-1}-\frac{1}{e^{a x} -1} = \frac{e^{a x}-e^{b x}}{(e^{a x}-1)(e^{b x }-1)}=\frac{\sinh\left(\frac{x(a-b)}{2}\right)}{2\sinh\left(\frac{x a}{2}\right)\sinh\left(\frac{x b}{2}\right)}\,,
\end{eqnarray}
and making use of that the combined integrand is even under $x\rightarrow -x$, 
we can write $\ln Z$ as 
\begin{eqnarray} \label{FVpreFinal}
	\ln Z=\frac{ r+y(dim-r)}{2} \log \left(ty/\delta\right)- \\
	\frac{1}{4}\int_{R_+} \frac{dx}{x} \frac{\sinh\left(x(ty-\delta)\right)}{\sinh\left(x ty \right)\sinh\left(x \delta\right)} \left(F_X(2x,y)- r-y(dim-r)\right)\,,
\end{eqnarray}

where the integration range passes the origin by an infinitesimal semi-circle in the upper (or lower) half of the complex plane. We denote the 
corresponding contour $R_+ $. We also take $x\rightarrow 2x \delta$. 

Due to the following identity
\begin{eqnarray}
	\frac{1}{4}\int_{R_+} \frac{dx}{x} \frac{\sinh\left(x(t-\delta)\right)}{\sinh\left(x t\right)\sinh\left(x \delta\right)}=-\frac{1}{2}\log\left(\frac{t}{\delta}\right)\,,
\end{eqnarray}
proved in \cite{KS} the integral of the $r+y(dim-r)$ term in fact cancels against the $\log$ term in (\ref{FVpreFinal}), so that we obtain the final expression:

\begin{eqnarray}\label{FintRep}
	\ln Z=-\frac{1}{4}\int_{R_+} \frac{dx}{x} \frac{\sinh\left(x(ty-\delta)\right)}{\sinh\left( x ty \right)\sinh\left(x \delta\right)} F_X(2x,y)
\end{eqnarray}

With a corresponding representation of $F_X(x,y)$ functions as a ratio of polynomials over $q=\exp x$, which is shown below in section 6, the latter 
expression can be transformed into a product of multiple sine functions (see, e.g. \cite{SGV, KM}), which then hopefully will make the further correspondence of it with the 
refined topological strings possible.

\section{The partition function of the refined CS for simply laced algebras}
In the non-refined case, i.e. at $y=1$ (when the sum over $m$ disappears), the partition function rewrites in terms of the Vogel's universal parameters.
The corresponding $F_X(x,1)$ coincides with the quantum dimension of the adjoint representation, which is the character
 $\chi_{ad}(x\rho)$, restricted to the $x\rho$ line, collinear with the Weyl vector $\rho$: 
\begin{eqnarray}
	F_X(x,1)=r+\sum_{\alpha_+}\left( e^{x(\alpha,\rho) }+e^{-x(\alpha,\rho)}\right)=	\chi_{ad}(x\rho)
\end{eqnarray}

So $F_X(x,y)$ can be called the refined quantum dimension.

The quantum dimension of the adjoint representation has been presented in the universal form in \cite{W3,MV}: 

\begin{eqnarray}\label{dim}
		\chi_{ad}(x\rho) \equiv
		f(x)		&=& \frac{\sinh(x\frac{\alpha-2t}{4})}{\sinh(x\frac{\alpha}{4})}\frac{\sinh(x\frac{\beta-2t}{4})}{\sinh(x\frac{\beta}{4})}\frac{\sinh(x\frac{\gamma-2t}{4})}{\sinh(x\frac{\gamma}{4})}
	\end{eqnarray}

Note that the notation $\alpha$ is used both for the root(s) of an algebra and for one of the Vogel's parameters. Since these objects are very different, hopefully no interpretation problem
appears.

Finally, the partition function in the non-refined case takes the following universal form

\begin{eqnarray}
Z(\kappa)=	Z(\kappa,1)= \left(\frac{t}{\delta}\right)^{\frac{dim}{2}} exp \left( -\int_{0}^{\infty} \frac{dx}{x(e^{x}-1)} \left( f\left(\frac{x}{\delta}\right)-f\left(\frac{x}{t}\right)  \right)      \right)
\end{eqnarray}
first given in \cite{M13}

In the refined case there is not a similar universal answer for the double sum over $m$ and $\alpha_+$. However, for $A_n$ and $D_n$ algebras 
Krefl and Schwarz \cite{KS} have made a statement, equivalent to 

\begin{eqnarray} \label{fxd}
\sum_{m=0}^{y-1}	\sum_{\alpha_+}\left( e^{x(y(\alpha,\rho) -m (\alpha,\alpha)/2)}+e^{-x(y(\alpha,\rho) -m (\alpha,\alpha)/2)}-2\right)=  \\ f\left(x,y\right)-dim(y) 
\end{eqnarray}

with
\begin{eqnarray}
	f(x,y)= \frac{\sinh(x\frac{\alpha-2ty}{4})}{\sinh(x\frac{\alpha}{4})}\frac{\sinh(xy\frac{\beta-2t}{4})}{\sinh(xy\frac{\beta}{4})}\frac{\sinh(xy\frac{\gamma-2t}{4})}{\sinh(xy\frac{\gamma}{4})},  \\
	dim(y)= \lim_{x\rightarrow 0} f(x,y) = y \,\, dim X-(y-1)\frac{(\beta-2t)(\gamma-2t)}{\beta \gamma} \\
	f(x,1)=f(x)
\end{eqnarray}
where it is assumed that $\alpha$ is the only negative parameter (equal to $-2$ in the minimal normalization of the scalar product).

The  $dim(y)$ can be further transformed. Indeed, consider the dimension formula for the simple Lie algebras:

\begin{eqnarray}
	dim=\frac{(\alpha-2t)(\beta-2t)(\gamma-2t)}{\alpha \beta \gamma} =\frac{\alpha-2t}{\alpha}  \frac{(\beta-2t)(\gamma-2t)}{ \beta \gamma}
\end{eqnarray}
In the last expression both fractions are independent of normalization. In the minimal normalization the
 first fraction is equal to $1+h^\vee$ (where $h^\vee$ is the dual Coxeter number) so we conclude that the second one is the rank of the algebra

\begin{eqnarray}
	 \frac{(\beta-2t)(\gamma-2t)}{ \beta \gamma} =r
\end{eqnarray}

since the following relation holds for all simply-laced algebras:

\begin{eqnarray}
	dim = (1+h^\vee) r
\end{eqnarray}

Finally, we have 
\begin{eqnarray}
	dim(y)=y(dim-r)+r
\end{eqnarray}

With this relation we see that (\ref{fxd}) is equivalent to 

\begin{eqnarray} \label{F=f}
	F_X(x,y)=f(x,y)
\end{eqnarray}

Then, with the use of (\ref{fxd}), the partition function (\ref{Zintg}) becomes:

\begin{eqnarray} \label{zky}
	Z(\kappa,y)= \left(\frac{ty}{\delta}\right)^{y\frac{dim-r}{2}+\frac{r}{2}} exp \left( -\int_{0}^{\infty} \frac{dx}{x(e^{x}-1)} \left( f\left(\frac{x}{\delta},y\right)-f\left(\frac{x}{ty},y\right)  \right)      \right)
\end{eqnarray}

As mentioned, this result has first been proven for $A_n$ and $D_n$ series in \cite{KS}. 
 In the next section we prove the relation (\ref{fxd}) (and hence (\ref{F=f}))  for the remaining simply laced algebras, namely, for $E_n$, thus generalizing (\ref{zky}) to all the
 simply-laced simple Lie algebras. 

\section{On the universality of the refined CS for all simply-laced algebras}
In this section, we prove the statement of the previous section, i.e. generalize the relation (\ref{fxd}) to all simply-laced algebras. 
We claim that

\begin{eqnarray}\label{FXf}
	F_X(x,y)=f(x,y)
\end{eqnarray}
for any simply-laced  Lie algebra $X$. 

Take e.g. the $E_6$ algebra, for which the corresponding universal parameters in the minimal normalization are:
 $\alpha=-2, \beta=6, \gamma=8, t=12$.
We should calculate the sum  

\begin{eqnarray}\label{F}
F_{E_6} (x,y)= 6+	\sum_{m=0}^{y-1}	\sum_{\alpha_{+}} e^{x(y(\alpha,\rho) -m)}+e^{-x(y(\alpha,\rho) -m) }
\end{eqnarray}

First note the number of roots $n_L$ with a given height $L=(\alpha,\rho)$ among all roots. The set of couples $(L,n_L)$ with a non-zero $n_L$ is 

\begin{eqnarray}
 (-11,1),(-10,1),(-9,1),(-8,2),(-7,3),(-6,3),(-5,4),(-4,5),\\ \nonumber (-3,5),(-2,5),(-1,6),(0,6),(1,6),(2,5),(3,5),(4,5),(5,4),(6,3), \\ \nonumber
 (7,3),(8,2),(9,1),(10,1),(11,1)
\end{eqnarray}
which of course is symmetric w.r.t. the $L \leftrightarrow -L$. We also include the element $(0,6)$ in this list, which is just the first term $6$ in (\ref{F}).
Then, using this data, we note that the sum in (\ref{F}) is given by  

\begin{eqnarray}
F_{E_6}=	\phi(11y)+\phi(8y)+\phi(7y)+\phi(5y)+\phi(4y)+\phi(y) \\
	\phi(n)=\sum_{i=-n}^n q^i = \frac{q^{2n+1}-1}{q^n (q-1)} \\
	q=e^x
\end{eqnarray}

Combining the sums $\phi(11y)+\phi(8y)+\phi(5y)$ and $\phi(7y)+\phi(4y)+\phi(y)$, we get

\begin{eqnarray}
	\phi(11y)+\phi(8y)+\phi(5y)= \frac{(q^{9y}-1)(q^{5y+1}-q^{-11y})}{(q-1)(q^{3y}-1)} \\
	\phi(7y)+\phi(4y)+\phi(y) = \frac{(q^{9y}-1)(q^{y+1}-q^{-7y})}{(q-1)(q^{3y}-1)} \\
	F_{E_6}=\frac{(q^{9y}-1)}{(q-1)(q^{3y}-1)}(q^{4y}+1)(q^{y+1}-q^{-11y}) = \\
	\frac{(q^{9y}-1)(q^{8y}-1)(q^{y+1}-q^{-11y})}{(q-1)(q^{3y}-1)(q^{4y}-1)}
\end{eqnarray}
which can be easily checked to coincide with $f(x,y)$ for the universal parameters corresponding to $E_6$ algebra.

Literally similar calculations can be carried out for the remaining $E_7, E_8$ algebras, as well as for Krefl-Schwarz cases $A_n, D_n$, leading to the same conclusion. 

\section {A universal-type presentation of the partition function for the non-simply laced algebras} 

Equations (\ref{fxd}), (\ref{F}) do not hold for the non-simply laced algebras. However, one can present the corresponding sum in a 
similar form, appropriate for the further duality considerations \cite{M13, RM20, KM}. The latter means that it can be presented as a ratio of a sum of exponents of 
$x$ (i.e. powers of $q=\exp x$) in the numerator and some sines in the denominator. We aim to represent $F_X$ as follows: 

\begin{eqnarray}\label{FX}
	F_X=    r+	\sum_{m=0}^{y-1}	\sum_{\alpha_{+}} \left( e^{x(y(\alpha,\rho) -m (\alpha,\alpha)/2)}+e^{-x(y(\alpha,\rho) -m (\alpha,\alpha)/2)}\right) 
	= \frac{A_X}{B_X}
\end{eqnarray}
where $X$ denotes an algebra of type $B,C,F$ or $G$, $r$ is its rank, $B_X$ is a product of a number of terms of the form $q^a-1$, and $A_X$ is a polynomial in $q$. 

One subtlety regarding the formulae (\ref{FX}), which makes them different from the (\ref{FXf}), is that in (\ref{FX}) one should explicitly mention the
 normalization of the scalar product. In (\ref{FXf}) both sides are invariant under the rescaling of the scalar product in the l.h.s.
  (with the corresponding rescaling of the universal parameters in the r.h.s.), and the simultaneous appropriate rescaling of $x$. 
  However, in (\ref{FX}) a similar rescaling of the scalar product and $x$ leaves invariant only the l.h.s., whilst the ratio $A_X/B_X$ in the r.h.s is dependent only on $x$,
   thus changing under its rescaling. This means that when substituting the r.h.s. of (\ref{FX}) into the partition function \ref{Zintg} one should 
take the parameters $t$ and $\delta$ in the same normalization. The normalizations below are chosen to avoid the appearance of fractional powers of $q$. 

Now we present $F_X$ for all non-simply laced algebras. 

Let us consider the $B_n$ algebras. Normalization corresponds to $\alpha=-4$, i.e. the square of the long root is $4$.
 The corresponding representation we mentioned above is

\begin{eqnarray}\label{F2}
	F_{B_n}(x,y)= 
	\frac{A_{B_n}}{B_{B_n}} \\
	A_{B_n}=q^{4 n y+2}+q^{-4 (n-1) y}+   \\
	(q+1) \left(q^y-1\right) \left(q^{2 y}+1\right) \left(q^{2 n y}-1\right) \left(q^{y-2 n y}+q\right)-q^{4 y}-q^2 \\ 
	B_{B_n}=\left(q^2-1\right) \left(q^{4 y}-1\right),
\end{eqnarray}

For the $C_n$ algebras we also choose the same normalization with the square of the long root being $4$. Then $F_X$ writes as

\begin{eqnarray}
	F_{C_n}= \frac{A_{C_n}}{B_{C_n}} \\
	B_{C_n}=(q^2-1)  \left(q^{2y}-1\right) \\
	A_{C_n}=	(q+1) q^y \left(q^{2 n y}-1\right) \left(q^{2 n y+1}-1\right)+\\
	\left(q^{2 y}-1\right) \left(q^{n y}-1\right) \left(q^{n y+1}-1\right) \left(q^{2 n y+1}-1\right)
\end{eqnarray}

For the $F_4$, with the same normalization, we have 

\begin{eqnarray}
	F_{F_4}= \frac{A_{F_4}}{B_{F_4}} \\
	B_{F_4}=(q^2-1)   \\
	A_{F_4}=q^{-16 y} \left(q^{2 y}+1\right) \left(-q^{2 y}+q^{4 y}+1\right) \left(q^{12 y+1}-1\right) \times \\ \left(q^{5 y+1}-q^{8 y+1}+q^{9 y+1}+q^{14 y+1}+q^{5 y}-q^{6 y}+q^{9 y}+1\right)
\end{eqnarray}

For the $G_2$ we use the normalization corresponding to the square of the long root to be equal to $6$. The corresponding $F_{G_2}$ function is

\begin{eqnarray}
	F_{G_2}= \frac{A_{G_2}}{B_{G_2}} \\
	B_{G_2}=q^3-1   \\
	A_{G_2}=q^{-9 y} \left(q^{6 y+1}-1\right) \times \\
	\left(q^{4 y+1}+q^{8 y+1}+q^{4 y+2}-q^{6 y+2}+q^{8 y+2}+q^{12 y+2}+q^{4 y}-q^{6 y}+q^{8 y}+1\right)
\end{eqnarray}

\newpage

\chapter{Summary}
The puzzle of Vogel's universal description of simple Lie algebras has been filled in by new pieces in this thesis.

New universal formulae for quantum dimensions of $(X_2)^k(g)^n$ representations and
for the second Casimir eigenvalues on them have been discovered. 

A new property of this new quantum dimension formula, that is linear resolvability has been revealed.

A remarkable connection between the universal formulae for the simple Lie algebras and some geometrical configurations of points and lines has been formulated.

Finally, a step forward has been taken in the understanding of the dualities between the refined Chern-Simons theories based on arbitrary simple gauge groups and some topological strings.

Construction of the refined partition functions for all simple gauge groups, ready to be transformed into Gopakumar-Vafa type partition functions of topological strings, has been carried out.

The more you dig in, the more there is to explore. The direct routes of the development of the present results would be the following.

First is the possible derivation of new universal formulae both in the scope of the representation theory of the simple Lie algebras and the physical theories based on them.
In fact, the complete universality of all representations appearing in the decompositions of the fourth and higher tensor powers of the adjoint representation remains open at the moment.
Although, identifyng several universal representations in this higher powers the newly derived series of dimensions does not cover all of them. 
So, a lot of interesting questions in this direction still remain open.

Second is the final solution to the problem of the uniqueness of universal dimensions.
The geometrical interpretation of this problem provided us with a quite solid toolbox taken from the theory of configurations of points and lines.
Particularly, now we understand that the existence of a realizable $(144_3, 36_{12})$ configuration is essential for the complete solution of the problem.
Hopefully, incomprehensive study of this configuration will foster collaborations between professionals working in various fields for studying this intriguing puzzle.
Evidently, any success in this direction will have considerable potential for uncovering new properties of universal formulae, consequently, for deepening the
understanding of Vogel's theory.

And third is the complete understanding of the Chern-Simons/topological strings dualities.
Vogel's universality demonstrated itself as an outstanding tool for studying the dualities of physical theories based on any simple gauge groups, more importantly – the exceptional ones.
Particularly, the partition functions presented in Chapter 5 are very promising for identifying possible dualities between the refined Chern-Simons theories and
some topological strings. The complete solution to this problem would definitely be an impressive result in the current understanding of nature. To this end, 
considerable efforts are now being taken in this direction \cite{AKM-21-1, AKM-21-2}, and some preliminary results have already been obtained shortly after the
 preparation of this thesis.
We hope that the complete understanding of these dualities is not that far and will be reached in the near future.
\newpage

\cleardoublepage

\end{document}